\documentclass[a4paper,fleqn,usenatbib]{mnras}
\usepackage[T1]{fontenc}
\usepackage{ae,aecompl}

\usepackage{graphicx}	
\usepackage[export]{adjustbox}
\usepackage{amsmath}	
\usepackage{amssymb}	
\usepackage{subfig}
\usepackage{fancyhdr}
\usepackage{url}
\usepackage{float}
\usepackage{multirow}
\usepackage{appendix}
\usepackage[portuges]{babel}
\usepackage[utf8]{inputenc}
\usepackage{framed}
\usepackage{tcolorbox}
\usepackage{adjustbox}
\usepackage{scalerel}
\usepackage{tikz}
\usetikzlibrary{svg.path}

\definecolor{orcidlogocol}{HTML}{A6CE39}
\tikzset{
  orcidlogo/.pic={
    \fill[orcidlogocol] svg{M256,128c0,70.7-57.3,128-128,128C57.3,256,0,198.7,0,128C0,57.3,57.3,0,128,0C198.7,0,256,57.3,256,128z};
    \fill[white] svg{M86.3,186.2H70.9V79.1h15.4v48.4V186.2z}
                 svg{M108.9,79.1h41.6c39.6,0,57,28.3,57,53.6c0,27.5-21.5,53.6-56.8,53.6h-41.8V79.1z M124.3,172.4h24.5c34.9,0,42.9-26.5,42.9-39.7c0-21.5-13.7-39.7-43.7-39.7h-23.7V172.4z}
                 svg{M88.7,56.8c0,5.5-4.5,10.1-10.1,10.1c-5.6,0-10.1-4.6-10.1-10.1c0-5.6,4.5-10.1,10.1-10.1C84.2,46.7,88.7,51.3,88.7,56.8z};
  }
}

\newcommand\orcidicon[1]{\href{https://orcid.org/#1}{\mbox{\scalerel*{
\begin{tikzpicture}[yscale=-1,transform shape]
\pic{orcidlogo};
\end{tikzpicture}
}{|}}}}

\usepackage{hyperref} 

\title[Asteroid triple system 2001 SN$_{263}$: surfaces characteristics and dynamical environment]
{Asteroid triple system 2001 SN$_{263}$: surfaces characteristics and dynamical environment}

\author[O. C. Winter; G. Valvano; T. S. Moura; G. Borderes-Motta; A. Amarante; R. Sfair]
        {O. C. Winter$^{1}$\thanks{E-mail:  othon.winter@unesp.br}\orcidicon{0000-0002-4901-3289}\,
        G. Valvano$^{1}$\thanks{E-mail:  giulia.valvano@unesp.br}\orcidicon{0000-0002-7905-1788}\,
        T. S. Moura$^{1}$\thanks{E-mail: santos.moura@unesp.br}\orcidicon{0000-0002-3991-8738}\,
        G. Borderes-Motta$^{3}$\thanks{E-mail: gabriel.borderes@uc3m.es}\orcidicon{0000-0002-4680-8414}\,
    \newauthor
        A. Amarante$^{2}$\thanks{E-mail: andre.amarante@ifsp.edu.br}\orcidicon{0000-0002-9448-141X}\,
        R. Sfair$^{1}$\thanks{E-mail: rafael.sfair@unesp.br}\orcidicon{0000-0002-4939-013X}\
\\
$^{1}$ Grupo de Din\^amica Orbital e Planetologia, S\~ao Paulo State University, Guaratinguet\'{a}, CEP 12516-410, 
  S\~{a}o Paulo, Brazil\\
$^{2}$ Laboratório Maxwell, Instituto Federal de Educação, Ciência e Tecnologia de São Paulo - IFSP, Cubatão, CEP 11533-160, 
  S\~{a}o Paulo, Brazil\\
  $^{3}$ Bioengineering and Aerospace Engineering Department, Universidad Carlos III de Madrid, Leganés, 28911, Madrid, Spain}

\date{Accepted XXX. Received YYY; in original form ZZZ}

\pubyear{2019}

\begin{document}
\label{firstpage}
\pagerange{\pageref{firstpage}--\pageref{lastpage}}
\maketitle

\begin{abstract}
The (153591) 2001 SN$_{263}$ asteroid system, target of the first Brazilian interplanetary space mission, is one of the known three triple systems within the population of NEAs. One of the mission objectives is to collect data about the formation of this system. The analysis of these data will help in the investigation of the physical and dynamical structures of the components (Alpha, Beta and Gamma) of this system, in order to find vestiges related to its origin. In this work, we assume the irregular shape of the 2001 SN$_{263}$ system components as uniform density polyhedra and computationally investigate the gravitational field generated by these bodies. The goal is to explore the dynamical characteristics of the surface and environment around each component. Then, taking into account the rotational speed, we analyze their topographic features through the quantities geometric altitude, tilt, geopotential, slope, surface accelerations, among others. Additionally, the investigation of the environment around the bodies made it possible to construct zero-velocity curves, which delimit the location of equilibrium points. The Alpha component has a peculiar number of 12 equilibrium points, all of them located very close to its surface. In the cases of Beta and Gamma, we found four equilibrium points not so close to their surfaces. Then, performing numerical experiments around their equilibrium points, we identified the location and size of just one stable region, which is associated with an equilibrium point around Beta. Finally, we integrated a spherical cloud of particles around Alpha and identified the location on the surface of Alpha were the particles have fallen. 

\end{abstract}

\begin{keywords}
Near-Earth objects, asteroids: triple system: (153591) 2001 SN$_{263}$ -- methods: numerical --  celestial mechanics.
\end{keywords}

\section{Introduction}

Asteroids are remnant bodies from the beginning of the solar system, and, as such, they preserve information that might help to understand the planetary formation process better.
Primitive asteroids might harbor complex organic molecules and water  that could have been delivered to the early Earth through collisions, contributing to the origin of life on our planet \citep{Morb2000,izi2013}.

Collisions of asteroids with the Earth occurred in the past and certainly will happen again at a yet unknown moment in the future. Then, knowing the composition and the internal structure of the various types of asteroids becomes of fundamental importance 
for a well planned strategy to avoid the collision and/or reduce its consequences.

Beyond all that, asteroids are also getting the attention of the mining industry, since they could be profitably exploited \citep{lewis1996}.  Therefore, asteroids are on the top of the list of relevant subjects to be studied and explored.

With the current technological development, an excellent way to significantly deepen our knowledge about asteroids is through space missions. Thus, an increasing number of missions to NEAs is being planned or is underway these days \citep{haya2015,lau2017}. The easier (closer and cheaper) space mission targets are Near-Earth asteroids (NEAs), since they get close to Earth's  orbit periodically. 
 
Collisional and other disruptive mechanisms made many of the minor bodies to become multiple systems. Among the NEA population, there are 72 asteroids currently known as multiple systems. From those, there are only three known triple systems, namely (153591) 2001 SN$_{263}$, (136617)1994 CC, and  
15
 (3122) Florence. All the remaining  known multiple NEA systems are binary systems. 

The triple system 2001 SN$_{263}$ was chosen to be the target of the Brazilian space mission named ASTER \citep{su2010}. One of the main reasons for this choice is associated with the sizes of  the three bodies. The largest component of this system (called Alpha) has a diametre $D_{\alpha}\sim 2.5$ km, while the second largest component (Beta)   
has diametre $D_{\beta}\sim D_{\alpha}/3$ and the smallest one (Gamma) has $D_{\gamma}\sim D_{\alpha}/6$. In this case, the bodies have comparable sizes, none of them has a size hundreds of times larger than the other. If they are remnants of a parent body that was broken by a collision, for instance, one could expect to see part of the parent body core exposed on the surface of one of the components, allowing the spacecraft to easily get information about the internal structure of the parent body. Also, note that none of the components is too small to be explored by a spacecraft. The smaller component (Gamma) has a diametre larger than the mean diametre of (25143) Itokawa, the target of the successful  Hayabusa mission \citep{haya2015}.

On September 20th of 2001 the LINEAR survey discovered the asteroid (153591) 2001 SN$_{263}$ \citep{Stokes2000}. Radar observations made by  the Arecibo Observatory on February 12th of 2008, when it made its closest approach to Earth at a distance of 0.06558 au, revealed it is composed of three components \citep{Nolan2008}. Early spectroscopic observations indicated that it is a B-type asteroid in the Bus-DeMeo taxonomy \citep{DeMeo2009}.

Later, \citet{perna2014} investigated the surface composition of the asteroid 2001 SN$_{263}$ and compared its reflectance spectra with laboratory spectra of meteorites and minerals. They found a featureless convex spectrum (Polana- or Themis-like), confirming it as a B-type. The observations showed that, among the small  solar system bodies, the system has the bluest visible spectrum ever observed. An organic- and magnetite-rich surface composition, like that of heated CI carbonaceous chondrites, was suggested by the spectra. A large abundance of Phyllosilicates is possible, as well. They also found indications of a composition variety within the triple system and a coarse-grained surface.

A first work on the dynamical solutions of the triple system was made by \citet{Fang2011}.
They analyzed possible dynamic solutions in order to describe the properties, origin, and evolution of the system 2001 SN$_{263}$. The orbital evolution of this system was configured through numerical simulations using an N-body integrator package that considers the gravitational interactions between the three components. Then, \citet{Fang2011} obtained estimates of masses and orbital elements for the three bodies, as indicated in Table \ref{table: infos}. They also verified that the observed values of the smaller bodies' eccentricities and mutual inclinations could be easily generated by planetary close encounters. 

The announcement of the ASTER mission motivated studies to characterize regions of stability of the system \citep{Araujo2012, Araujo2015}. The knowledge of the structure of such regions is crucial in the design of the mission. Regions of stable prograde trajectories \citep{Araujo2012} might indicate the location of possible remnant debris of the system. Then, regions that are unstable for the prograde cases, but stable for the retrograde cases, would be safer to place a spacecraft in a retrograde trajectory \citep{Araujo2015}.

There are other works devoted to get information and look for orbits that might help in the definition of the spacecraft trajectories plan (see, for instance \citet{Bellerose2011}; \citet{Prado2014}; \citet{Masago2016}; \citet{Santos2018}).

More recently, \citet{Becker2015} analyzed a data set composed of 240 delay-Doppler images from radar observations of the Arecibo Observatory and light curves of different observatories of the system 2001 SN$_{263}$, and obtained a three-dimensional model that characterized the shape of the three bodies and relevant physical aspects. Initially, several models were investigated and studied to obtain a refined and detailed format of each object and coherent with radar images and light curves. After delimiting good approximations of shape, the period of rotation, and pole orientation, \citet{Becker2015} presented a model based on a polyhedron composed of 1148 vertices and 2292 faces that best described the irregular shape of each body of the system 2001 SN$_{263}$\footnote{PDS. Website: https://sbn.psi.edu/pds/shape-models/}. Moreover, these polyhedra models contributed to the calculation of some physical parameters, among them the equivalent diametre of Alpha, Beta, and Gamma whose values are $2.5 \pm 0.2$, $0.77 \pm 0.12$ km and $0.43 \pm 0.14$ km, respectively. \citet{Becker2015} also computed the densities for the three components of the system, see Table \ref{table: infos}, in relation to the masses derived by \citet{Fang2011}. Associating the low densities with the spectral observations of the triple asteroid, \citet{Becker2015} concluded that the objects of the system 2001 SN$_{263}$ are dark and belong to type B (carbonaceous), implying a porous internal structure, similar to rubble.

Therefore, besides the peculiarity of being a triple system of asteroids with interesting features, the system 2001 SN$_{263}$ is the target of a space mission \citep{su2010}, which motivated us to explore the dynamic environment and the topographic characteristics of each component.

Thus, in the current work, we use the detailed shape models presented in \citet{Becker2015} to make further explorations on this system. In the next section, we present the three-dimensional model of the shape of each object constituting the triple system 2001 SN$_{263}$, in addition to its physical characteristics. Section \ref{grav_geop} introduces the definition of gravitational potential and geopotential, based on the polyhedra method. The topographic features of the triple system are presented in relation to the geometry of the bodies, Section \ref{geom}, and in relation to the behavior of the geopotential on the surface of the bodies, Section \ref{geo}. The investigation of the environment around the Alpha, Beta, and Gamma bodies, including the location and stability of the equilibrium points, is discussed in Section \ref{dynamical environment}. 
Considering that small bodies, such as asteroids, are periodically impacted, in Section \ref{mapa} we integrated a cloud of massless particles orbiting each one of the 2001 SN$_{263}$ system bodies to investigate the impacts against the surface of the three components. Then, we statistically investigated whether these impacts are evenly distributed or whether there are preferential regions. The last section encompasses the final comments.


\section{Shape model and physical properties}
\label{shape}

\begin{figure}
\begin{center}
\subfloat{\includegraphics*[trim = 0mm 8cm 0mm 0mm,width=\columnwidth, frame]{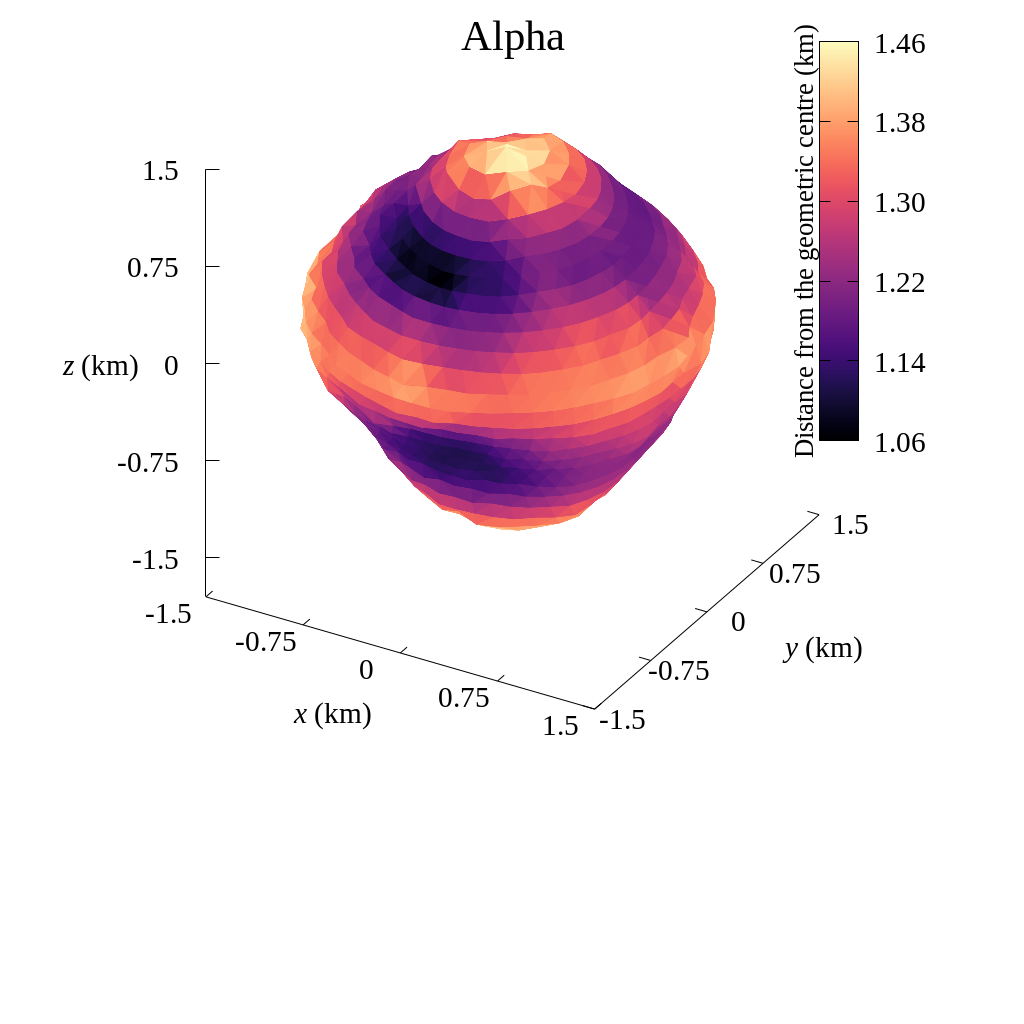}}\\
\subfloat{\includegraphics*[trim = 0mm 8cm 0mm 0mm,width=\columnwidth, frame]{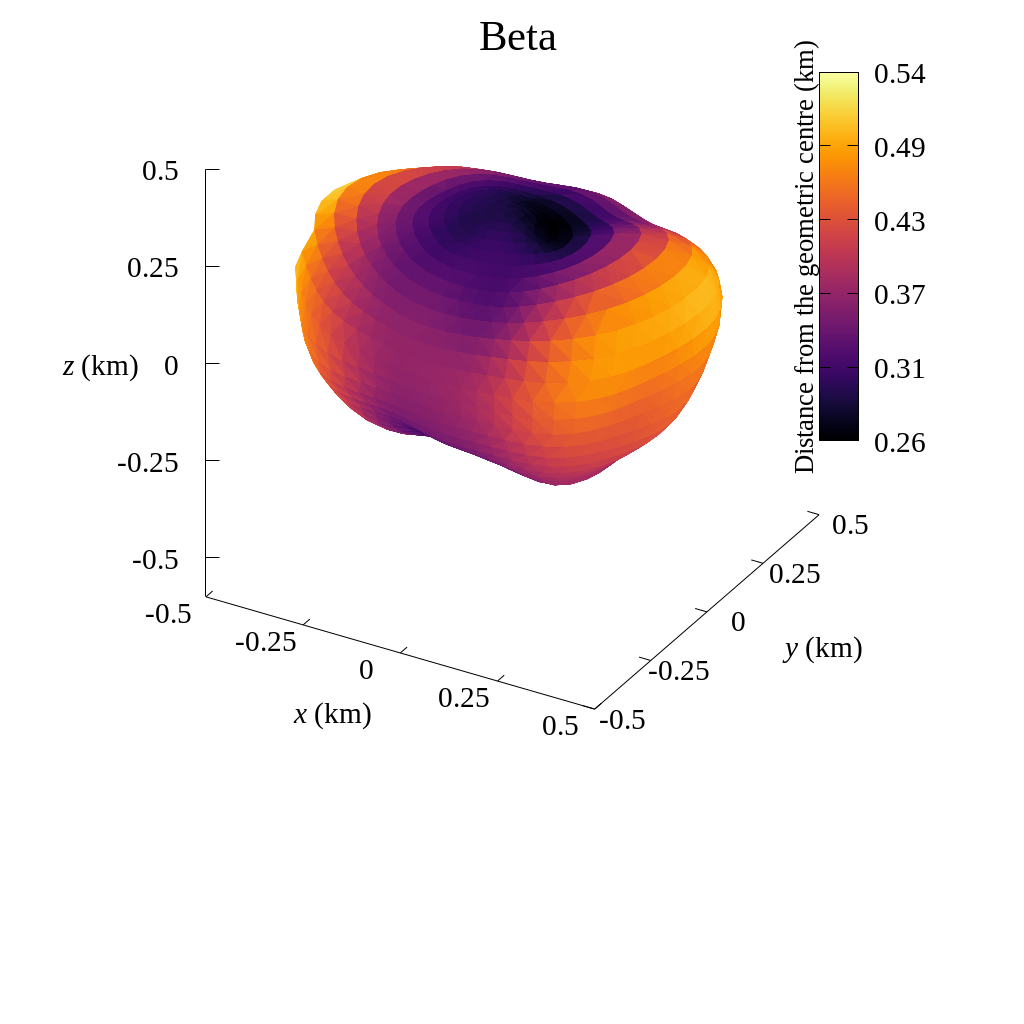}}\\
\subfloat{\includegraphics*[trim = 0mm 8cm 0mm 0mm,width=\columnwidth, frame]{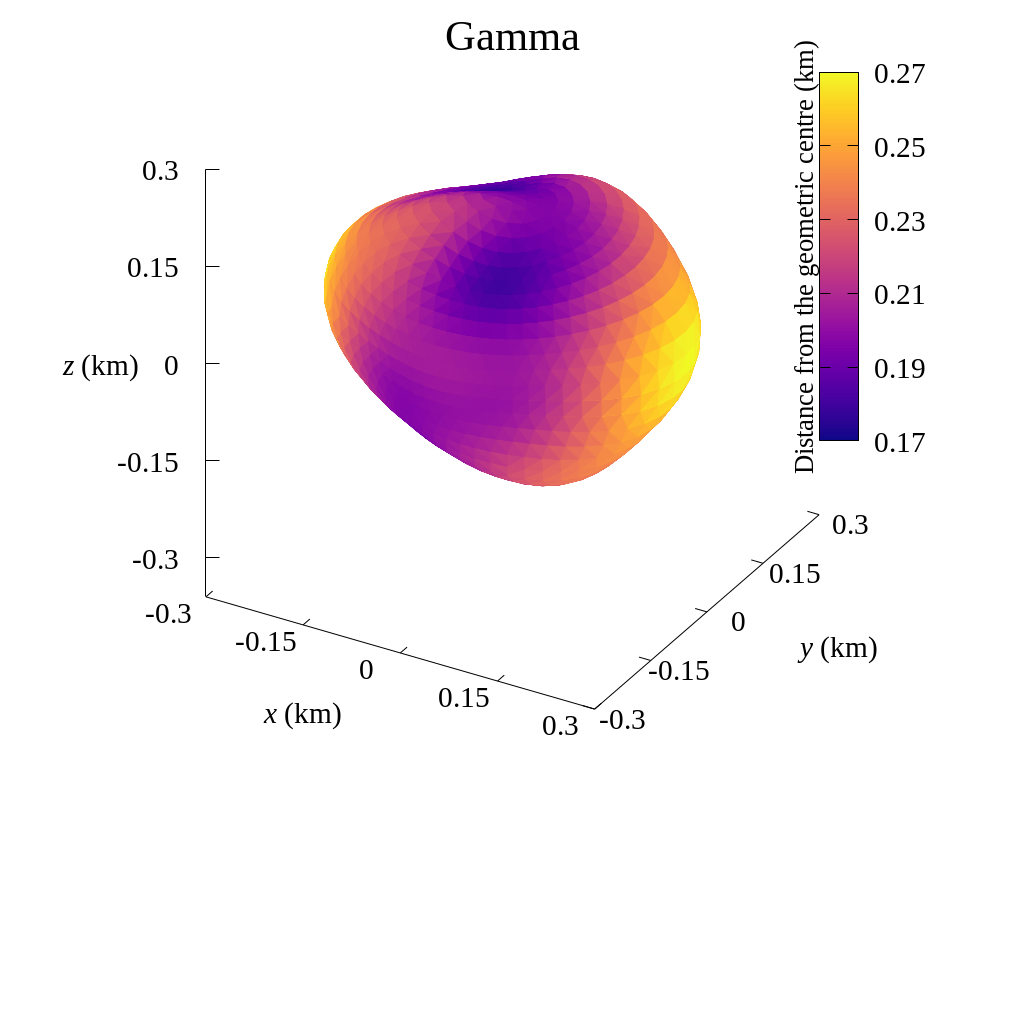}}
\end{center}
\caption{Three-dimensional polyhedral model representing the shape of the Alpha, Beta and Gamma bodies, respectively. All three forms were constructed based on polyhedra with 1148 vertices and 2292 faces. The color code provides the distance from the geometric centre of the body, in km. The origins of the systems are the geometrical centre of the bodies and the axes $x$, $y$ and $z$ are aligned with the smallest, the intermediary and the largest principal axes of inertia.}
\label{fig:shape}
\end{figure}

\begin{table*}
\centering
\caption{Physical and orbital properties of the triple asteroid system (153591) 2001 SN$_{263}$}
\label{table: infos} 
\begin{tabular}{c|c|c|c|c|c|c|c|c|c}
\hline\hline
  Body & Orbit & Semi-major & Eccentricity$^{a}$ & Inclination$^{a}$ & Orbital & Spin$^{b}$ & Mass$^{a}$ & Density$^{b}$& Volumetric\\  
   &  & axis $^{a}$ &  & (deg) & period & (hours) & (10$^{10}$ kg) & (kg$\cdot$m$^{-3}$)& radius (km) \\ 
\hline
Alpha & Sun & 1.99 au & 0.48  & 6.7  & 2.8 years$^{c}$ & 3.4256 & 917.466 $\pm$ 2.235 & 1,100 & 1.25\\
Beta & Alpha & 16.63 km & 0.015 & 0.0 & 6.23 days$^{b}$ & 13.43 & 24.039 $\pm$ 7.531 & 1,000& 0.39\\
Gamma & Alpha &3.80 km & 0.016 & 14 & 16.46 hours$^{b}$ & 16.40 & 9.773 $\pm$ 3.273 & 2,300& 0.22\\
\hline\hline
\end{tabular}
  \begin{flushleft}
  	\quad  {\footnotesize $^{a}$ \citet{Fang2011}}
  				
  	\quad  {\footnotesize $^{b}$ \citet{Becker2015}}
  				
  	\quad  {\footnotesize $^{c}$ JPL. Website: https://ssd.jpl.nasa.gov/}
\end{flushleft}
\end{table*}

We present a brief discussion regarding the irregular shape of the triple system 2001 SN$_{263}$ components.
Fig. \ref{fig:shape} illustrates the three-dimensional shape of the Alpha, Beta, and Gamma bodies, according to the model constructed by \citet{Becker2015}, which adjusts the surface of each object to a polyhedron with 1148 vertices and 2292 faces. We can observe that Alpha has a protuberance at the equator whose neighborhood is bounded by two regions with low altitude, one near the north pole and the other near the south pole. The difference in altitude between these regions and the salience at the equator is no more than 400 meters. This relatively low difference occurs because we have a certain similarity between the altitudes of the equator and the poles, which does not occur for Beta and Gamma. Although the total variation in altitude of the main body is $\sim4$ times that of Gamma and $\sim2$ times greater than that of Beta, are the satellites that present a more elongated and flattened format, resulting in high altitudes in the extremities of the equatorial region and low altitudes near the southern and northern hemispheres.
	
With this information, we use an algorithm proposed by \citet{mir1996} and transfer each polyhedra model to a fixed coordinate system in the body and with origin in the centre of mass of the object. We also align the axes $x$, $y$, and $z$ to the principal axes of the smallest, intermediate, and largest moment of inertia, respectively. We emphasize that all our calculations consider that the bodies of the triple system have a constant density (Table \ref{table: infos}) and rotate uniformly on the axis of greatest moment of inertia. Although we adopted the polyhedra model and the density estimated by \citet{Becker2015} and still the value of the mass proposed in \citet{Fang2011}, we had to apply a correction factor to adjust the volume of Alpha, Beta, and Gamma, to be consistent with our considerations. We obtained for Alpha the volume of 8.34 km$^{3}$, for Beta 0.24 km$^{3}$, while for Gamma 0.04 km$^{3}$.

Still using the algorithm of \citet{mir1996}, we compute the values of the principal moments of inertia, normalized by the body mass, in order to determine the values of the second degree and order gravity coefficients of the Alpha, Beta and Gamma objects (Table \ref{table: eixos}), according to \citet{huscheeres2004}:

\begin{equation}
 C_{20}=-\frac{1}{2R_{n}^2}(2I_{zz}-I_{xx}-I_{yy}),
\label{eq:C20}  
\end{equation}
\begin{equation}
C_{22}=\frac{1}{4R_{n}^2}(I_{yy}-I_{xx}),
\label{eq: C22}  
\end{equation}
where $R_{n}$ is the volumetric radius of the body used for normalization.

The coefficient $C_{20}$ is commonly represented by $J_{2}$, such that $J_{2}=-C_{20}$. The gravitational coefficients express the irregularities of the mass distribution of a body. From this perspective, the coefficients $C_{20}$ and $C_{22}$ determine how flat and elongated the body is, respectively. We observed that among the three objects, Beta has the most flattened and elongated form, while the low values, in modulus, of the coefficients $C_{20}$ and $C_{22}$ of the Alpha, expressed in Table \ref{table: eixos}, indicate a less prominent flattening and ellipticity. Note also that the value of the coefficient $C_{20}$, in modulus, of all bodies is greater than that of the coefficient $C_{22}$, characterizing an object whose shape presents a flattening at the poles more accentuated when compared to the elongation of its structure towards the equatorial region. By comparison criterion, the coefficient $J_{2}$ of Alpha is consistent with the numerical simulations of \citet{Fang2011}, which converged at a value of $J_{2}=0.013 \pm 0.008$. In Section \ref{geom}, we will give a more detailed discussion of the shape and geometry of the components that form the triple system 2001 SN$_{263}$.

\begin{table}
\centering
\caption{Values of the principal moments of inertia, normalized by the mass of the body, and of the gravitational coefficients $C_{20}$ and $C_{22}$ calculated for Alpha, Beta and Gamma.}
\label{table: eixos} 
\begin{tabular}{c|c|c|c|c}
\hline\hline
   & Alpha & Beta & Gamma\\  
\hline
 $I_{xx}/M$ (m$^{2}$)  & 629,300 & 44,720 & 14,810 \\
 $I_{yy}/M$ (m$^{2}$) & 660,360 & 68,580 & 20,070 \\
 $I_{zz}/M$ (m$^{2}$) & 671,290 & 77,530 & 20,270 \\
  &  &  &  \\
$C_{20}$ & $-$0.01672 & $-$0.14053 & $-$0.06042 \\
$C_{22}$ & 0.00491 & 0.04013 & 0.02803 \\
\hline\hline
\end{tabular}
\end{table}

\section{Gravitational field and geopotential}
\label{grav_geop}

\citet{wernerscheeres1996} have developed a method that describes the gravitational field of small bodies taking into account their irregular format. The shape of the body is modeled as a polyhedron with $f$ faces and $v$ vertices, assuming uniform density. This method provides a much more accurate calculation than the Legendre Polynomials and triaxial ellipsoids methods, often used to compute the gravitational potential of asteroids, for example. Thus, by applying the method of polyhedra, we can write the energy and attraction of a body, respectively, as \citep{wernerscheeres1996}:
\begin{equation}
 U=\frac{G\rho}{2} \sum_{e\in edges}{\pmb r_e} \cdotp {\pmb E_e} \cdotp {\pmb r_e} \cdotp L_e - 
 \frac{G\rho}{2} \sum_{f\in faces}{\pmb r_f} \cdotp {\pmb F_f} \cdotp {\pmb r_f} \cdotp \omega_f,
 \label{eq: potencial}  
\end{equation}
\begin{equation}
 \nabla U=-G\rho \sum_{e\in edges} {\pmb E_e} \cdotp {\pmb r_e} \cdotp L_e + G\rho \sum_{f\in faces} {\pmb F_f} \cdotp {\pmb r_f} \cdotp \omega_f,
 \label{eq: gradpotencial}  
\end{equation}
such that the gravitational constant is given by $G=6.67428\times 10^{-11}\ {\rm m}^{3}\ {\rm kg}^{-1}\ {\rm s}^{-2}$ and $\rho$ is the density of the body; ${\pmb r_e}$ and ${\pmb r_f}$ are the vectors from a field point  to any point in the edge $e$ plane and in the face $f$ plane, respectively; ${\pmb E_e}$ and ${\pmb F_f} $ are tensors of the edges and faces; $L_ e$ is the integration factor associated with each edge, and $\omega_f$ is the signed solid angle subtended by planar region (face of a polyhedron) when viewed from field point.

By adding the effect of uniform rotation of a body to its gravitational potential, we obtain the expression of the geopotential, which provides the relative energy according to the location in the body-fixed frame. Thus, when we consider the movement of a massless particle on the surface of one of the components of the triple system, we associate the intensity of the geopotential with the amount of energy necessary for this particle to be able to move along the surface of the body. In addition, the geopotential and its derivatives make it possible to compute the accelerations, the slope angle, among other quantities, that act on the surface of the body and assist in the investigation of the topographic characteristics (Section \ref{geo}).

In this paper, we are assuming that the body rotates uniformly with velocity $\omega$, and the polyhedral model representing its shape is centred in the centre of mass and in aligned with the main axes of inertia. Then, the geopotential, in this case, takes the form \citep{Scheeres2016}:
\begin{equation}
V(\pmb r) = -\frac{1}{2} \omega ^{2}(x^2+y^2) - U(\pmb r),
\label{eq:geopotencial}
\end{equation}
where ${\pmb r}$ is the position vector of a particle in the body-fixed frame relative to the centre of mass of the body and $U({\pmb r})$ is the gravitational potential computed by the polyhedra method.

\subsection{Equation of motion}
\label{equation of motion}

The movement of a particle in the vicinity of the triple system 2001 SN$_{263}$ suffers the effects of the gravitational potential of bodies with irregular mass distribution. However, perturbations of the Sun, such as gravity and radiation pressure force, can also influence and produce changes in the orbit of the particle. Both perturbations become more evident as the particle moves away from the system. In addition, the ratio of area to mass of the particle influences the intensity of the solar radiation pressure. Despite this, when we compare the effects of the perturbative forces that influence the motion of a particle, we often verify that a single force dominates the entire system, the others being neglected without loss of information in the analysis of the results.

From this perspective, in this paper, we consider that the perturbation of the gravitational field of the triple system components is much more effective than perturbations from the Sun and the gravitational field of other bodies. Moreover, we are not interested in the dynamics of very small particles ($ \ll 1$ cm), which could produce significant contributions from the solar radiation pressure.

Then, the equations that determine the motion of a particle orbiting a body with uniform rotation on its axis of maximum moment of inertia are described as \citep{Scheeres2016}:
\begin{equation}
 \ddot{\pmb r} +2{\pmb\omega} \times \dot{\pmb r} =-\frac{\partial V}{\partial \pmb {r}},
 \label{eq: equações do movimento}
\end{equation}
where $\dot{\pmb r}$ and $\ddot{\pmb r}$ indicate velocity and acceleration of the particle in the body-fixed frame.

When we assume that the body rotates uniformly, the equations (\ref{eq: equações do movimento}), in the body-fixed reference frame, are time invariant, implying the existence of a motion integral similar to the energy in the rotating frame. This conserved quantity is called the Jacobi constant ($J$) and is given by \citep {Scheeres2016}:
\begin{equation}
J = \frac{1}{2}v^2+V(\pmb r),
\label{eq: jacobi}
\end{equation}
where $v$ is the magnitude of the velocity vector with respect to the body-fixed, rotating frame.

\section{Geometric topography}
\label{geom}

The analysis of some topographic features on the surface of an object is in relation to the attributes of geometric character, that is, without the involvement of its properties and physical phenomena. Considering this perspective, \citet{Scheeres2016} defined and investigated the characteristics on the surface of the (101955) Bennu asteroid by \textit{Geometric Altitude} and \textit{Tilt}.

\begin{figure}
\begin{center}
\subfloat{\includegraphics*[trim = 0mm -0.3cm 0mm 0mm, width=\columnwidth, frame]{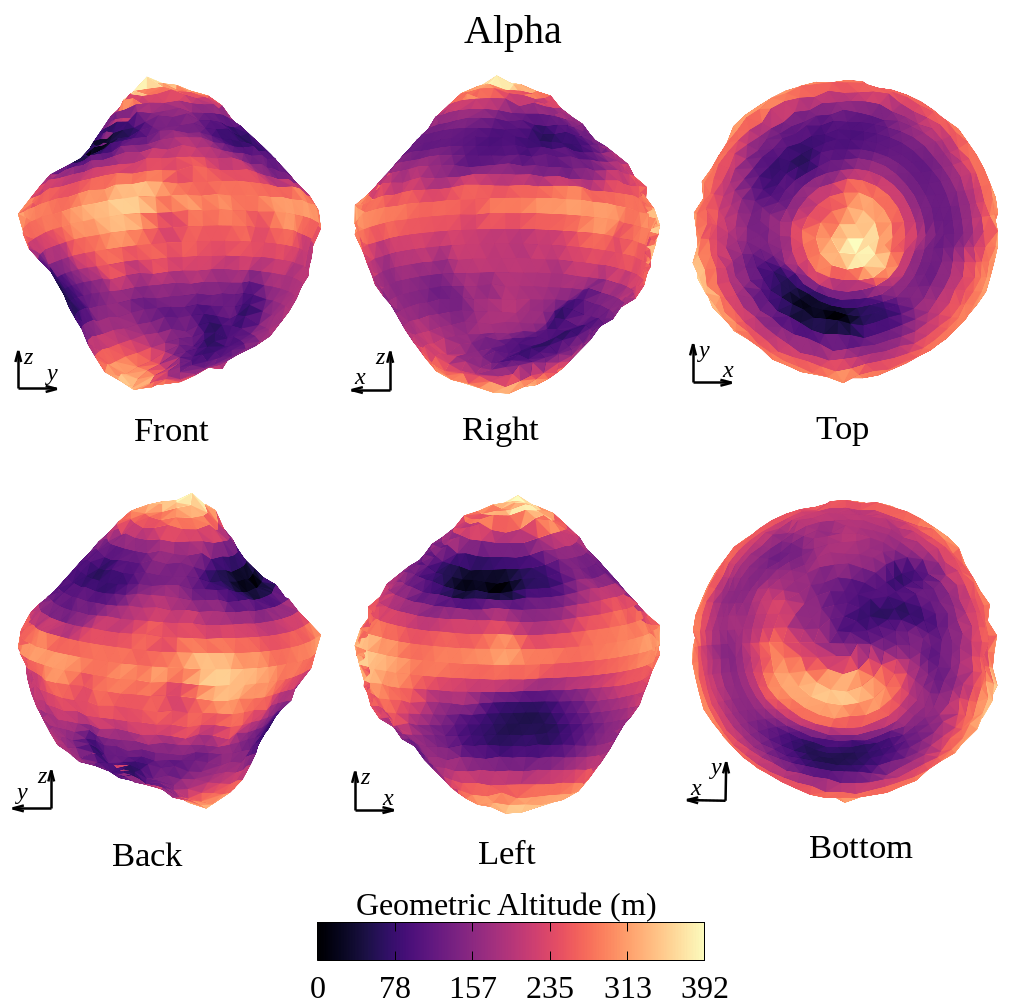}}\\
\subfloat{\includegraphics*[trim = 0mm 7cm 0mm 0mm, width=\columnwidth, frame]{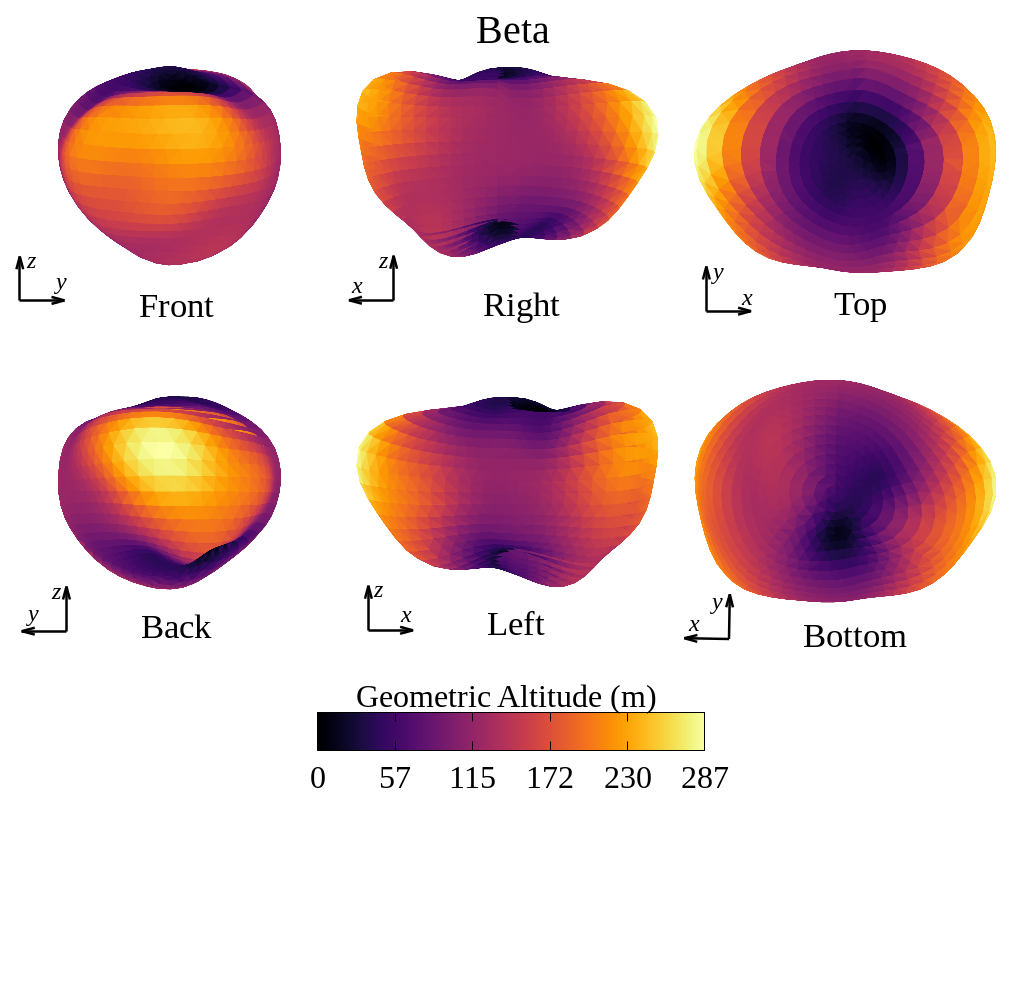}}\\
\subfloat{\includegraphics*[trim = 0mm 5.3cm 0mm 0mm, width=\columnwidth, frame]{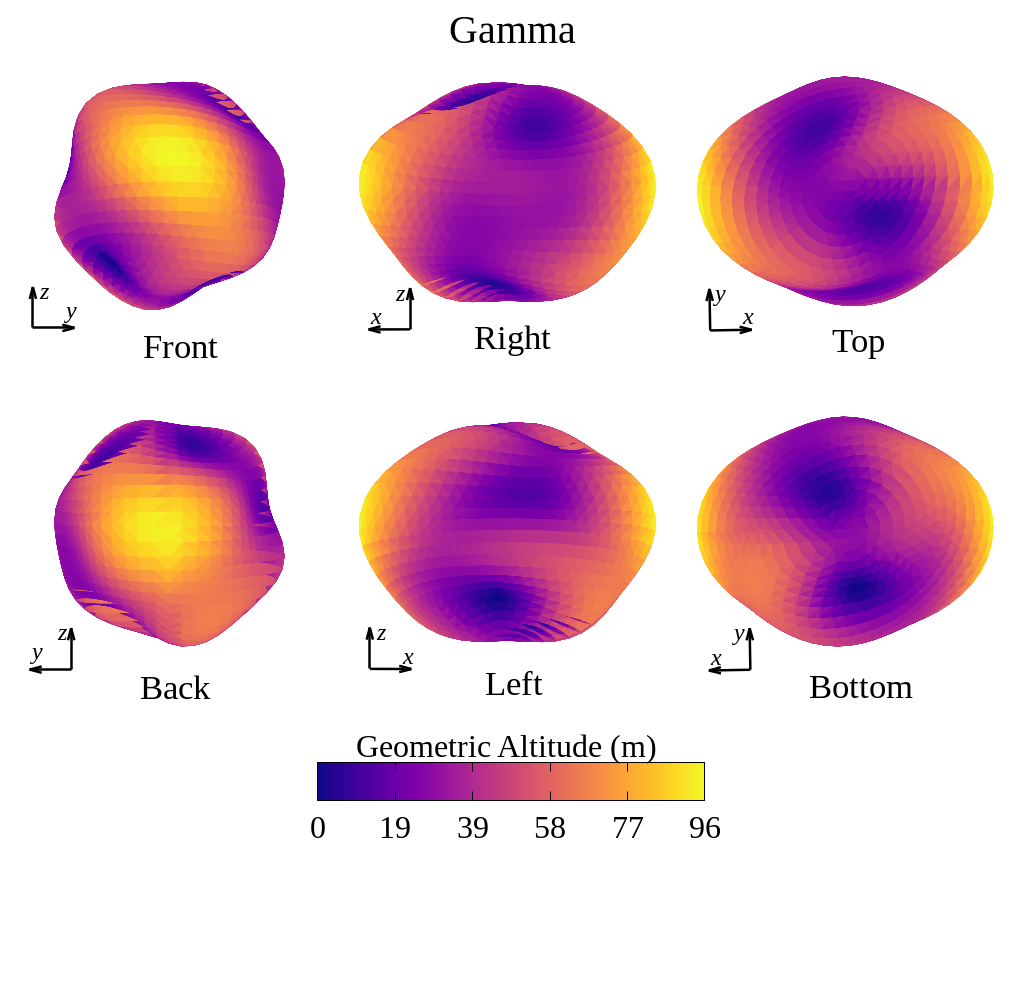}}
\end{center}
\caption{Map of the geometric altitude computed across the surface of Alpha, Beta and Gamma, respectively. Note that the colour scales are different for each body.}
\label{fig:geom}
\end{figure}

First, we measure the radius of the body given by the distance between the geometric centre of the object and the barycentre of each of the triangular faces delimiting its surface. By definition, the Geometric Altitude mapping expresses the variation of the body radius in relation to a minimum radius, among those computed. The values of this smaller magnitude of the vector radius for Alpha, Beta and Gamma are 1.06 km, 0.26 km and 0.17 km, respectively. Fig. \ref{fig:geom} shows the geometric altitude variation for the three system 2001 SN$_{263}$ components.

Analyzing the results for the Alpha body, we noticed a similar altitude distribution between the poles and the equator. This feature becomes more evident if Alpha were modeled as an ellipsoid, where parameters $a$, $b$, and $c$ (Table \ref{table:abc}) would determine their size and shape. It is apparent that these parameters have similar measurements, indicating that the distance from the centre of mass of the body to the poles is comparable to the distance to the equator. Also, the highest altitudes are located precisely at the poles and in some regions of the equator, as shown by the different views in Fig. \ref{fig:geom} for Alpha. However, between these higher altitude locations, some depressions may be present, as illustrated in the Left view for Alpha (Fig. \ref{fig:geom}), where the lowest values of radius were computed (the minimum distance of the surface to its geometric centre is 1.06 km). Although the altitude difference of these depressions is less than 400 m, they can not be ignored since they represent a relatively high unevenness on the surface of a body whose volumetric radius is about 1.25 km.

In the case of Beta, which has a shape similar to an ellipsoid, its poles are flattened and have the lowest altitudes (Top and Bottom views of Fig. \ref{fig:geom} - Beta) and its elongate equatorial region carries the highest altitudes located at the extremities (Front and Back views of Fig. \ref{fig:geom} - Beta). And, just as occurred for Alpha, the ellipsoidal form of Beta is compatible with the values of parameters $a$, $b$, and $c$ of Table \ref{table:abc}. The value of $a$ is the largest of the parameters, showing a flattened shape at the poles and elongated toward the equator. This is in accordance with the gravitational coefficients $C_{20}$ and $C_{22}$ of Beta (Table \ref{table: eixos}).
We note that Beta has a kind of ``waist'', which from the ends of the equatorial region tapers towards the centre of the body, as observed in the Right and Left views of Fig. \ref{fig:geom} - Beta. Among the values computed, in this ``waist'' are located the average altitudes. Moreover, the ends of the equatorial region are not symmetrical, the extreme given by the Front view (Fig. \ref{fig:geom} - Beta) has a more uniform altitude distribution with intermediate values, whereas the other extreme, Back view (Fig. \ref{fig:geom} - Beta), has the highest altitudes. There are two regions that delimit the ``waist'' of the body and are close to the poles, exposed in the Right view of Fig. \ref{fig:geom} - Beta, which presents rough irregularities with respect to the rest of the surface. These irregularities possibly culminate in valleys of up to 200 m, and obviously must be taken into account, since that value is about half the volumetric radius of the Beta body.

\begin{table}
\centering
\caption{Values of the semi axes ($a$, $b$ and $c$) representing the dimensions of an equivalent ellipsoid for each component of the triple system 2001 SN$_{263}$.}
\label{table:abc} 
\begin{tabular}{ccccc}
\hline\hline
 Body & $a$ (m) &  $b$ (m) &  $c$ (m) \\       
\hline
Alpha & 1,314.79& 1,255.30& 1,233.69\\
Beta & 515.27& 374.91& 306.02\\
Gamma & 261.46& 200.45& 197.70\\
\hline\hline
\end{tabular}
\end{table}

As opposed to Alpha and Beta, which present brief patterns in its formats, such that for Alpha, we observe a slight symmetry between the northern and southern hemispheres, while for Beta a format a little similar to that of an ellipsoid, Gamma has a much more irregular shape. The poles, Top, and Bottom views of Fig. \ref{fig:geom} - Gamma, delimit regions with low and medium altitudes next to each other, evidencing the diversity and irregularity of the relief in the surface of the body. Although Gamma has a shape indeed irregular, it is clear that in the equatorial region, precisely at the extremes, are the highest altitudes, as shown in the Front and Back views of Fig. \ref{fig:geom} - Gamma. Again, this feature can be verified by analyzing parameters $a$, $b$, and $c$ of Table \ref{table:abc}, since the measure of $a$ is approximately 30\% greater than that of parameters $b$ and $c$. Then, just as in the case of Beta, Gamma has the equatorial region more elongated than the region of the poles.

\begin{figure}
\begin{center}
\subfloat{\includegraphics*[trim = 0mm -0.3cm 0mm 0mm,width=\columnwidth, frame]{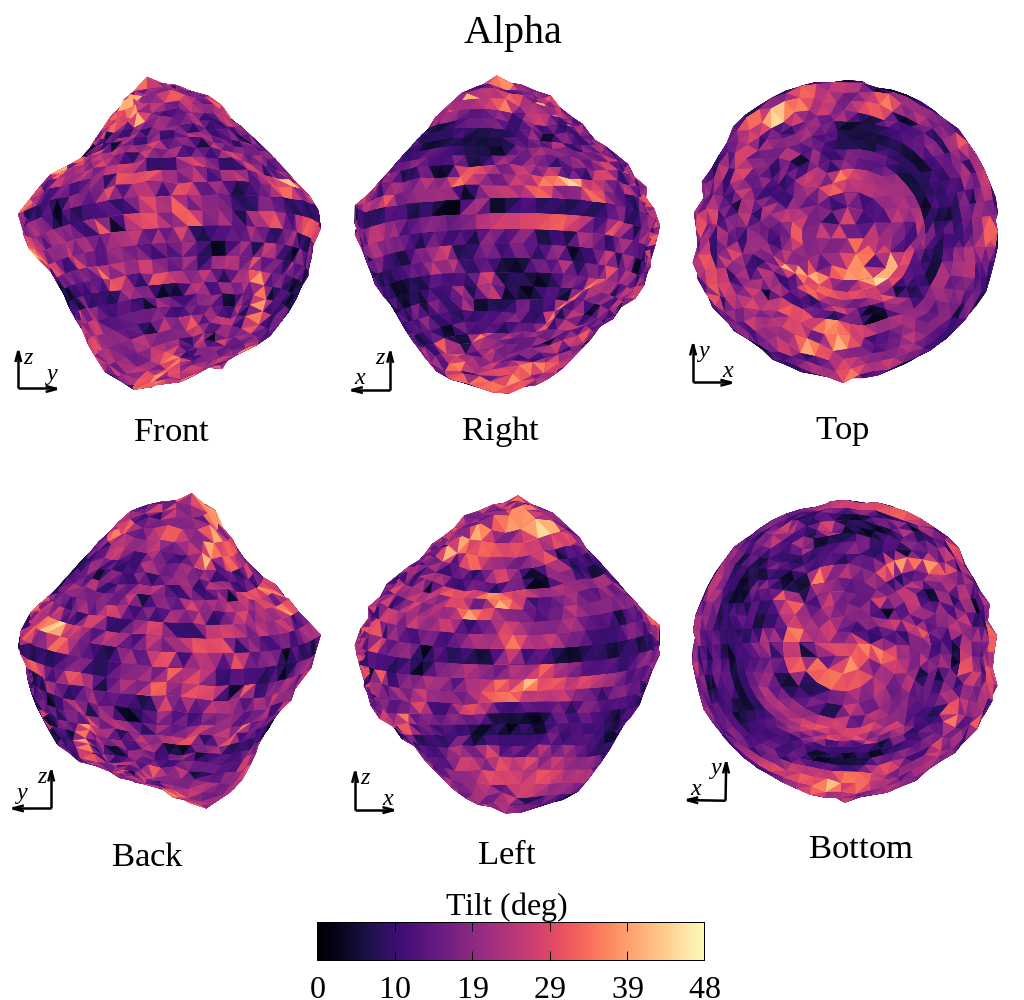}}\\
\subfloat{\includegraphics*[trim = 0mm 7cm 0mm 0mm, width=\columnwidth, frame]{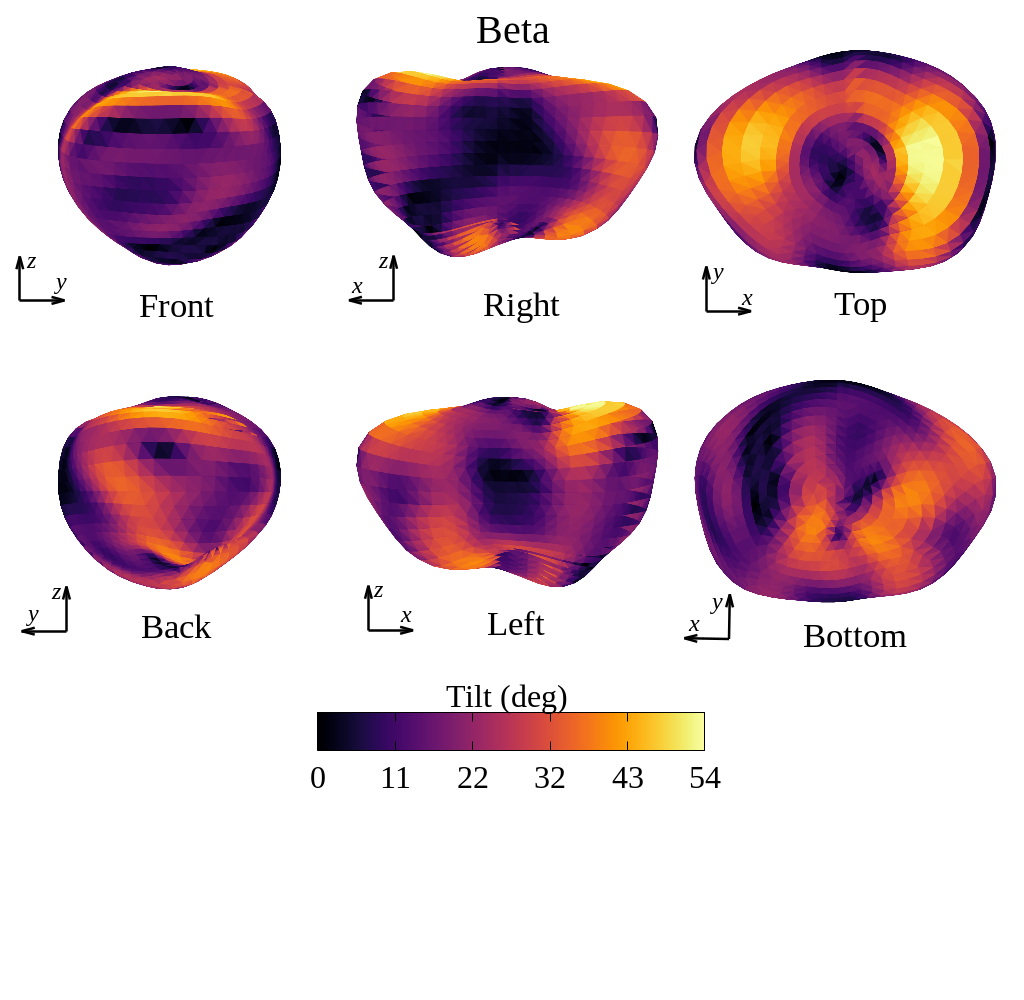}}\\
\subfloat{\includegraphics*[trim = 0mm 5.3cm 0mm 0mm, width=\columnwidth, frame]{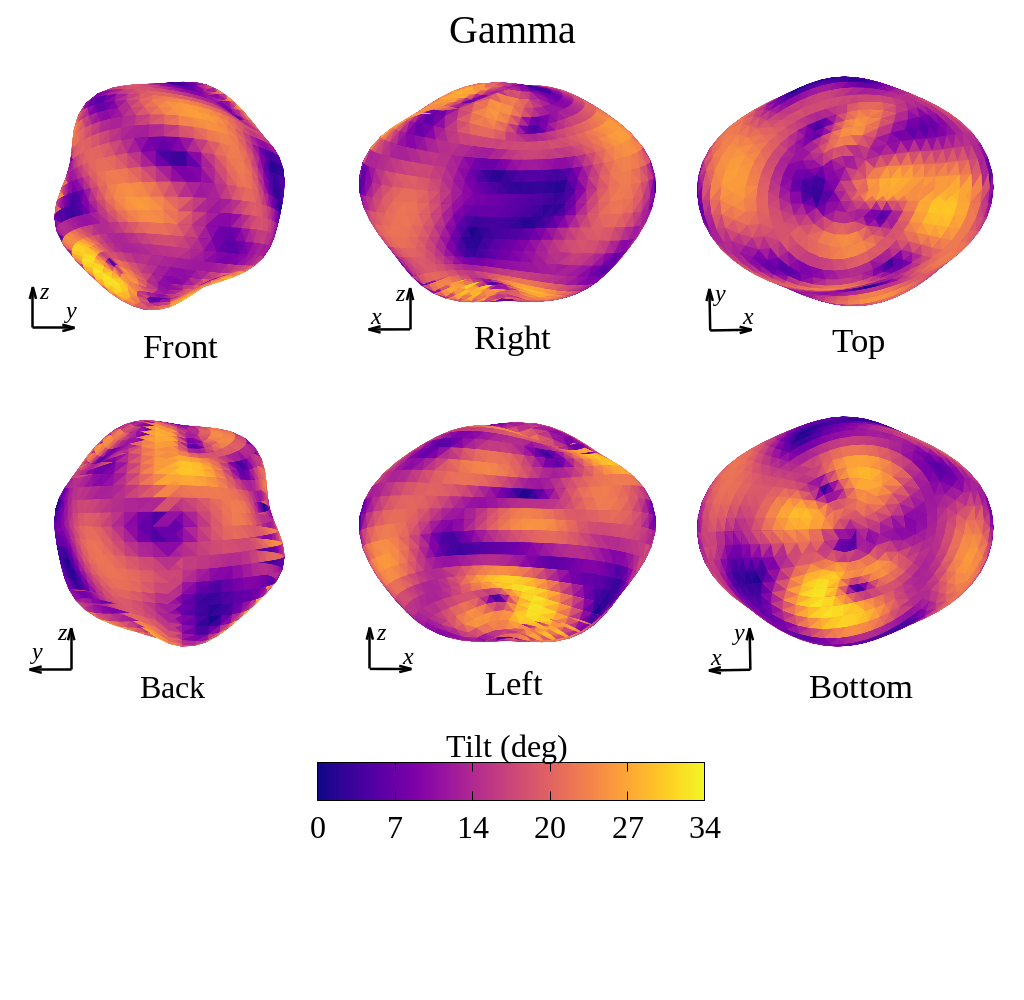}}
\end{center}
\caption{Map of the tilt angle computed across the surface of Alpha, Beta and Gamma, respectively. Note that the colour scales are different for each body.}
\label{fig:tilt}
\end{figure}

Still in the concept of geometric topography, \citet{Scheeres2016} defined the Tilt angle, in the body-fixed frame. Let us consider a point on one of the triangular faces that cover the surface of the body. The tilt angle at this location is given as the angle between the normal vector of the surface at that point and the radius vector, which originates from the body centre of mass and extends through the surface at the point.
Then, we map the tilt angle for the three objects of the system 2001 SN$_{263}$, so that we compute the normal vector to the surface and the radius vector for each triangular face of the polyhedral shape model of these objects. The result is expressed in Fig. \ref{fig:tilt} and is useful in understanding the orientation of the surface of the object, that is, we have a sketch of how each triangular face is oriented in the structure of the body. For example, if the ASTER mission plans to land on the surface of one of the objects to collect or analyze material, the tilt angle could orient the spacecraft and provide the direction relative to the normal vector of the landing region. Besides, of course, assisting in the choice of a possible landing region.

The variation of the tilt angle on the Alpha surface did not exceed $48^\circ$, and analyzing the Right and Left views of Fig. \ref{fig:tilt} - Alpha, we can conclude that the equatorial region has a uniform distribution with some locations whose tilt angle is below $10^\circ$. While in the regions between the equator and the poles, there are larger bands constituting the regions with the lowest values of the tilt angle.

On the surface of Beta, the minimum value of the tilt angle is $0.6^\circ$ and the total variation is $54^\circ$, slightly higher compared to Alpha. We can note that one of the poles, Top view of Fig. \ref{fig:tilt} - Beta,  presents a central region with low values of the angle, and the location around this region is mostly filled with the highest values. The other pole, Bottom view of Fig. \ref{fig:tilt} - Beta, has locations with intermediate values of the tilt angle, close to an extensive band with low values. At both poles, it is clear how the geometric orientation of each triangular face relative to its normal vector varies angularly. And consequently, we can relate these results with the fact that we have two regions that delimit the ``waist'' of Beta and are very close to the poles, Right view of Fig. \ref{fig:geom} - Beta, that have a clearly irregular relief.
In addition, the central region of this ``waist'' presents mostly low values of the angle tilt, and the rest of the equatorial region concentrates values a little lower than the maximum, according to Right and Left views of Fig. \ref{fig:tilt} - Beta.

The lowest variation range of the tilt angle occurs precisely for the smallest object, Gamma. The mapping of the tilt angle on its surface does not exceed $34^\circ$.
The Right and Left views (Fig. \ref{fig:tilt} - Gamma) point to a track in a curve format located in the centre of the equatorial region that delimits the lower values of the tilt angle ($\sim 0.2^\circ$). Something similar occurs for the ends of the equator, Front and Back views of Fig. \ref{fig:tilt} - Gamma, with the highlight of three locations whose tilt angle values, are also low.

We conclude that, although the variation range of the tilt angle on the surface of Alpha is $48^\circ$, which is equivalent to only $6^\circ$ of difference of the variation range for Beta, the faces of Alpha are generally less inclined with respect to their normal vector than the faces of Beta. By analyzing the behavior of the tilt angle for the three system 2001 SN$_{263}$ components and a possible landing of a spacecraft on the surface of one of them, we believe that the Alpha body is the most viable for such an objective. Since it presents a practically uniform distribution of the tilt angle on its surface, with the exception of the poles. And this behavior does not occur on the Beta surface, which clearly has isolated regions where the tilt angle is either small or large. We exclude Gamma by the fact that its entire surface alternates between regions with high and low tilt angles.

\section{Geopotential topography}
\label{geo}

Simultaneously with the geometric topography analyzes, we investigated the geopotential and its derivatives, computed across the surface of the bodies of the system 2001 SN$_{263}$, to obtain information and to identify topographic characteristics of the objects, based on the format and the physical data. The geopotential, as defined in equation (\ref{eq:geopotencial}), is the contribution of the gravitational potential plus the effects of the uniform rotation of the body. And the result allows estimating the relative energy anywhere on the surface of the studied body. Thus, variations in the behavior of the geopotential across the surface of the object encompass analyzes that are dependent on the size, density, and rotation speed of each body. For the system 2001 SN$_{263}$, these variations will be monitored by means of the surface geopotential, surface accelerations, potential speed, and slope angle, according to the definitions presented in \citet{Scheeres2016}.

\begin{figure}
\begin{center}
\subfloat{\includegraphics*[trim = 0mm -0.3cm 0mm 0mm,width=\columnwidth, frame]{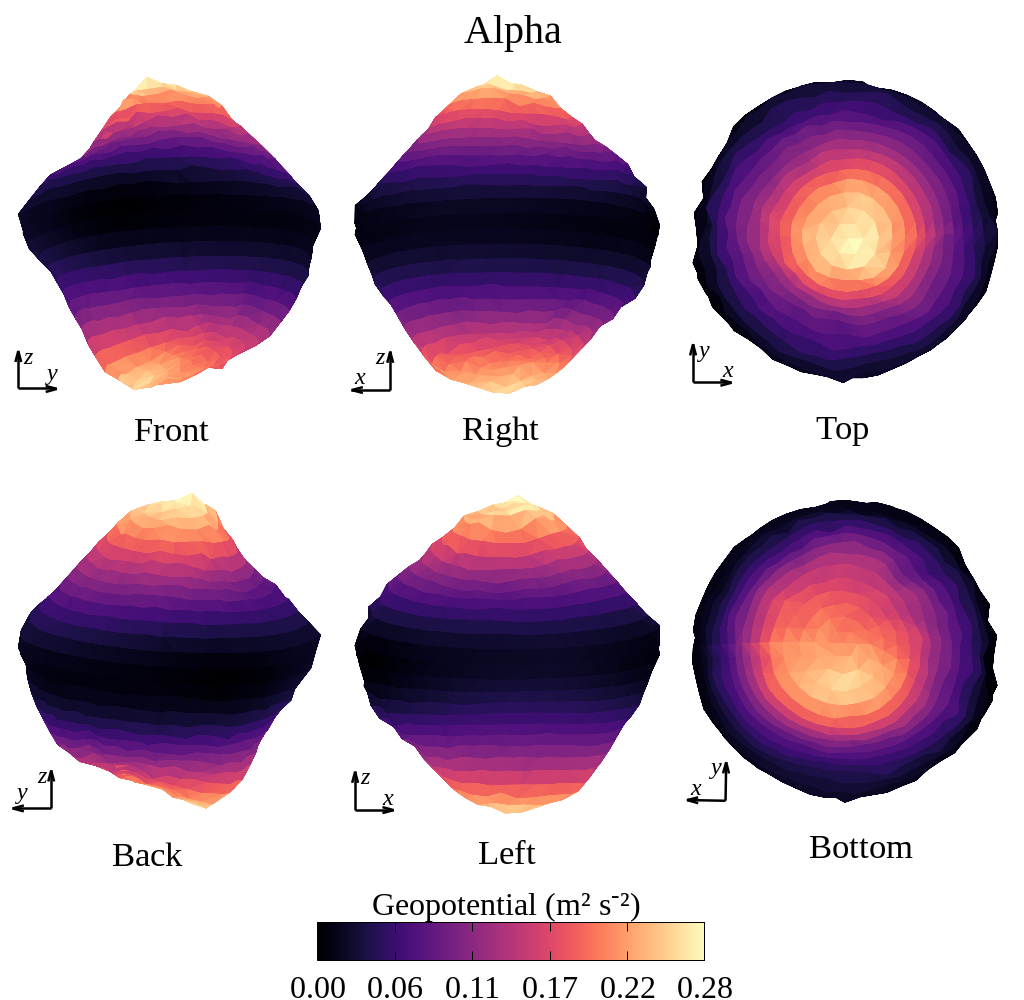}}\\
\subfloat{\includegraphics*[trim = 0mm 7cm 0mm 0mm, width=\columnwidth, frame]{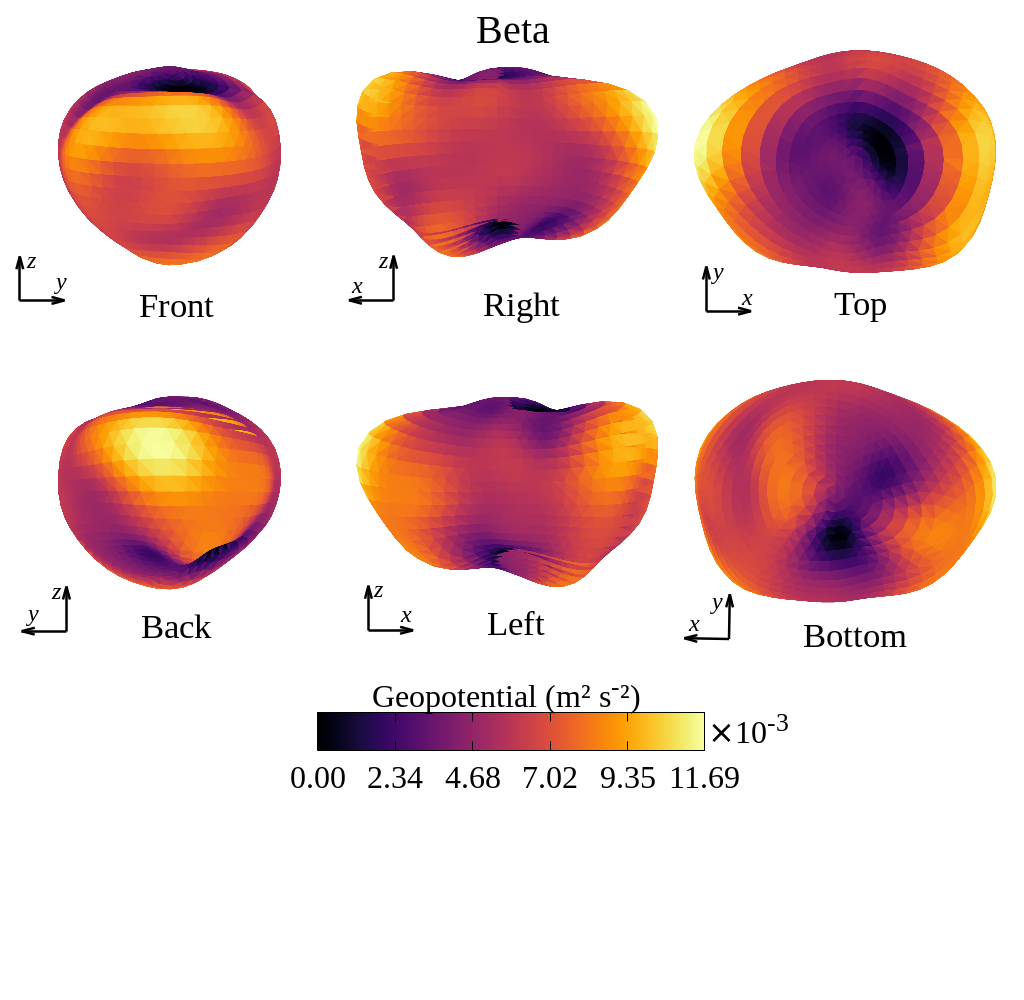}}\\
\subfloat{\includegraphics*[trim = 0mm 5.3cm 0mm 0mm, width=\columnwidth, frame]{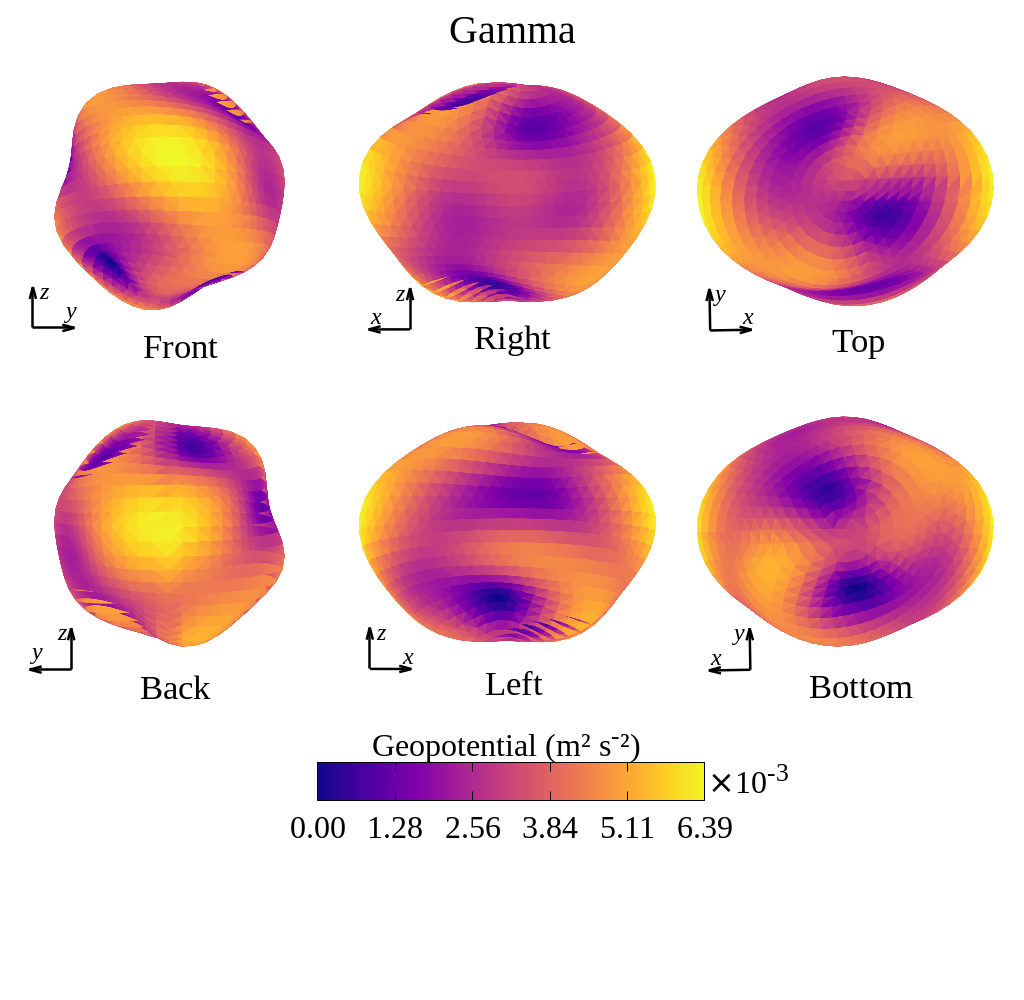}}
\end{center}
\caption{Map of the geopotential computed across the surface of Alpha, Beta and Gamma, respectively. Note that the colour scales are different for each body.}
\label{fig:geo}
\end{figure}

As Alpha, Beta, and Gamma have a polyhedral shape with triangular faces, the gravitational potential, equation (\ref{eq: potencial}), was computed in the barycentre of each face. This plus the rotational contribution, considering that all bodies rotate uniformly on the axis of greatest moment of inertia, provided the value of the geopotential on the surface.
Fig. \ref{fig:geo} shows the mapping of this relative energy in the surface of the body in relation to a reference value, denominated “sea-level” value, or, more precisely, the smallest value of the geopotential computed across the surface of the object \citep{Scheeres2012, Scheeres2015}. This “sea-level” value on the surface of Alpha is $-0.70995$ m$^2$ s$^{-2}$, $-0.048285$ m$^2$ s$^{-2}$ on Beta and $-0.033797$ m$^2$ s$^{-2}$ on Gamma.

The equatorial region of the largest component of the system, Right and Left views of Fig. \ref{fig:geo} - Alpha, suffers the influence of a minimum geopotential with respect to the poles, Top and Bottom views of Fig. \ref{fig:geo} - Alpha, which have the highest values. This influence is almost three times stronger at the poles than at the equator. It is common to observe in planetary bodies the relationship between altitude change, in relation to surface geometry, and variations in geopotential. That is, as there is an increase in the altitude of the body surface, there is also an increase in potential energy.
\citet{Scheeres2016} did not find this correlation when investigating the behavior of the geopotential on the surface of the (101955) Bennu asteroid, in function of being a very small body with a high spin rate (mean radius = 246 m and rotation period = 4.29746 h). This occurs because the rotational component interferes significantly in the effect of the geopotential across the surface. Such behavior is also observed for Alpha, whose radius is about 5 times the radius of Bennu if we compare Figs \ref{fig:geom} and \ref{fig:geo}. However, although it is not expressively large, its rotation speed is 25$\%$ higher than that of Bennu, a high rate in relation to its size, and this causes a considerable effect on the performance of the geopotential on the surface. Therefore, even for high altitudes in the equatorial region (Right and Left views of Fig. \ref{fig:geom} - Alpha), the rotational component of the geopotential is stronger than the component of
the gravitational potential, and consequently, it changes the geopotential in that location (Right and Left views of Fig. \ref{fig:geo} - Alpha), reflecting in low values. And at the poles, where the spin axis is located, such a contribution is not as significant, and the gravitational potential stands out, so that at high altitudes (Top and Bottom views of Fig. \ref{fig:geom} - Alpha) the geopotential is high (Top and Bottom views of Fig. \ref{fig:geo} - Alpha). With the exception of the equatorial region, almost all locations on the Alpha surface obey the relation: as altitude increases, the geopotential also increases.

For the Beta body, which has a longer rotational period, the geopotential parcel related to its rotation was not so relevant. Thus, we observed a behavior opposite to that identified across the surface of Alpha, with the maximum values of the geopotential located through the equator (Right and Left views of Fig. \ref{fig:geo} - Beta) and the lowest values concentrated at the poles (Top and Bottom views of Fig. \ref{fig:geo} - Beta). Even Beta being smaller than Alpha, its low speed of rotation (290$\%$ lower than that of Alpha) motivates the change in altitude correlated to the change in the geopotential across its surface, as shown in Figs \ref{fig:geom} and \ref{fig:geo} of this object. We highlight that the intensity of the geopotential at one end of the equatorial region (Back view of Fig. \ref{fig:geo} - Beta) is about 12 times stronger than the intensity at the poles. A fairly great difference if we consider that the altitude variation between these two locations is less than 300 m, as identified in the color palette of Fig. \ref{fig:geom} - Beta.

Recalling that Gamma is the smallest component of the triple system and also carries the longest rotational period, Figs \ref{fig:geom} and \ref{fig:geo} of this body indicate the variation of the geopotential across its surface related to the change in altitude. The result is contrary to the pattern suggested in \citet{Scheeres2016} for small bodies, since the rotation speed of Gamma is almost 4 times larger than that of Bennu, although the radius of the first is only 26 m smaller than the radius of the second. Due to its expressively irregular shape, we noticed the presence of regions with intermediate to high values of the geopotential distributed throughout its surface, especially at the extremities of the equator (Front and Back views of Fig. \ref{fig:geo} - Gamma), where the highest altitudes are located (Front and Back views of Fig. \ref{fig:geom} - Gamma). In addition, there are well defined locations that delimit the lowest values of the geopotential, as is the case of the poles (Top and Bottom views of Fig. \ref{fig:geo} - Gamma) and regions near the equator (Left view of Fig. \ref{fig:geo} - Gamma).

Ultimately, the geopotential across the surface of Alpha is about 100 times greater than across the surface of Beta and Gamma, as expected. However, it is also the component that has the lowest total variation of the geopotential across its surface, less than 3$\%$ in relation to the minimum value. While the Beta surface had the highest total variation, almost 12$\%$ above the lowest geopotential value.

\subsection{Surface accelerations}
\label{acel}

In this subsection we compute the total acceleration (gravitational plus centrifugal) on each face of the surfaces. The most striking feature found is a wide range of values in the case of Alpha.
 
Given the geopotential at any location $\pmb r$ on the surface of the asteroid, when we calculate its gradient ($-\partial V/ \partial \pmb r$) we obtain the total acceleration in that region. Thus, the total acceleration intensity is measured as a function of gravitational and centrifugal accelerations. Its variation can affect the movement of loose material on the surface of the body, and help in the prediction of possible preferential regions for an accumulation of this material when associated with the slope angle (Section \ref{slope}).

\begin{figure}
\begin{center}
\subfloat{\includegraphics*[trim = 0mm -0.3cm 0mm 0mm,width=\columnwidth, frame]{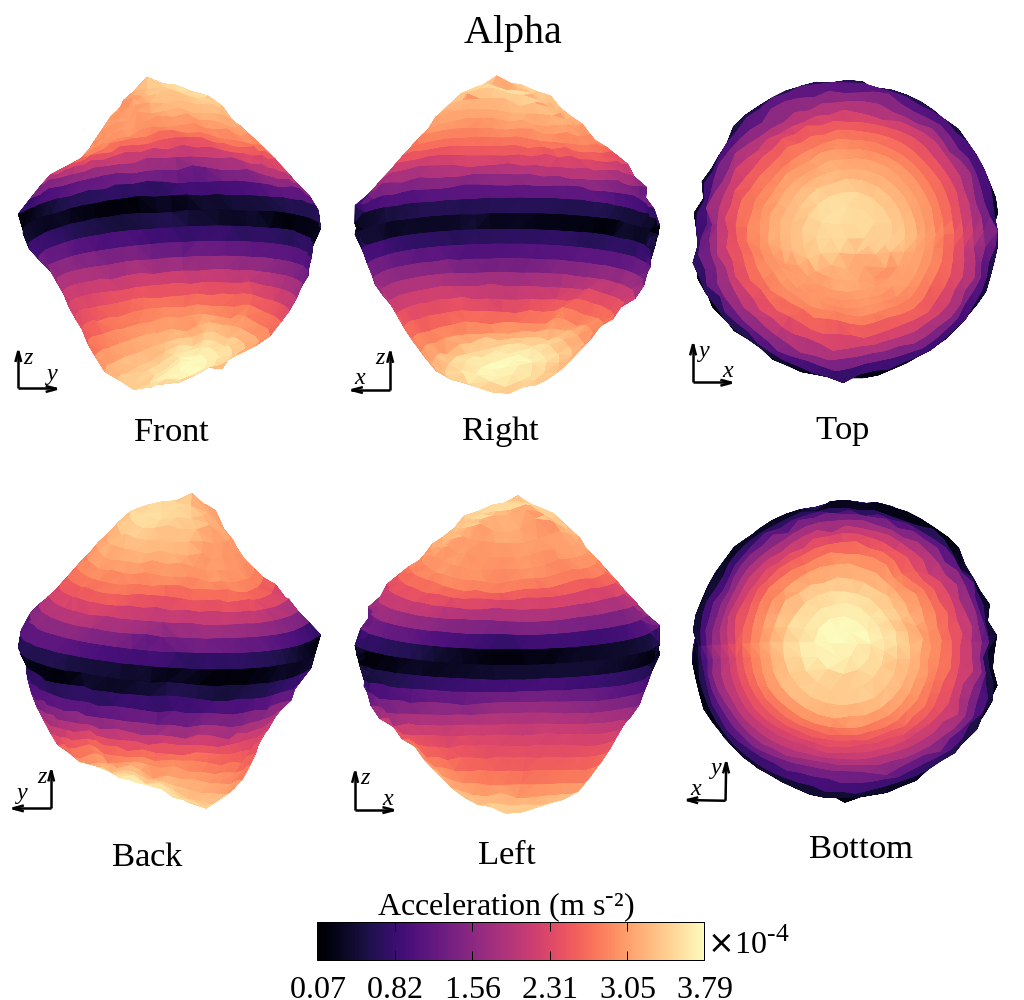}}\\
\subfloat{\includegraphics*[trim = 0mm 7cm 0mm 0mm, width=\columnwidth, frame]{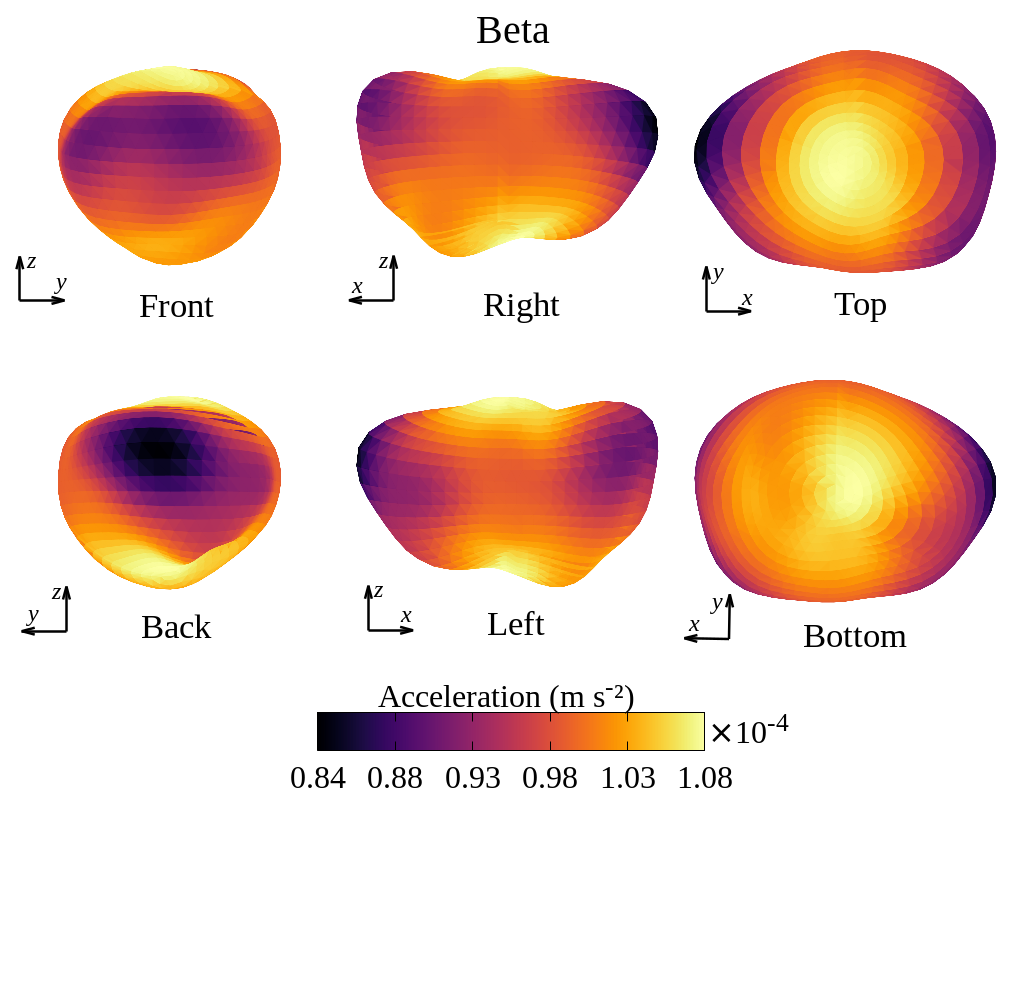}}\\
\subfloat{\includegraphics*[trim = 0mm 5.3cm 0mm 0mm, width=\columnwidth, frame]{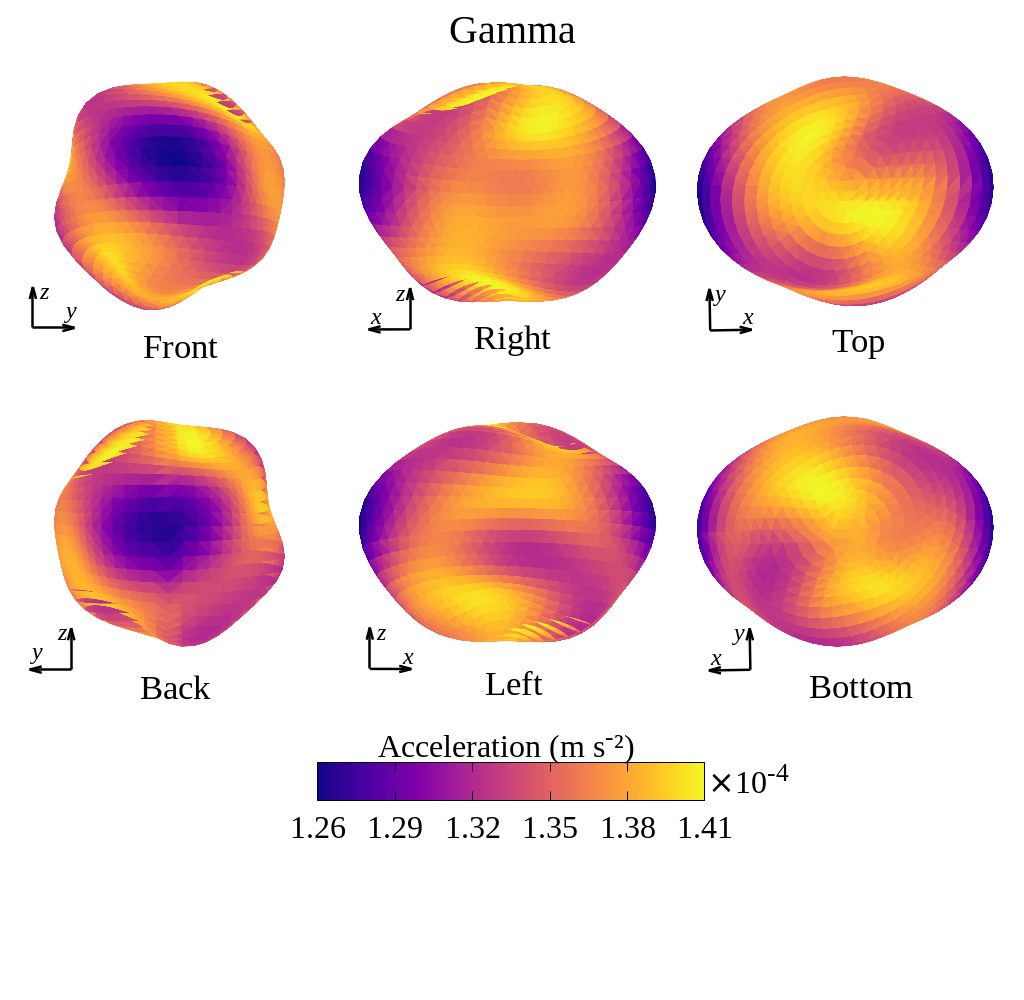}}
\end{center}
\caption{Map of the total acceleration computed across the surface of Alpha, Beta and Gamma, respectively. Note that the colour scales are different for each body.}
\label{fig:acel}
\end{figure}

Fig. \ref{fig:acel} shows the total acceleration surfaces for the objects of the triple system 2001 SN$_{263}$. For this, the total acceleration was computed in the barycentre of each triangular face of the polyhedral shape model of the body. We observed, by the color code shown in Fig. \ref{fig:acel} of each component, that the greatest variation of the total acceleration occurs on the surface of Alpha, while the lowest on the Gamma surface ($\sim2.6$ times weaker than that of Alpha). These results are in agreement with the analysis presented in Section \ref{geo}, where the effect of the geopotential on the surface of Gamma is about 100 times smaller than that of Alpha.

It is clear that the distribution of the total acceleration across the surface of Alpha suggests a symmetry between the northern and southern hemispheres. The intensity of the total acceleration at the poles (Top and Bottom views of Fig. \ref{fig:acel} - Alpha) is almost 54 times greater than the intensity throughout the equatorial region (Right and Left views of Fig. \ref{fig:acel} - Alpha), the same behavior described for the geopotential (Fig. \ref{fig:geo} - Alpha).

The variation of the total acceleration across the surfaces of Beta and Gamma occurs in a similar way. The poles of both Top and Bottom views of Fig. \ref{fig:acel} - Beta and Gamma delimit central regions with the largest magnitudes of the total acceleration, and in the vicinity of these regions, we have intermediate magnitudes. The equatorial region of the two bodies also presents median magnitudes, except the extremities (Front and Back views of Fig. \ref{fig:acel} - Beta and Gamma), which delimit sites with the lowest intensities of the total acceleration. Opposite to what we observe for Alpha, the minimum and maximum intensity regions of the total acceleration for Beta and Gamma occur where the geopotential is maximum and minimum, respectively, describing an inverse pattern. Such behavior occurs due to direct competition between gravitational and centrifugal accelerations.

\citet{Scheeres2016} computed that the magnitude of the total acceleration at the surface of the asteroid Bennu does not exceed $1.00 \times 10^{-4}$ m s$^{-2}$. If we compare with Alpha, whose rotation speed is 25$\%$ higher than that of Bennu, this magnitude is $\sim4$ times smaller. While on the surface of Gamma, the component closest to the size of Bennu, but with much slower rotational speed, the maximum magnitude of the total acceleration is below $1.41 \times 10^{-4}$ m s$^{-2}$. This allows us to conclude that body rotation exerts a significant influence on the acceleration experienced through the surface.

\subsection{Slope}
\label{slope}

In this subsection we compute the slope, which is a measure on how much inclined a region on the surface of the object is in relation to its acceleration vector, and is important in order to estimate the movement of material across the surfaces.

Consider the acceleration vector computed in each triangular face of the polyhedral shape model of the body, as identified in Section \ref{acel}. Let us analyze how the surface, that is, each face, is oriented in relation to this vector. Let $\theta$ be the angle between the normal vector the surface and the total acceleration vector of a given face. The angle slope, defined as the $\theta$ angle supplement \citep{Scheeres2012,Scheeres2016}, is a measure that indicates how much inclined a region on the surface of the object is in relation to its acceleration vector.

\begin{figure}
\begin{center}
\subfloat{\includegraphics*[trim = 0mm -0.3cm 0mm 0mm,width=\columnwidth, frame]{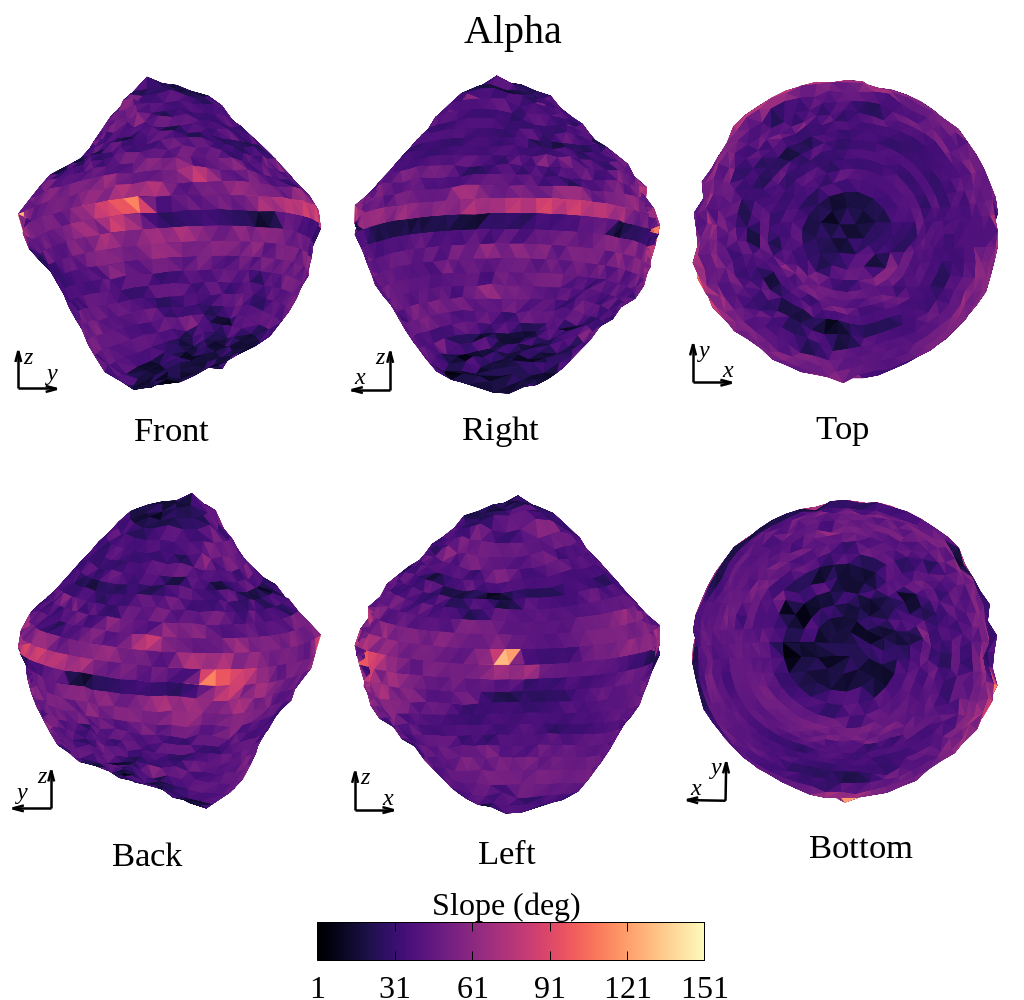}}\\
\subfloat{\includegraphics*[trim = 0mm 7cm 0mm 0mm, width=\columnwidth, frame]{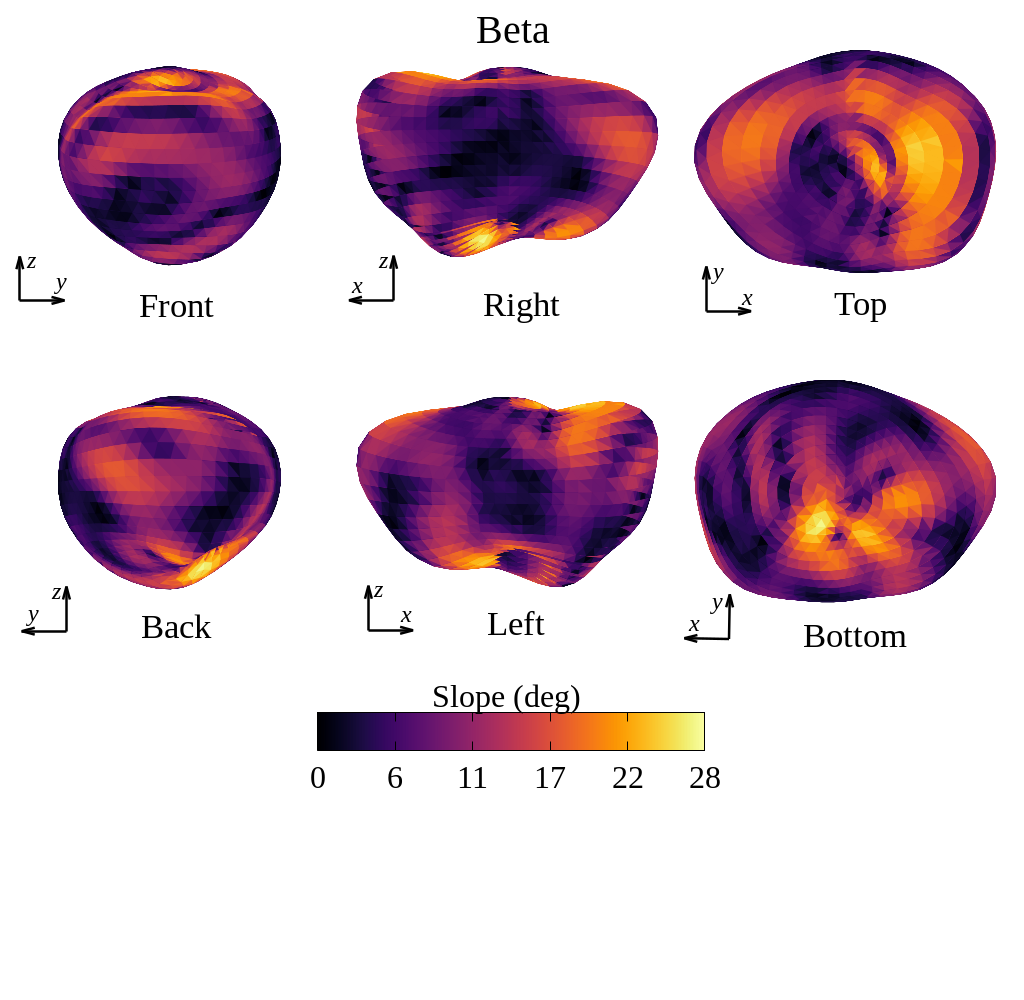}}\\
\subfloat{\includegraphics*[trim = 0mm 5.3cm 0mm 0mm, width=\columnwidth, frame]{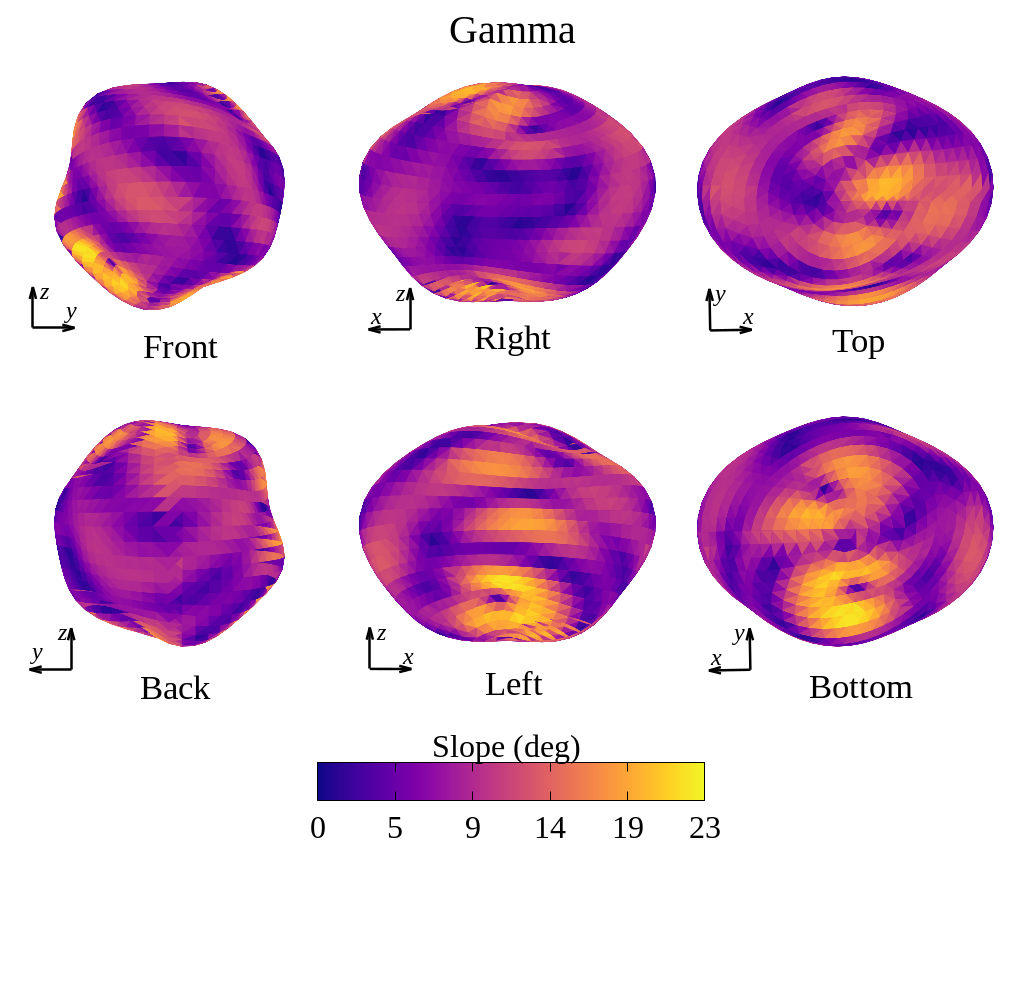}}
\end{center}
\caption{Map of the slope angle computed across the surface of Alpha, Beta and Gamma, respectively. Note that the colour scales are different for each body.}
\label{fig:slope}
\end{figure}

\begin{figure}
\begin{framed}
\begin{center}
\subfloat{\includegraphics*[trim = 0mm 7cm 0mm 0cm,width=\columnwidth]{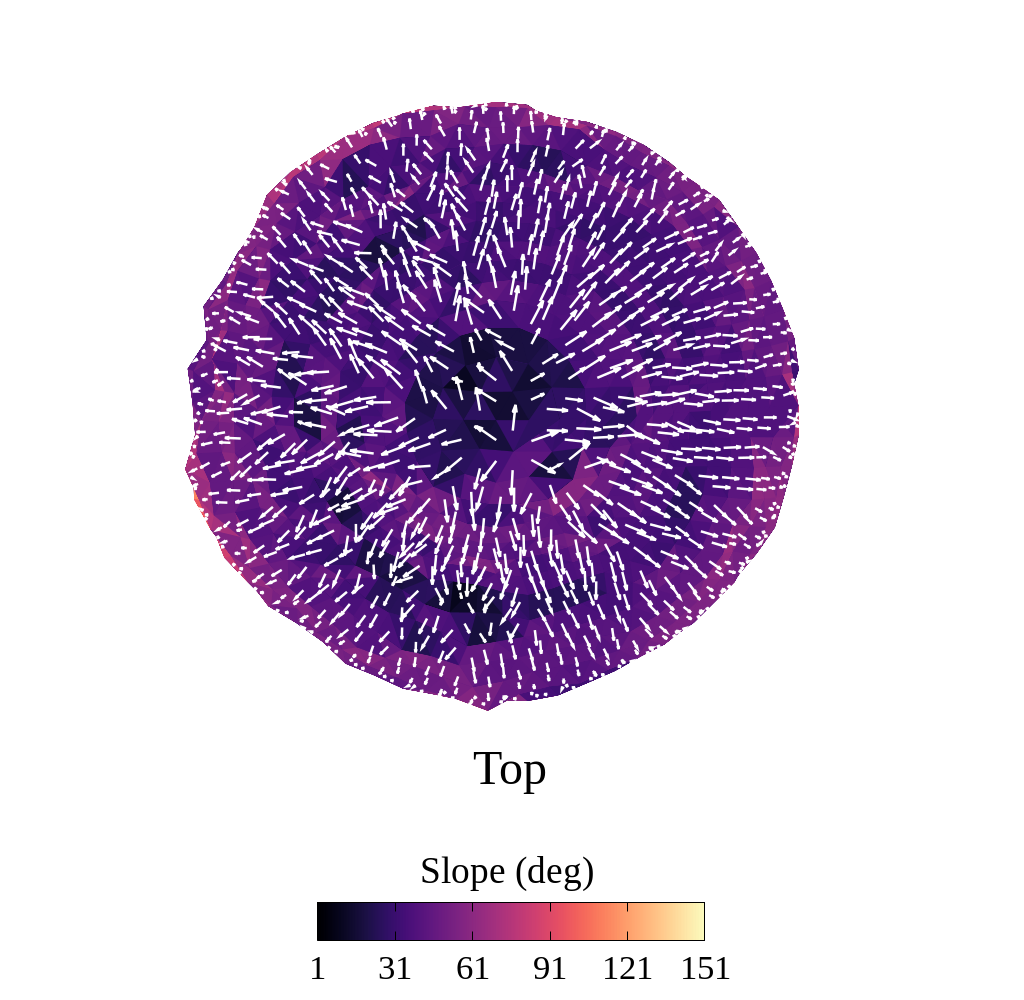}}\\
\subfloat{\includegraphics*[trim = 0mm 7.3cm 0mm 1cm, width=\columnwidth]{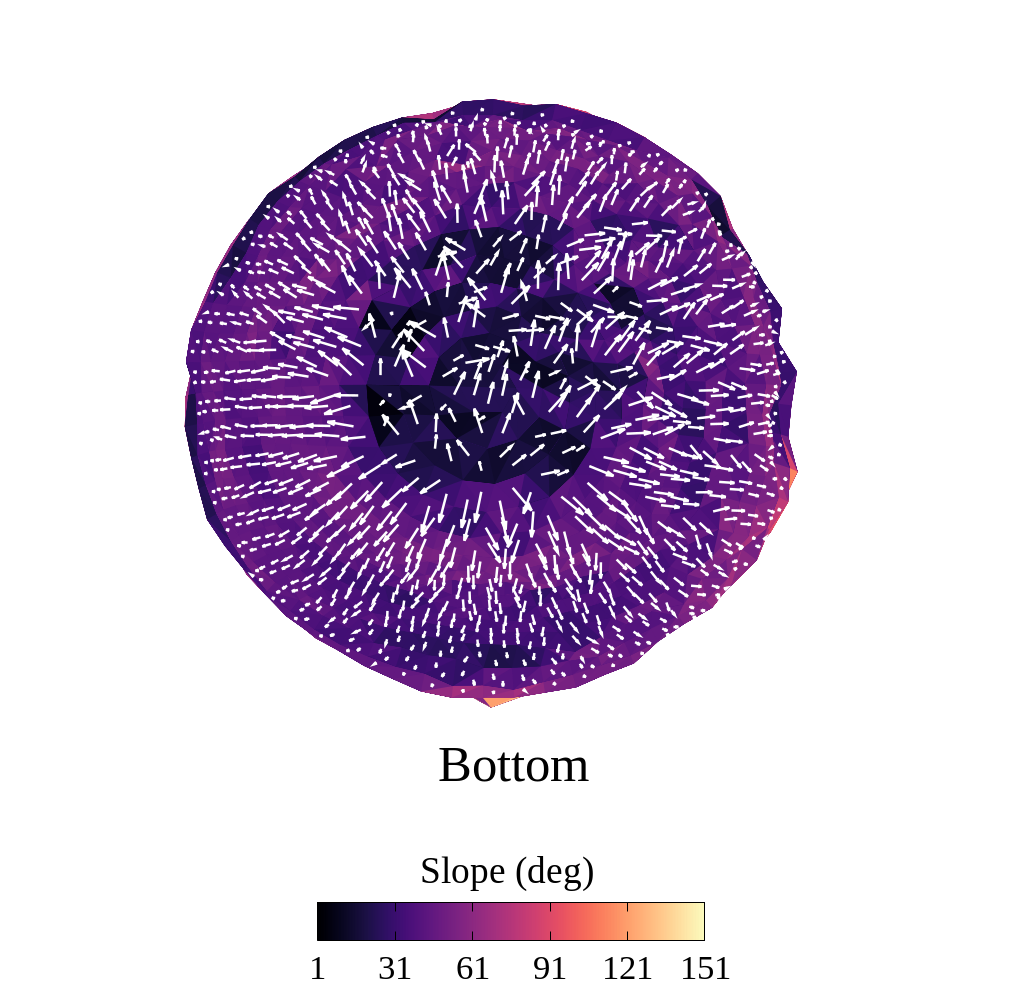}}\\
\subfloat{\includegraphics*[trim = 0mm 0cm 0mm 0cm, width=\columnwidth]{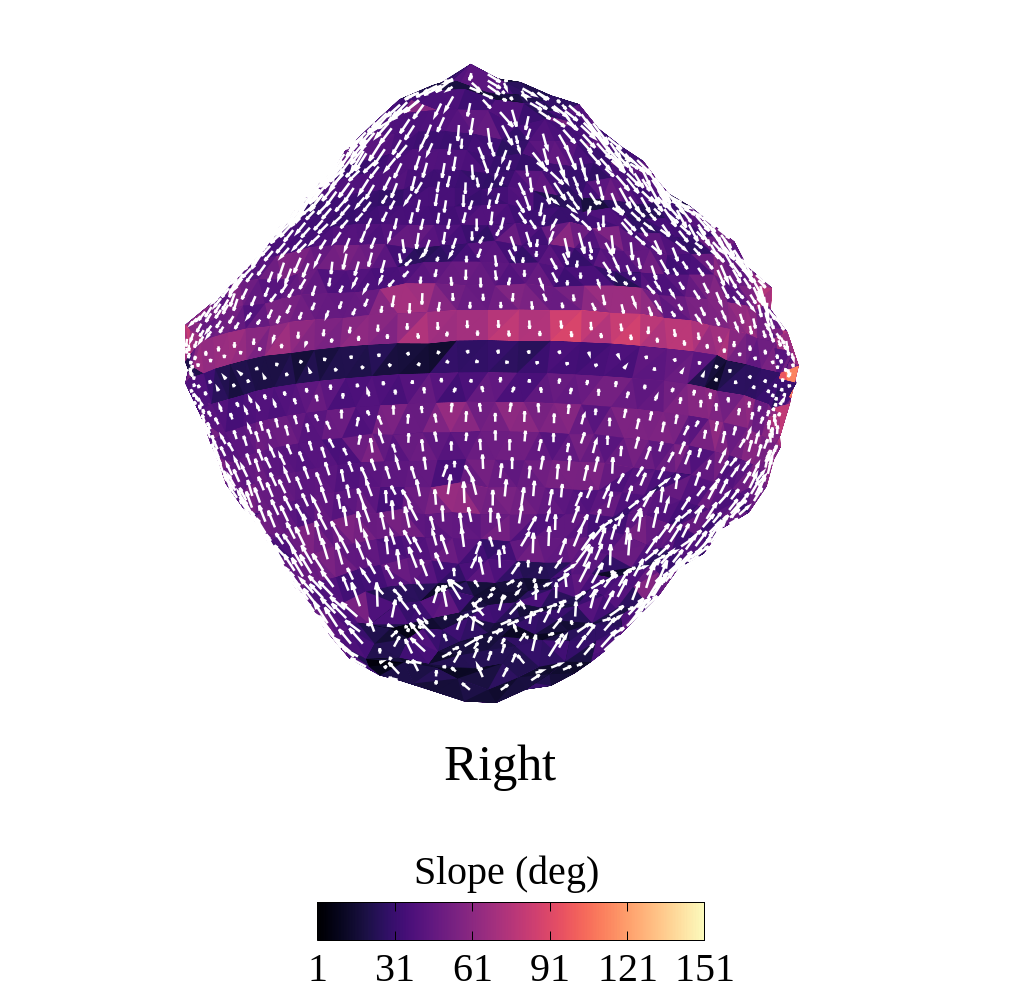}}
\end{center}
\end{framed}
\caption{Directions of the local acceleration vectors tangent to the surface of Alpha, under the Top, Bottom and Right views.}
\label{fig:slope_seta_alpha}
\end{figure}

\begin{figure}
\begin{framed}
\begin{center}
\subfloat{\includegraphics*[trim = 2cm 8cm 0mm 5.6cm,width=\columnwidth]{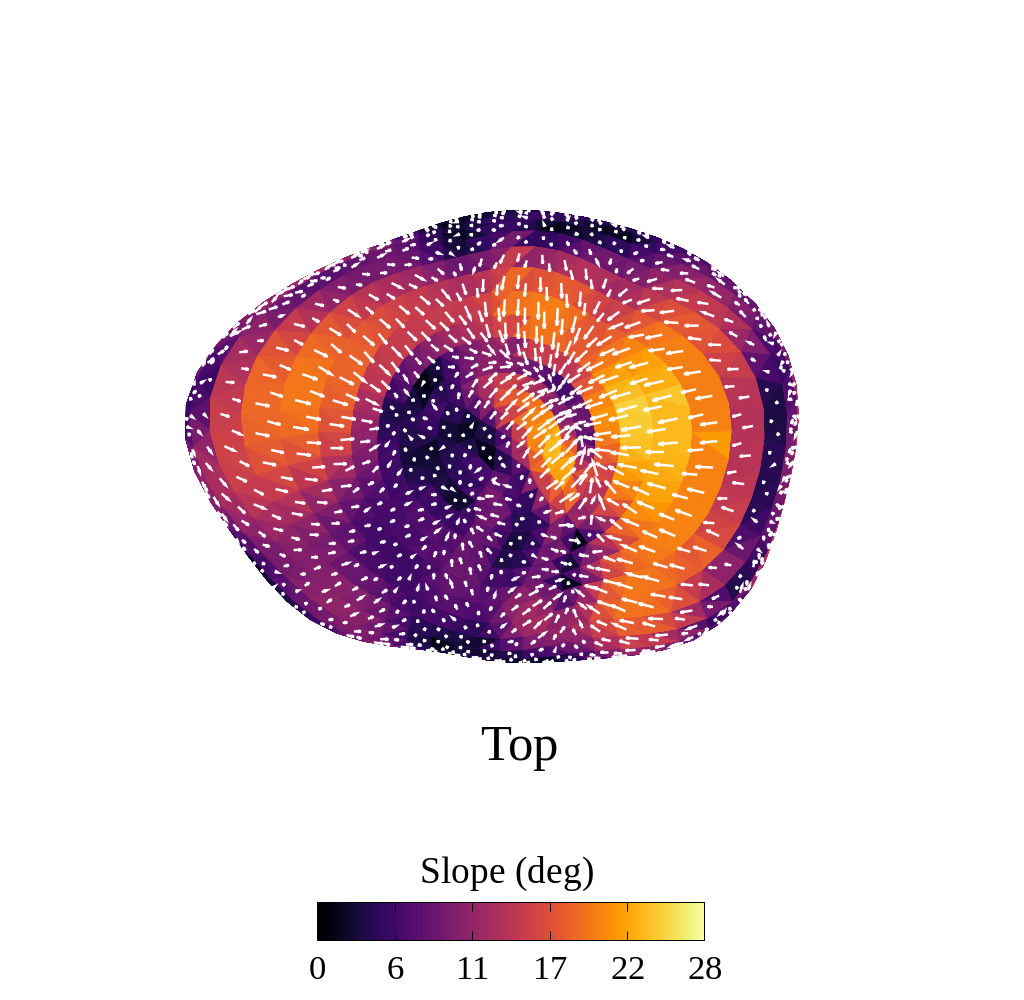}}\\
\subfloat{\includegraphics*[trim = 0mm 8cm 0mm 5cm, width=\columnwidth]{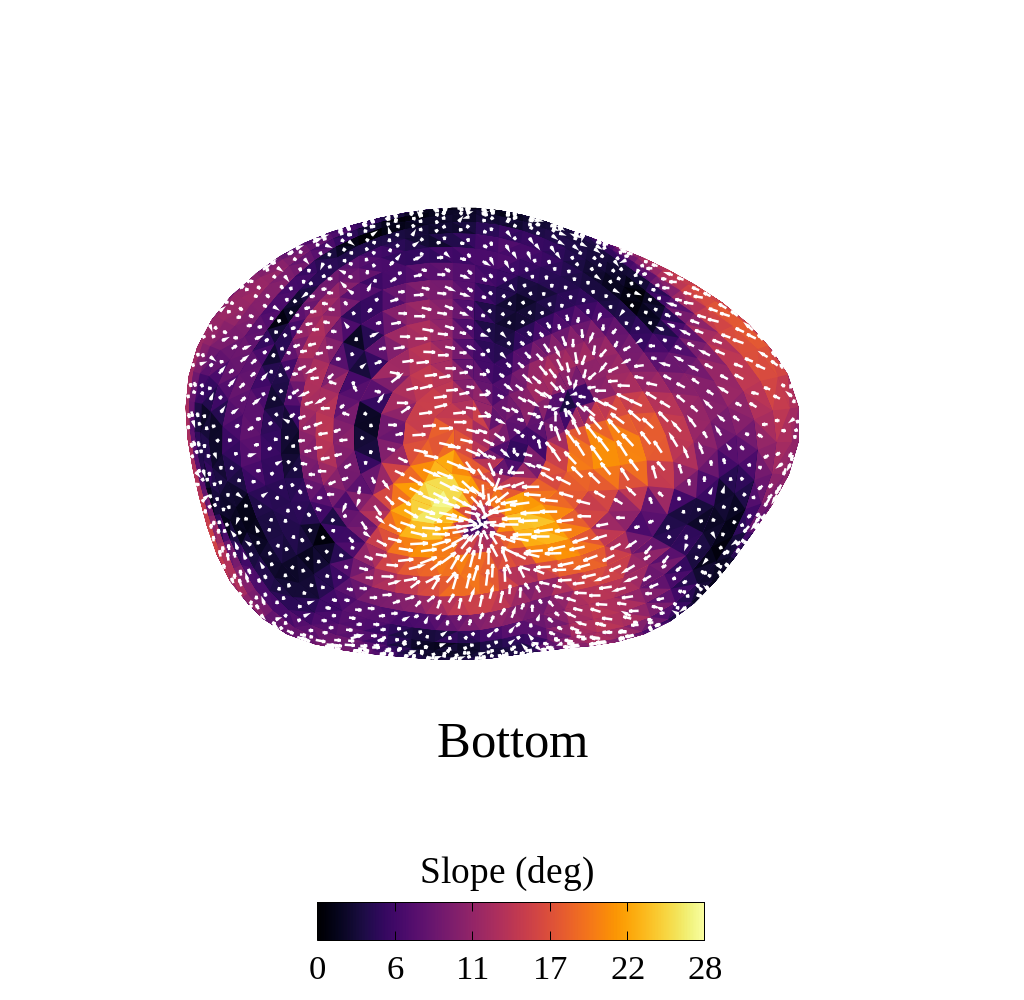}}\\
\subfloat{\includegraphics*[trim = 0mm 0cm 0mm 7cm, width=\columnwidth]{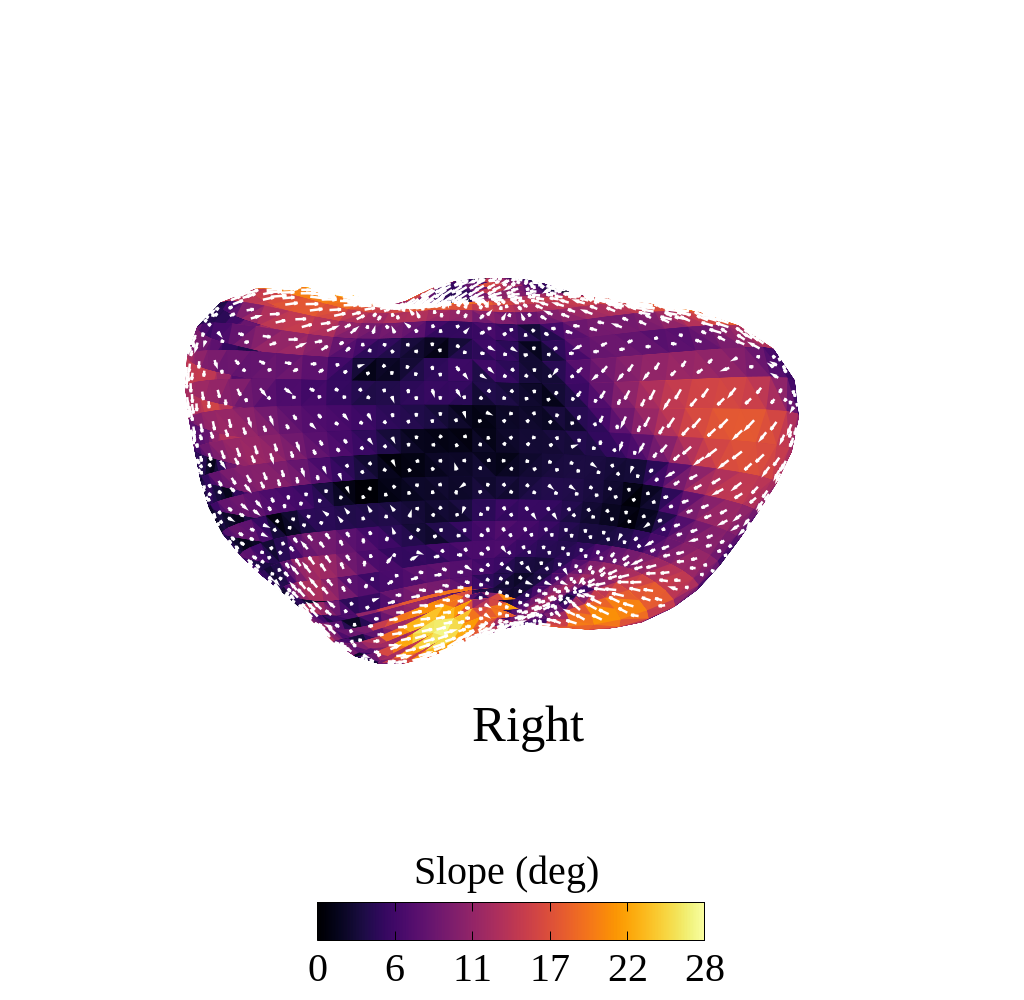}}
\end{center}
\end{framed}
\caption{Directions of the local acceleration vectors tangent to the surface of Beta, under the Top, Bottom and Right views.}
\label{fig:slope_seta_beta}
\end{figure}

\begin{figure}
\begin{framed}
\begin{center}
\subfloat{\includegraphics*[trim = 1.5cm 7.5cm 0mm 7.9cm,width=\columnwidth]{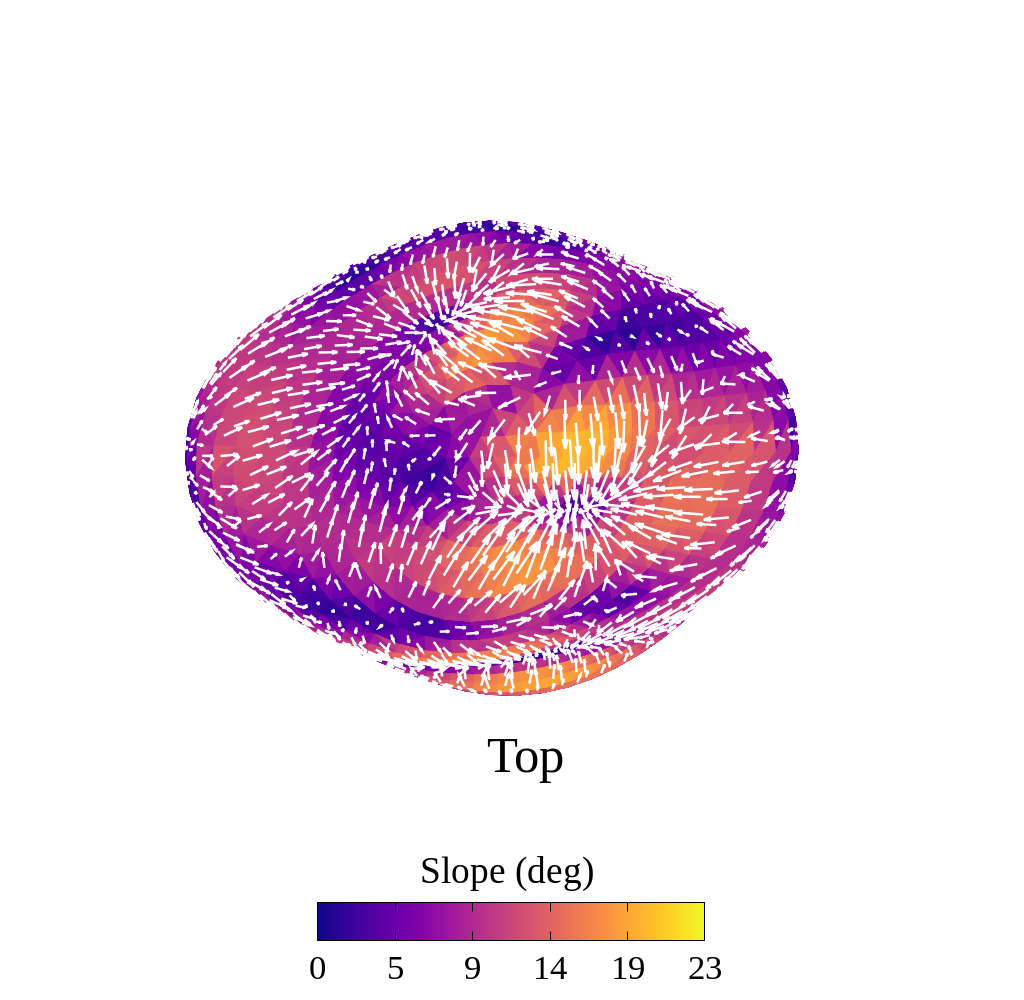}}\\
\subfloat{\includegraphics*[trim = 0mm 7.5cm 0mm 5.2cm, width=\columnwidth]{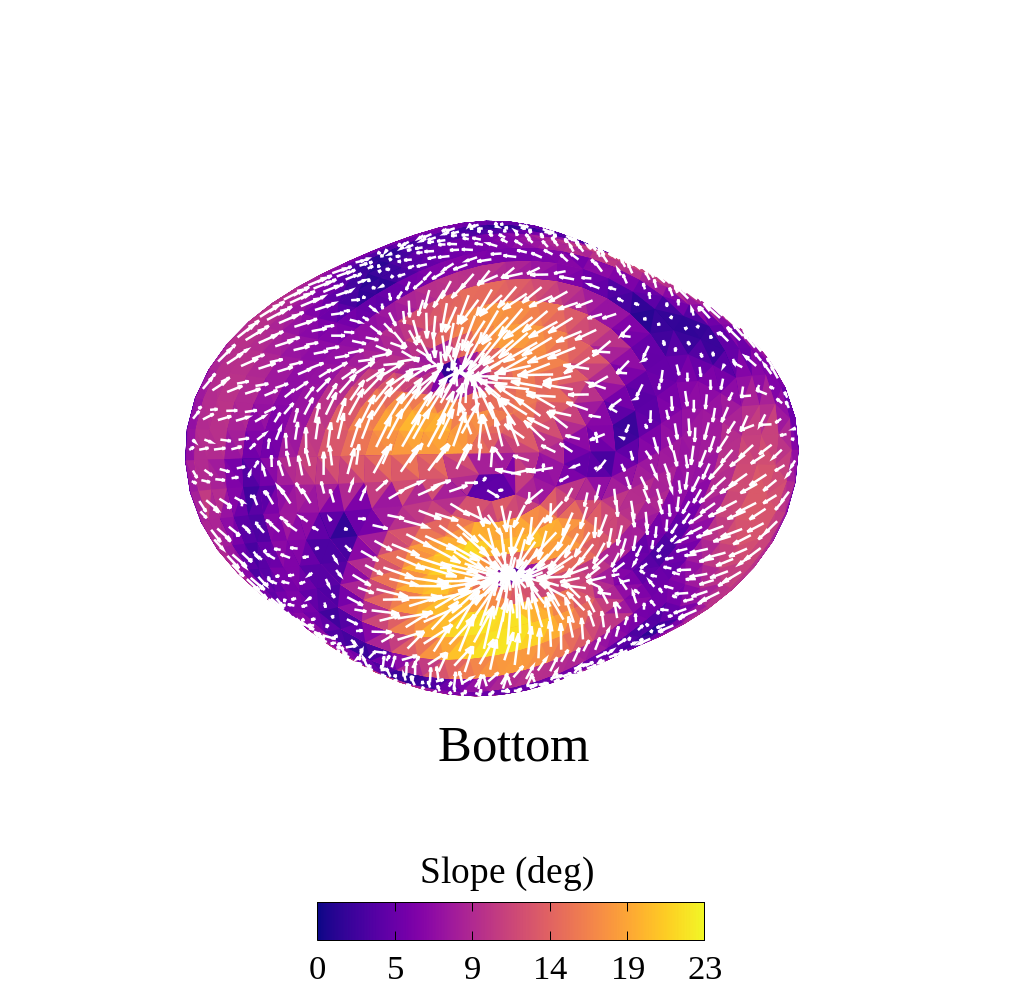}}\\
\subfloat{\includegraphics*[trim = 0mm 0cm 0mm 5.1cm, width=\columnwidth]{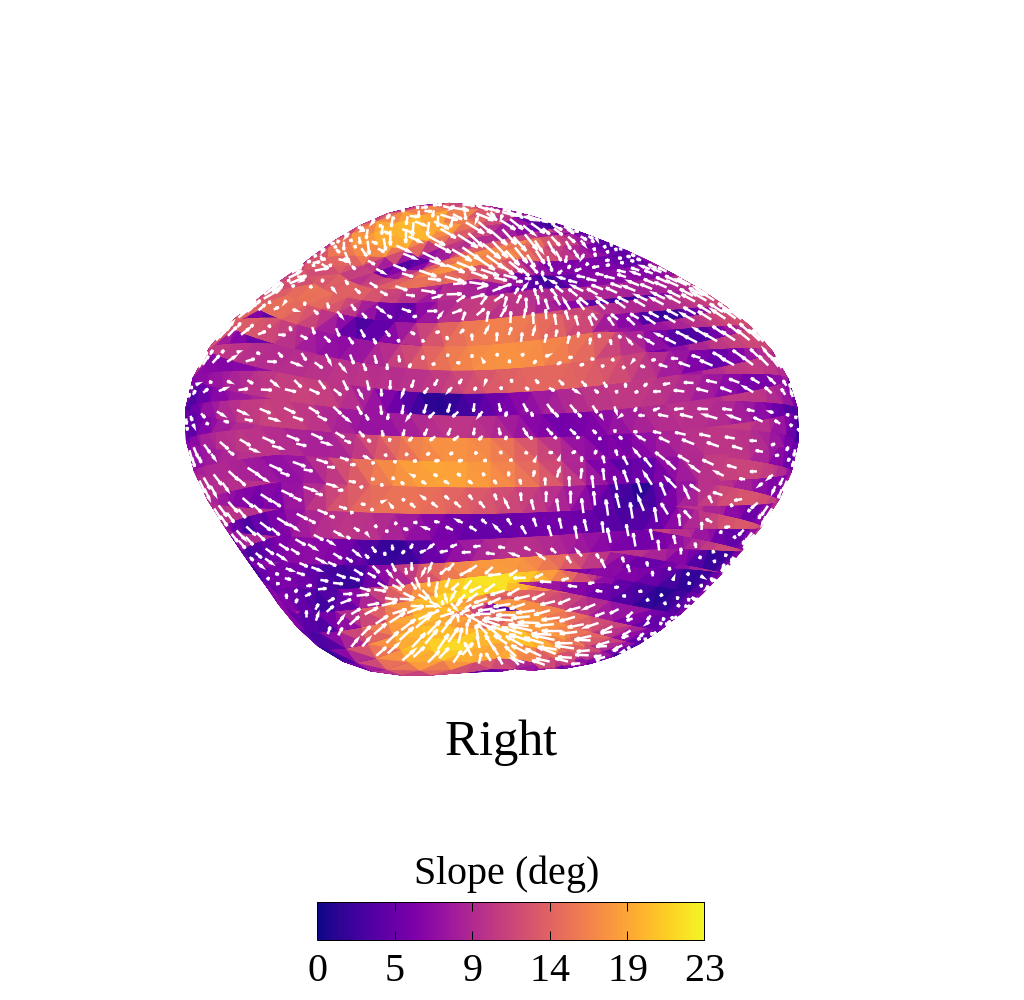}}
\end{center}
\end{framed}
\caption{Directions of the local acceleration vectors tangent to the surface of Gamma, under the Top, Bottom and Right views.}
\label{fig:slope_seta_gamma}
\end{figure}

The mapping of slope angle variation assists in understanding the movement of particles across the surface of the object. A particle initially at rest, positioned in any region on the surface of the body, has its movement tending towards the locations of the lower intensity of the slope angle. Consequently, these locations are prone to accumulating small materials. Therefore, due to the relevant character, we map the slope angle across the surface of the three system 2001 SN$_{263}$ components, as indicated in Fig. \ref{fig:slope}.
In advance, we can notice, by the color bar denoting the intensity of the slope angle of each component (Fig. \ref{fig:slope}), that the surface of Beta carries the highest values of this angle, while Gamma the smaller ones.

Although the maximum value of the slope angle mapped on the Alpha surface is $151^\circ$, there are only a few regions with this high peak of intensity, as can be seen in the Left view of Fig. \ref{fig:slope} (Alpha) a small area in the equatorial region, besides two slightly larger regions located at the ends of the equator of Alpha (Front and Back views of Fig. \ref{fig:slope} - Alpha). Practically all over the surface of Alpha, the slope angle is preferably distributed in a range of $0^\circ$ to $60^\circ$, which does not exclude the fact that it is a still high total variation. And the central region of the poles concentrates the lowest values, well below $30^\circ$. Another peculiarity is the existence of a thin strip (Right view of Fig. \ref{fig:slope} - Alpha) just below the ``waist'' that delimits the equator of Alpha, where the minimum values of the slope angle are also located.
We emphasize that it is precisely the equatorial region that comprises the lowest intensities of the geopotential (Right and Left views of Fig. \ref{fig:geo} - Alpha), while the poles delimit regions with greater geopotential (Top and Bottom views of Fig. \ref{fig:geo} - Alpha). Therefore, the flow tendency of free material across the surface of Alpha is to migrate towards the strip that covers the lowest intensities of the slope angle, located near the equator. To validate this analysis we illustrate the directions of the tangential acceleration vectors, which generally point in the direction of downslope motion. Fig.\ref{fig:slope_seta_alpha} shows the tangential acceleration vectors pointing mostly from the polar regions towards the equatorial region, precisely to the thin band below the ``waist'' of the body. Thus, this region represents a stable resting area, while the polar regions are unstable. This is an expected behavior pattern for small bodies with a high rotation rate, as was verified for the asteroids (1999) KW4 Alpha \citep{Scheeres2006, Scheeres2012} and (101955) Bennu \citep{Scheeres2016}. We verified that the shape model of (1999) KW4 Alpha is very close to that of the largest component of the 2001 SN$_{263}$ system, in addition to having a rotation speed only 1.24 times faster. As illustrated in \citet{Scheeres2012}, (1999) KW4 Alpha also restricts some peaks with high intensity of the slope angle located at its equator, as well as near-zero slopes located at the poles and on a strip just below the equator. Finally, the accumulation of material, as loose and unconsolidated regolith, will tend to form in regions with the lowest slopes and low energy, which in the case of Alpha would explain the salience, a kind of bulge, in its equatorial region. Furthermore, the low accelerations in this region (Right and Left views of Fig. \ref{fig:acel} - Alpha) also contribute to a minimum compression of the material, which may explain the high porosity rate of Alpha, equivalent to 68\% \citep{Becker2015}.

The surface of Beta presents, for the most part, slopes below $17^\circ$ and its maximum value reaches $28^\circ$. In the equatorial region (Right and Left views of Fig. \ref{fig:slope} - Beta), which demarcates minimum values of the slope angle, it was found that the energy values are intermediate in relation to the rest of the surface (Right and Left views of Fig. \ref{fig:geo} - Beta). As we mentioned in Section \ref{geom}, there are two regions that delimit the ``waist'' of Beta and are near the poles (Right view of Fig. \ref{fig:geom} - Beta), and that due to the variation of altitude possibly contain valleys. Precisely in these regions, there is a difference of almost $17^\circ$ among the computed values of the slope angle. By analyzing the variation of the slope angle around the polar regions (Top and Bottom views of Fig. \ref{fig:slope} - Beta), is evident their difference. The north (Top view) is dominated by higher values (more than $20^{\circ}$), while in the south there are only a few spots of high values, located close to the pole. At the north pole (Top view) we observed a large region with high slope values involving the central region, which has slopes close to zero. Such behavior indicates that the presence of loose material across the Beta surface tends to agglomerate at the poles, precisely in the areas that demarcated low slope values. And really the directions of the tangential acceleration vectors point to these regions at the poles that concentrate low slopes, as can be seen in Fig. \ref{fig:slope_seta_beta}. Again this result can be confronted with the environment around the poles (Right view of Fig. \ref{fig:geom} - Beta), since it is characterized by the presence of more protruding regions that are close to the regions that probably contain valleys. In addition to the fact that Beta also has a high porosity rate, about 72\% \citep{Becker2015}.

On the surface of the smallest component of the system 2001 SN$_{263}$, the maximum value of the slope angle does not exceed $23^\circ$, a relatively low value when compared to the values computed for Alpha. Throughout the surface of Gamma, there is a distribution of regions whose slopes are close to zero, with the exception of the poles that have a greater variation (Top and Bottom views of Fig. \ref{fig:slope} - Gamma). However, the variation of the slope angle across the surface of Gamma does not strictly follow the same pattern of energy variation. A clear example is the extremities of the equatorial region that delimit low slopes (Front and Back views of Fig. \ref{fig:slope} - Gamma) and high intensity of geopotential (Front and Back views of Fig. \ref{fig:geo} - Gamma). Especially in the north pole, there is also this opposite behavior, as ascertained in the Bottom view of Figs \ref{fig:geo} and \ref{fig:slope} for Gamma. Since the poles concentrate regions with higher slopes, we can conclude from Fig. \ref{fig:slope_seta_gamma} that the movement of loose material in these regions tends to move towards the near-zero slope locations, according to the pointing direction of the tangential acceleration vectors. As reported in \citet{Scheeres2012}, when a body presents slope values below $30^\circ$, it may be indicative of a relaxed surface. Thus, the possible migration of loose material across the surface of Gamma may have occurred in the past. So much that this body has a low rate of porosity, corresponding to 36\% \citep{Becker2015}.

Recalling that the mapping of the slope (Fig. \ref{fig:slope}) across the surface of the triple system  components differs notably from the values computed in the mapping of the angle tilt (Fig. \ref{fig:tilt}), since the first takes into account the period of rotation of the body and the second one is a purely geometric measure. However, we can still identify a slight pattern in which regions with low tilt angle values also have low slope angle values, and vice versa. Again, we consider the surface of Alpha to be the best option if a space mission decides to land and collect sample material. This selection, besides being based on the variation of the tilt angle, also takes into account that the slope angle variation across the surface of Alpha is not exaggeratedly high, predominantly below $60^\circ$. In addition to the possible existence of loose material accumulated around the equatorial region of Alpha, that delimits a species of bulge, that could be easily collected. Although Gamma has near-zero slopes, it is a body that presents a random variation of both the tilt angle and the slope angle.

\subsection{Potential speed}
\label{speed}

In this section we map the increment of velocity needed to move a particle from one location to another across the whole surface of each body.

Aiming to further understand the dynamics of particle motion across the surface of the asteroid (101955) Bennu, \citet{Scheeres2016} proposed the ``Jacobi speed'' calculation. This speed relates the kinetic energy required to move a particle from an initial location to a final, or more precisely, the amount of energy required for that purpose. Let us consider a particle at rest in the position $\pmb r_0$, which begins its trajectory towards the location $\pmb r_1$, on the surface of the body. Then, the values of the Jacobi constant (equation (\ref{eq: jacobi})) computed in positions $\pmb r_0$ and $\pmb r_1$ of this trajectory can be compared:
\begin{equation}
\frac{1}{2}v_1^2+V(\pmb r_1)=V(\pmb r_0),
 \label{eq: jacobi1}
\end{equation}
to find the speed demanded in this movement
\begin{equation}
\label{eq: velocity}
v_1 = \sqrt{-2(V(\pmb r_1)-V(\pmb r_0))}.
\end{equation}
The above expression determines whether the trajectory of a particle from $\pmb r_0$ to $\pmb r_1$ is possible or not. If the subtraction $(V(\pmb r_1)-V(\pmb r_0))$ returns a negative result, the movement between these two locations is allowed. Otherwise the movement will be prohibited. Equation (\ref{eq: velocity}) provides the velocity as a function of the relative energies evaluated at any two positions on the surface of the body. In this way, the displacement between two locations may generate a gain or loss of velocity depending on the region on the surface of the body that the particle will travel. From this idea, it is also possible to map the geopotential on the surface of the body and identify the lowest and highest energy locations.

\begin{figure}
\begin{center}
\subfloat{\includegraphics*[trim = 0mm -0.3cm 0mm 0mm,width=\columnwidth, frame]{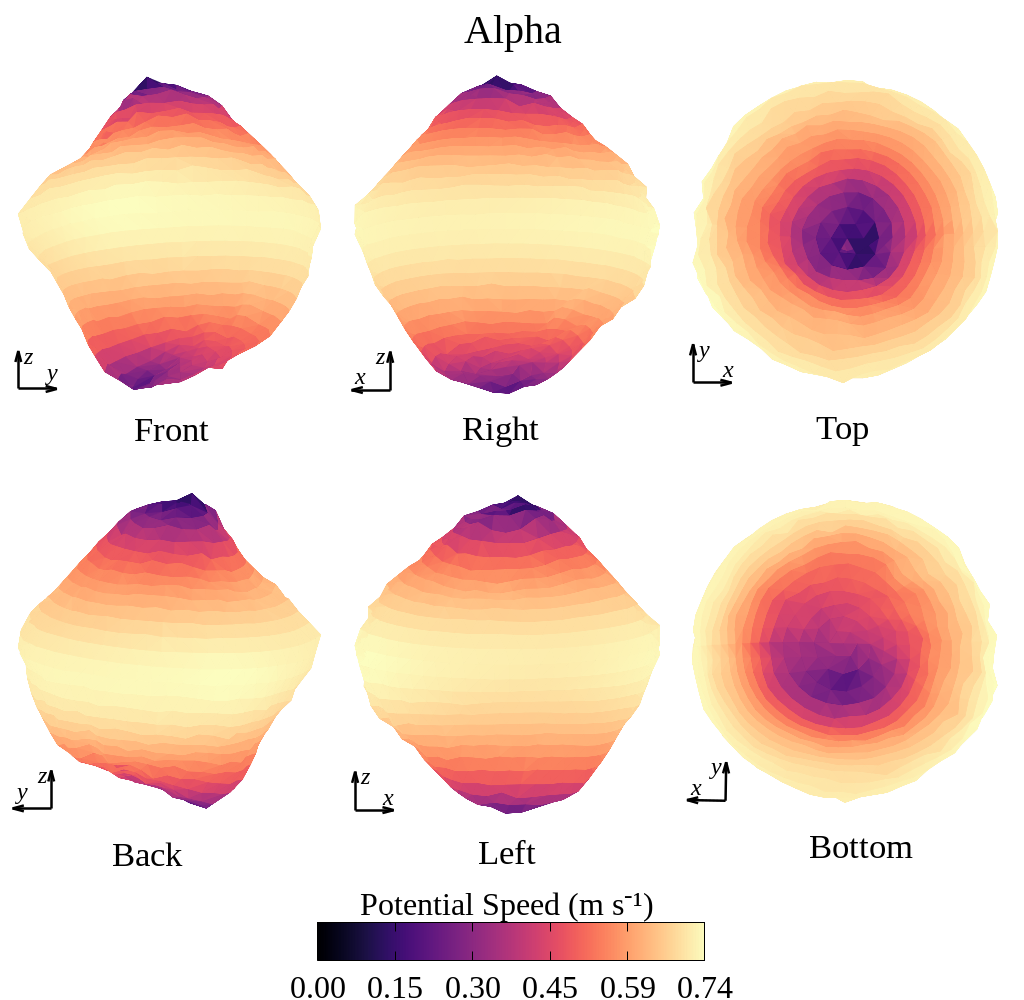}}\\
\subfloat{\includegraphics*[trim = 0mm 7cm 0mm 0mm, width=\columnwidth, frame]{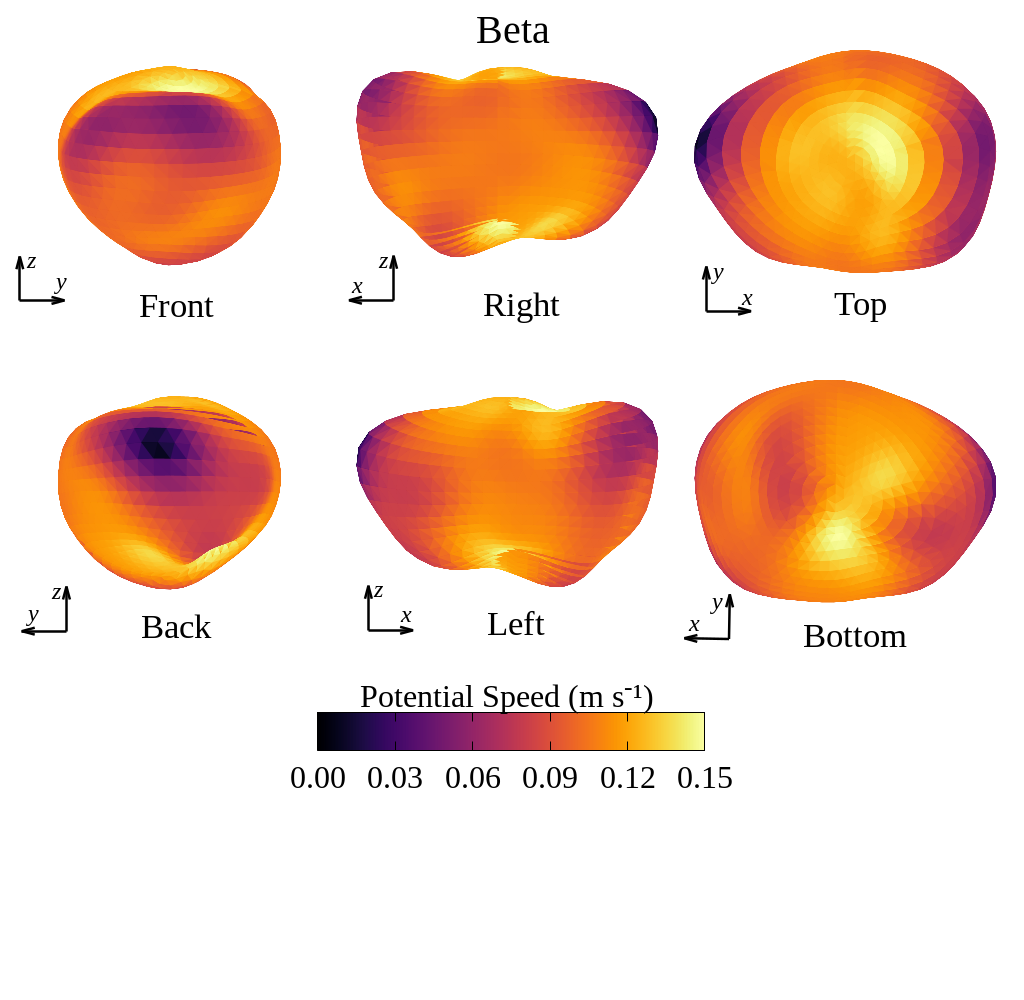}}\\
\subfloat{\includegraphics*[trim = 0mm 5.3cm 0mm 0mm, width=\columnwidth, frame]{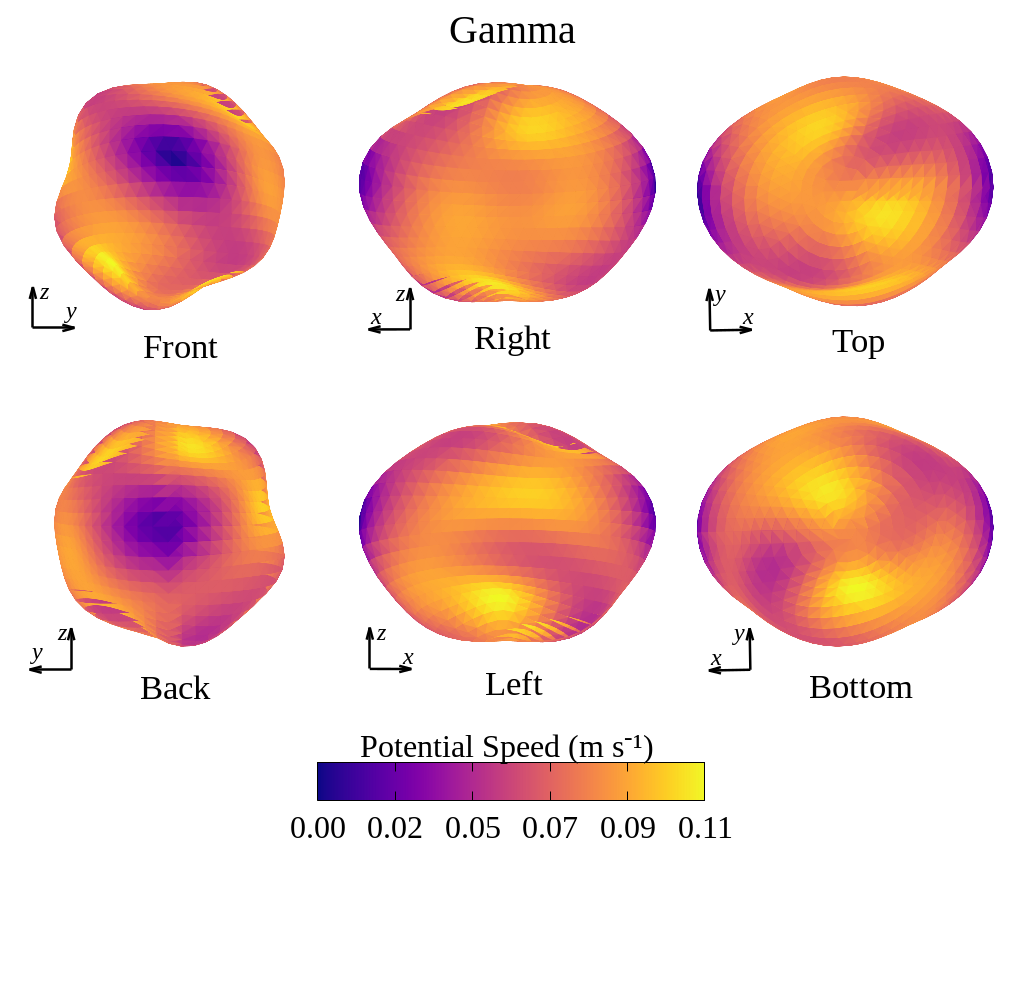}}
\end{center}
\caption{Relative Jacobi speeds $\Delta v_j$ computed across the surface of Alpha, Beta and Gamma, respectively. Note that the colour scales are different for each body.}
\label{fig:speed}
\end{figure}

Finally, this energy variation between any two positions of the particle in the body-fixed frame is expressed by means of the ``Jacobi speed'', a term adopted by \citet{Scheeres2016} to define the expression
\begin{equation}
v_j=\sqrt{-2V(\pmb r)}.
 \label{eq: velocity1}
\end{equation}
Note, in this expression, that $v_j$ will have its maximum value precisely at the point where we obtain the lowest geopotential, among the values computed on the surface of the body. So, we define $v_j^m$ as the minimum value of the Jacobi speed among those calculated on the surface of the body, which establishes the highest point in the geopotential. Consequently, the mapping of the relative values of Jacobi speed measured across the body surface is performed as follows
\begin{equation}
\label{eq: delta}
\Delta v_j(\pmb r) = \sqrt{(v_j(\pmb r))^2-(v_j^m)^2},
\end{equation}
where $\Delta v_j$ determines the necessary increase in the speed of a particle that travels from the highest point in the geopotential to another location on the surface of the body. On the other hand, $\Delta v_j$ can also represent the velocity required for a particle that starts its motion at any location $\pmb r$ toward the highest point in the geopotential. Since this amount provides an additional interpretation of the geopotential involving the movement of particles on the surface of the body, we map its value across the surface of Alpha, Beta, and Gamma, as shown in Fig. \ref{fig:speed}.

We observed a band around the equator (Right and Left views of Fig. \ref{fig:speed} - Alpha) of the largest component of the triple system highlighting the maximum value of $v_j$, that is, a region of minimum geopotential and where the lowest point in the geopotential is located. While the minimum value of $v_j$ is concentrated in the poles (Top and Bottom views of Fig. \ref{fig:speed} - Alpha), in regions of maximum geopotential and where the highest point in the geopotential is located. These results are in conformity with the behavior of the geopotential in the equator (Right and Left views) and the poles (Top and Bottom views), as shown in Fig. \ref{fig:geo} - Alpha. At the highest point (poles) of the surface of Alpha we have $v_j^m=0.93136$ m s$^{-1}$, and for a particle to complete its trajectory from that point towards the equatorial region, a value $\Delta v_j$ will be added to its velocity in order to boost its motion. That required amount is about $0.743$ m s$^{-1}$.

For the two smaller bodies of the triple system, the behavior of the relative values of Jacobi speed is contrary to that observed on the surface of Alpha. We noticed a similar pattern in the mapping of $\Delta v_j$ values on the surface of Beta and Gamma. The maximum value of $v_j$ is distributed in areas near the poles (Top and Bottom views of Fig. \ref{fig:speed} - Beta and Gamma), in regions of lower geopotential. The mean values of $v_j$ are preferably located in the equatorial region (Right and Left views of Fig. \ref{fig:speed} - Beta and Gamma). While it is at the extremities of the equator (Front and Back views of Fig. \ref{fig:speed} - Beta and Gamma) where $v_j$ is minimal and, consequently, are locations where the geopotential is highest. Again this analysis agrees with the mapping of the geopotential shown in Fig. \ref{fig:geo} for the Beta and Gamma bodies. In the location of the highest point in the geopotential of Gamma, the value of $v_j^m$  is about 13.5\% smaller than that of Beta, whose value is $0.27052$ m s$^{-1}$.

\section{Features of the dynamical environment}
\label{dynamical environment}
Since one of the goals of the ASTER mission is to launch a small probe to orbit and explore the triple system 2001 SN$_{263}$, it is essential to investigate the dynamic environment around these bodies. Thus, in this section, we present the mapping of the energy variation $J$, through zero-velocity curves, in order to identify the limits of physical motion so that a spacecraft does not have an impact on the surface of the body. Moreover, zero-velocity curves also allow the identification of external equilibrium points in the gravitational field of the body. Lastly, the location of the lower value of the Jacobi constant identifies the zero-velocity curve that surrounds the body and, by comparing all other values of $J$ to this, it is possible to identify whether the motion is confined to the internal region or the external  region to this curve. Thus, the non-occurrence of impacts on the body surface is conditioned to suitable $J$ values. This analysis can be applied to space missions planning to orbit asteroids, for example.

\subsection{Zero-velocity curves}
\label{zero curves}

\begin{figure*}
\begin{center}
\subfloat[]{\includegraphics*[trim = 1cm 0cm 0.5cm 0cm,
width=\columnwidth]{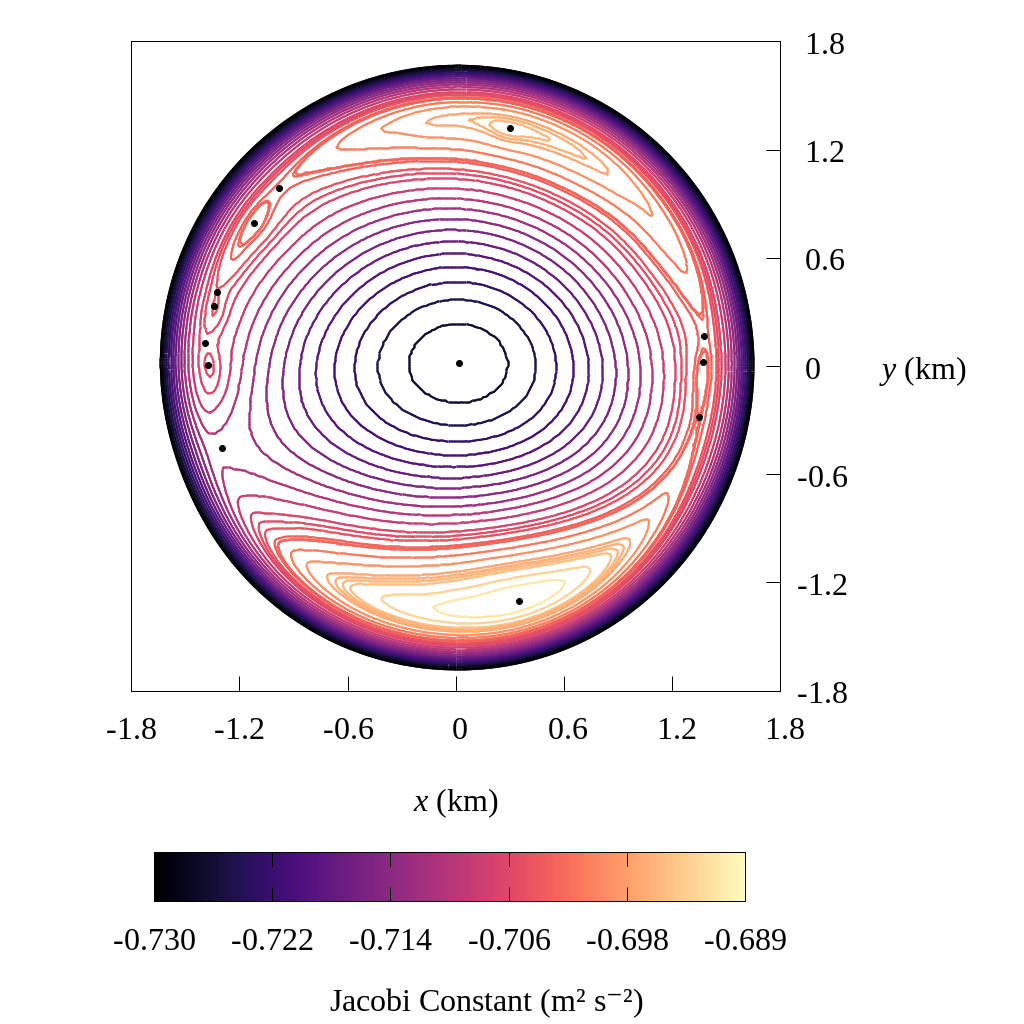}\label{fig: cvz alpha}}
\subfloat[]{\includegraphics*[trim = 1cm 0cm 0.5cm 0cm, width=\columnwidth]{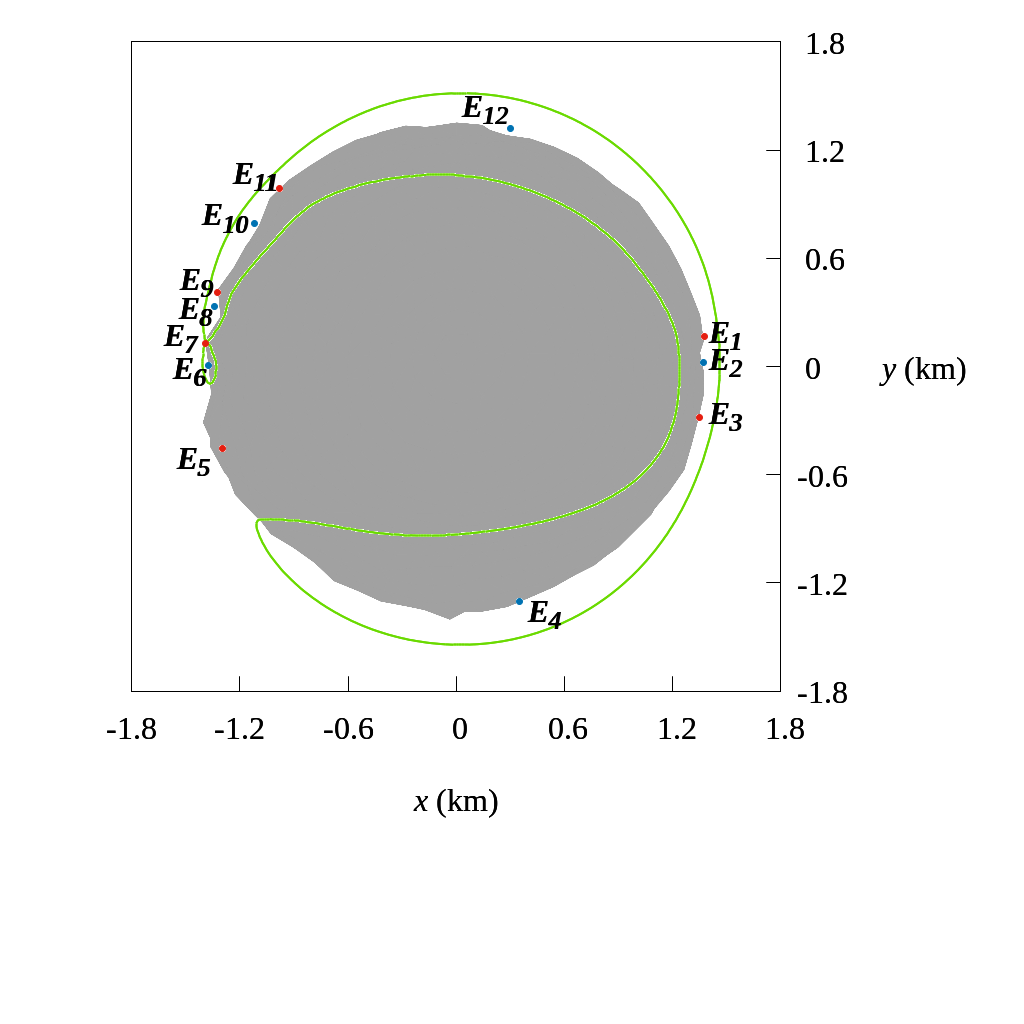}\label{fig: pontos alpha}}\\
\end{center}
\caption{(a) Contourplots of zero-velocity curves in the $xoy$ plane corresponding to the Alpha component of the triple system 2001 SN$_{263}$. Black dots represent equilibrium points. The color scale determines the value of the Jacobi constant. (b) Location of the equilibrium points of the Alpha component in the $xoy$ plane. Red dots are classified topologically as Saddle-Centre-Centre and blue dots as Sink-Source-Centre. The green line defines the rotational Roche Lobe of the Alpha body, projected on the $xoy$ plane.}
\label{fig: Alpha curves and points}
\end{figure*}

\begin{figure*}
\begin{center}
{\includegraphics*[trim = 1cm 0cm 0.5cm 0cm,
width=\columnwidth]{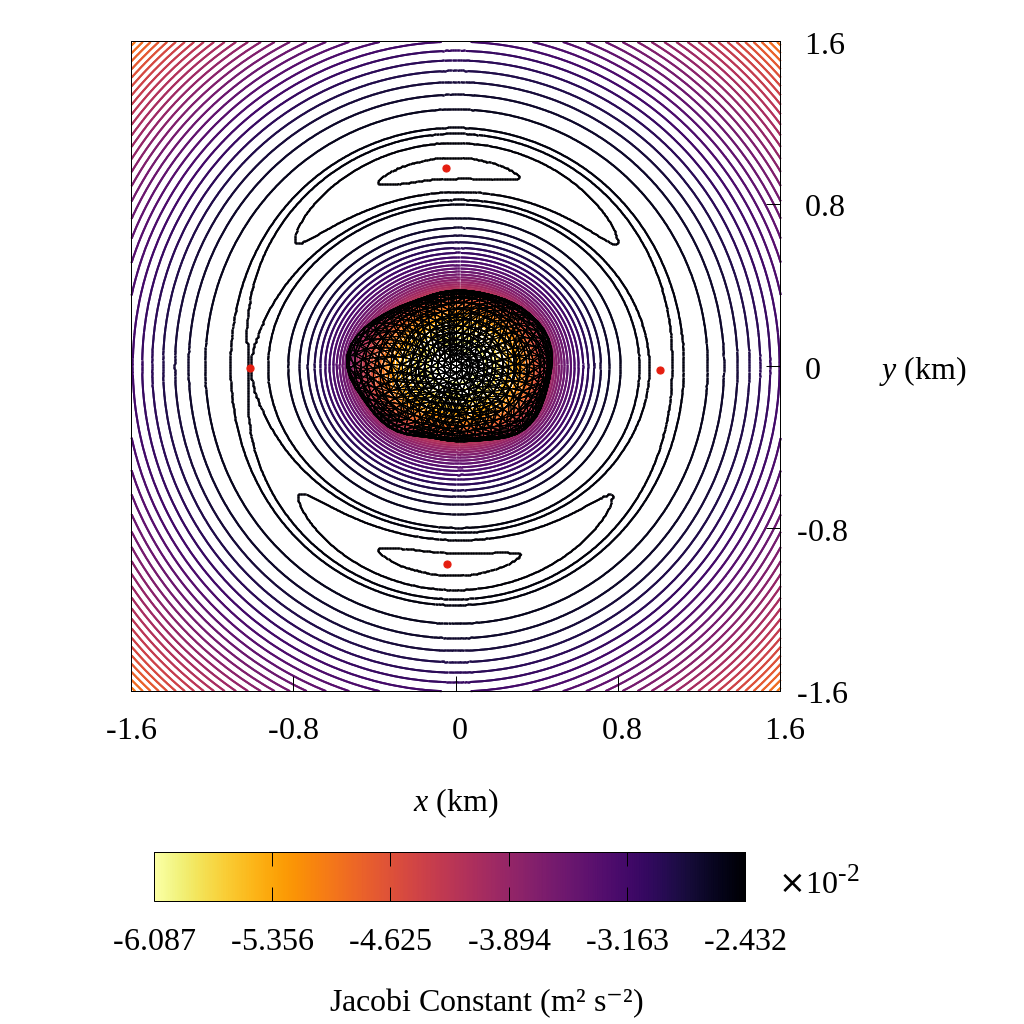}}
{\includegraphics*[trim = 1cm 0cm 0.5cm 0cm, width=\columnwidth]{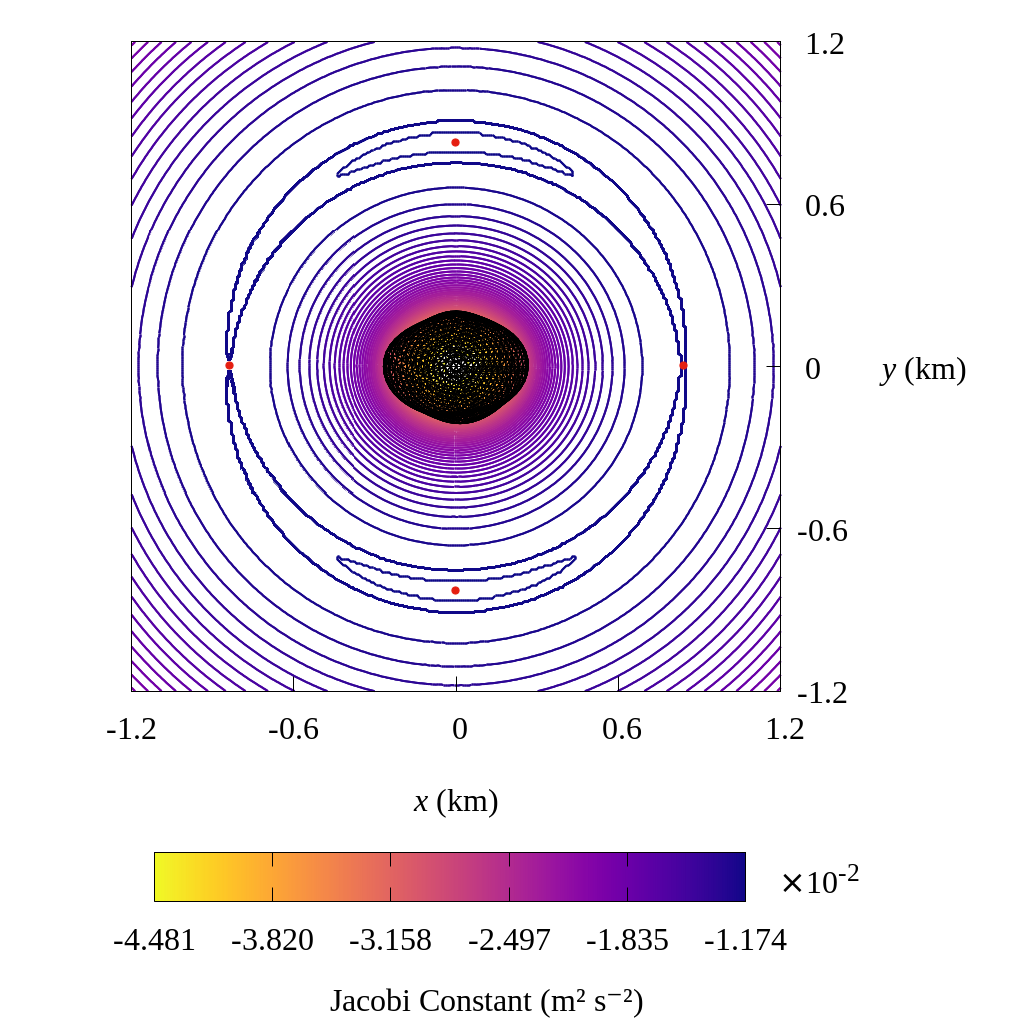}}\\
\end{center}
\caption{Contourplots of zero-velocity curves in the $xoy$ plane corresponding to the Beta and Gamma components of the triple system 2001 SN$_{263}$, respectively. Red dots represent equilibrium points. The color scale determines the value of the Jacobi constant.}
\label{fig: Beta and Gamma curves}
\end{figure*}
Equation (\ref{eq: jacobi}) presented in Subsection \ref{equation of motion} determines, fairly accurately, regions in the vicinity of the central body where the movement of a particle is permitted or prohibited. That is, the Jacobi constant establishes a restriction on the orbital motion of particles around the body. This restriction arises directly from the inequality below, since the term $\frac{1}{2}v^2 \geq 0$:
\begin{equation}
J-V(\pmb r) \ge 0.
\label{eq: zero}
\end{equation}
The above expression separates space into forbidden regions and regions accessible to the particle. Thus, regions whose $V(\pmb r) < J$ offer no impediment, at first, to the movement of a particle. While regions, where the trajectory of a particle is prohibited, violate inequality, since we have $V(\pmb r)> J$. If these forbidden regions divide into several disjoint regions in space, particle displacement between such regions will never occur. Finally, zero-velocity surfaces, or zero-velocity curves in the planes, determine the boundary of these regions, governed by $V(\pmb r)=J$.

The contour plots of the zero-velocity curves in the $xoy$ plane corresponding to the polyhedral model of the triple system 2001 SN$_{263}$ components are illustrated in Figs \ref{fig: cvz alpha} and \ref{fig: Beta and Gamma curves}. Note that the curves are plotted according to the intensity of the Jacobi constant. In addition, the arrangement of these curves delimits the location of equilibrium points around the body.

\subsection{Equilibrium points}
\label{point}

As mentioned earlier, zero-velocity curves assist in identifying equilibrium points in the gravitational field of the body. In Figs \ref{fig: cvz alpha} and \ref{fig: Beta and Gamma curves} it is possible to observe regions where these curves intersect or even close in upon themselves, delimiting the so-called equilibrium points, to critical values of $J$. For small irregular bodies, such as asteroids and comets, there are usually four external and one internal equilibrium points, according to the investigation of \citet{Wang2014} regarding the location of these points in the gravitational field of 23 bodies.

Since at the equilibrium point the resulting force is null, then its location satisfies the following condition \citep{Jiang2014}:
\begin{equation}
\frac{\partial V(x,y,z)}{\partial x}=\frac{\partial V(x,y,z)}{\partial y}=\frac{\partial V(x,y,z)}{\partial z}=0
\label{eq: condition}
 \end{equation}
The number of solutions of this equation depends on some factors such as the irregular shape,
density, and speed of rotation of the body. Therefore, the number of equilibrium points is not fixed, and their arrangement in the vicinity of the body may change, despite being common the existence of only four external points. 

Table \ref{table: loc} identifies the coordinates of equilibrium points in the gravitational field of the Alpha, Beta, and Gamma components of the system 2001 SN$_{263}$. For better visualization, Fig. \ref{fig: pontos alpha} shows the projection of the equilibrium points of the largest component in the $xoy$ plane. Note that the points were named $E_1$, $E_2$, and so on, from the $+x$ axis counterclockwise. Similarly, we apply this same procedure to Beta and Gamma, whose location of equilibrium points in the $xoy$ plane is illustrated in Fig. \ref{fig: Beta and Gamma curves}. Although all bodies have an internal central equilibrium point, we will not analyze it, since this escapes the objective of this paper. Comparing the $x$, $y$, and $z$ coordinates of Table \ref{table: loc}, we conclude that the equilibrium points of the three bodies are practiced in the equatorial plane. The external points of both Beta and Gamma are symmetrical about the $x$ and $y$ axes and are positioned away from the body surface. While for the primary component the opposite occurs, the points are very close to the body surface, and clearly there is no symmetry with respect to the $x$ and $y$ axes. Furthermore, Fig. \ref{fig: pontos alpha} presents in the perpendicular direction only one equilibrium point on each side, and a much higher concentration of points in the left side region compared to the opposite region, where only the points $E_1$, $E_2$ and $E_3$ are located. Indeed, the arrangement of these points correlates with the irregular gravitational potential of the body and its high rate of rotation. This investigation is the subject of a detailed study that is in progress.

Next, the equations of motion relative to the equilibrium point were linearized and using the same method as \citet{Jiang2014}, we analyzed the disposition of the six eigenvalues of the characteristic equation in the complex plane. This practice is performed in order to investigate the topological structure of the equilibrium point. The analysis of the eigenvalues exposed in Table \ref{table:autovalores} (appendix \ref{eigenvalues}) shows that the topological structure of the equilibrium points of Alpha alternates in Saddle-Centre-Centre and Sink-Source-Centre. The equilibrium points $E_1$, and $E_3$, located near the $x$ axis, of the Beta and Gamma components are Saddle-Centre-Centre type. Already the topological structure of the points $E_2$ and $E_4$ of Gamma is Centre-Centre-Centre, while for Beta $E_2$ it is Sink-Source-Centre and $E_4$ is Centre-Centre-Centre.

\begin{table}
\centering
\caption{Location of equilibrium points of Alpha, Beta and Gamma, respectively, and the value of the Jacobi constant, $J$.}
\label{table: loc} 
\begin{tabular}{c|r|r|r|c}
\hline\hline
\multicolumn{5}{c}{Alpha}\\
\hline\hline
  Point & $x$ (km) &  $y$ (km) &  $z$ (km) & $J$ (m$^{2}$ s$^{-2}$)\\  
\hline
$E_1$ & 1.375 & 0.166 & 0.024 & $-$0.7025 \\ 
$E_2$ & 1.372 & 0.020 & 0.044 & $-$0.7022 \\ 
$E_3$ & 1.350 & $-$0.282 & 0.064 & $-$0.7028 \\ 
$E_4$ & 0.354 & $-$1.305 & 0.099 & $-$0.6911 \\ 
$E_5$ & $-$1.294 & $-$0.452 & 0.021 & $-$0.7103 \\ 
$E_6$ & $-$1.373 & 0.004 & 0.041 & $-$0.7050 \\ 
$E_7$ & $-$1.388 & 0.127 & 0.025 & $-$0.7054 \\ 
$E_8$ & $-$1.338 & 0.331 & 0.014 & $-$0.7040 \\ 
$E_9$ & $-$1.325 & 0.413 & 0.013 & $-$0.7041 \\ 
$E_{10}$ & $-$1.116 & 0.795 & 0.016 & $-$0.7016 \\ 
$E_{11}$ & $-$0.980 & 0.989 & 0.002 & $-$0.7024 \\ 
$E_{12}$ & 0.304 & 1.318 & 0.013 & $-$0.6949 \\ 
\hline\hline
\multicolumn{5}{c}{Beta}\\
\hline\hline
  Point & $x$ (km) &  $y$ (km) &  $z$ (km) & $J$ (m$^{2}$ s$^{-2}$)\\  
\hline
$E_1$ & 1.009 & $-$0.021 & 0.008 & $-$0.0249 \\ 
$E_2$ & $-$0.041 & $-$0.974 &  0.000 &$-$0.0243 \\ 
$E_3$ & $-$1.016 & $-$0.009 & 0.006 & $-$0.0250 \\ 
$E_4$ & $-$0.048 & 0.974 & 0.000 & $-$0.0243 \\ 
\hline\hline
\multicolumn{5}{c}{Gamma}\\
\hline\hline
  Point & $x$ (km) &  $y$ (km) &  $z$ (m) & $J$ (m$^{2}$ s$^{-2}$) \\  
\hline
$E_1$ & 0.839 & 0.004 & 0.000 & $-$0.0118 \\ 
$E_2$ & $-$0.002 & $-$0.829 & 0.000 & $-$0.0117 \\ 
$E_3$ & $-$0.839 & 0.003 & 0.000 & $-$0.0118 \\ 
$E_4$ & $-$0.001 & 0.829 & 0.000 & $-$0.0117 \\ 
\hline\hline
\end{tabular}
\end{table}

Considering the topological structure of the equilibrium points of the system 2001 SN$_{263}$ bodies, we investigated the presence of stable regions around these points. We distributed massless particles in a region containing the points and performed numerical simulations to identify the size of these regions. For obvious reasons, we did not do this kind of simulation for Alpha, since the equilibrium points are too close to its surface.

For the simulations, we adopted the N-BoM package, which is an N-body integrator that uses the Bulirsch-Stoer algorithm and manipulates the gravitational potential of irregular bodies by a mass concentration (MASCONS) model \citep{Geissler1996}. This method assumes a number $N$ of equally spaced points distributed within the body volume, and the sum of the mass assigned to each point results in the total mass of the object. The MASCONS model is not as accurate as the polyhedra method but requires  much less time to integrate orbits, since it eliminates various calculations integral formalism. And the results ensure good accuracy compared to the polyhedron method. According to \citet{borderes2018}, the irregular gravitational potential of the body is given by the sum of the gravitational potential computed at each mass point:
\begin{equation}
 U (x, y, z)= \sum_{i=0}^{N} \frac{Gm}{r_i},
 \label{eq: mascon}  
\end{equation}
where $N$ is the number of mascons, $m=\frac{M}{N}$ is the mass of each mascon, such that $M$ is the total mass of the object, and $r_i$ computes the distance between the mascon and the orbiting particle. 

To reproduce the gravitational potential of Beta, we consider a 15.5 m spaced grid that generated 64259 mascons with $m \approx 3.73489 \times 10^{6}$ kg each. Then, we integrate 10 thousand massless particles randomly distributed in the region of $800 < r < 1200$ m, a range of a few meters from the Beta surface and encompassing all external equilibrium points. Initially, we assume all particles in circular orbits and in the equatorial plane. To create the flat cloud, we distributed among the particles random values from 0$^{\circ}$ to 360$^{\circ}$ for the mean anomaly. The equations of motion were integrated for a timespan of 10 years ($\sim 6600$ spin periods of Beta).

\begin{figure}
\begin{center}
\includegraphics*[trim = 0cm 7cm 0cm 0cm,width=\columnwidth]{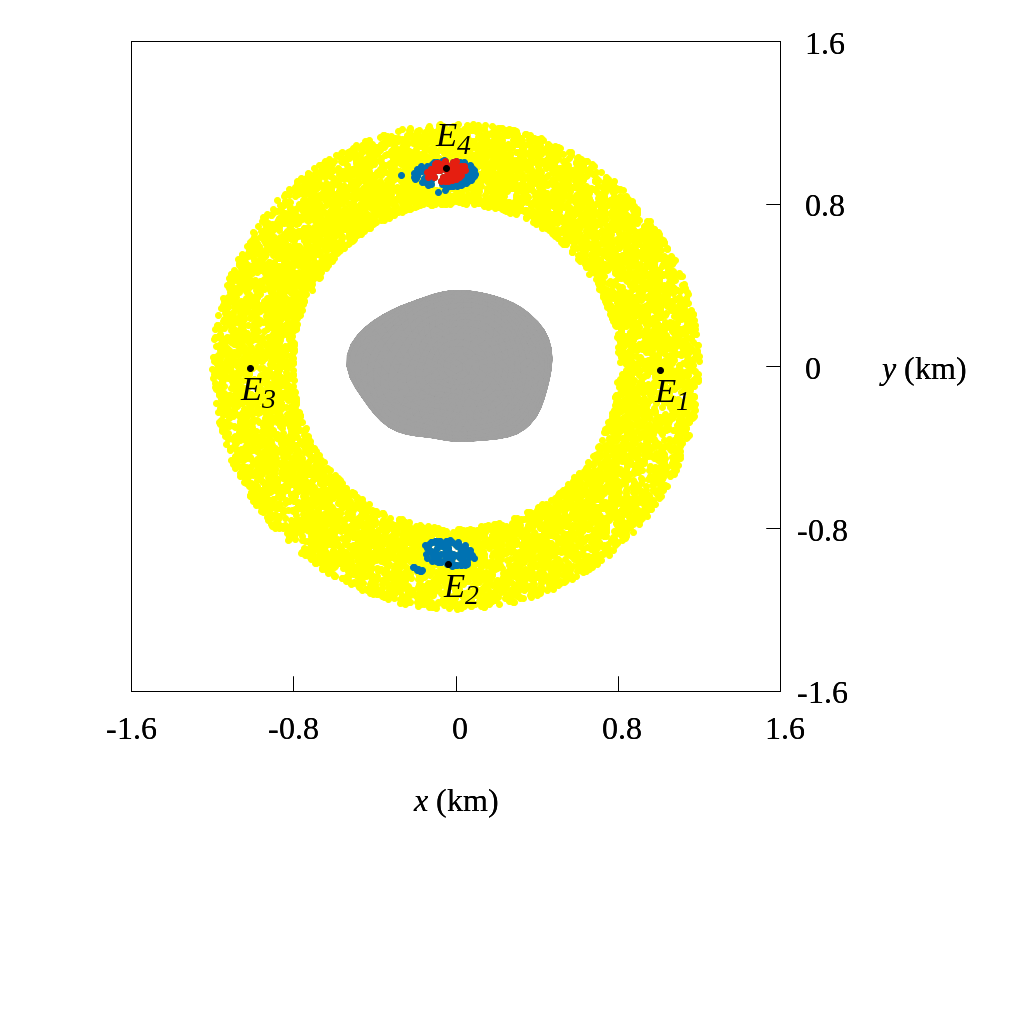}
\end{center}
\caption{The yellow dots represent the positions in the xy plane of 10 thousand massless particles randomly distributed around the Beta component of the system 2001 SN$_{263}$ at the initial time. The blue dots represent the initial positions of the particles that survived after 10 years of integration. The red dots represent the initial positions of the particles that survived after 10 years of integration under the additional gravitational influence of the Alpha and Gamma components. Black dots represent the position of equilibrium points in the vicinity of Beta.}
\label{fig:beta anel}
\end{figure}

\begin{figure}
\begin{center}
\includegraphics*[width=\columnwidth]{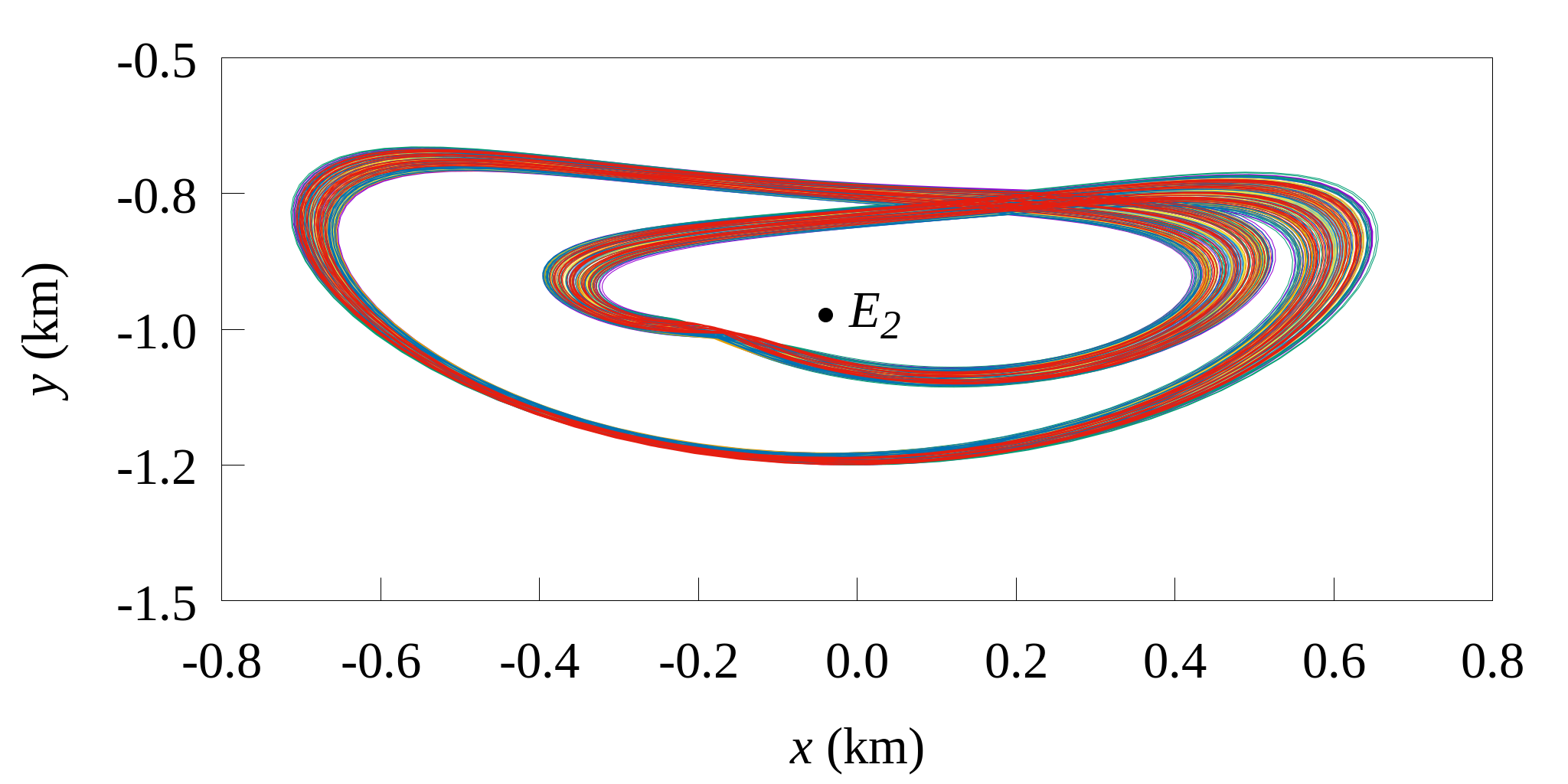}
\end{center}
\caption{Orbits of particles that survived after 10 years of integration and are initially located in the smaller disconnected region of the stable region around equilibrium point $E_2$ (Fig. \ref{fig:beta anel}, in blue).}
\label{fig: resonance}
\end{figure}

To analyze the results, we will consider that a stable region is one where the particles remained in orbit for the integration period considered, and obviously were not ejected or collided with the body.
Thus, Fig. \ref{fig:beta anel} illustrates the initial position of the 10 thousand particles flat cloud at the beginning of the simulation (yellow dots) and those that survived after the 10 years of integration (blue dots). Note the presence of stable regions only around equilibrium points $E_2$ and $E_4$, whose topological structures differ from points $E_1$ and $E_3$ (Saddle-Centre-Centre). In addition, the region around point $E_2$ has an angular amplitude of $\sim$ 15.1$^{\circ}$, while for $E_4$ it is $\sim$ 16.7$^{\circ}$, evidencing the no symmetry between them. Also note that near the stable region around point $E_2$ there is a very small disconnected region whose orbits of particles at this location exhibit unstable behavior (blue). In this displaced region, we identified a quasi-periodic behavior, suggesting a possible resonance, as shown in Fig. \ref{fig: resonance}. However, this question demands further studies.

In order to obtain a more realistic result, we performed another numerical simulation assuming the same initial conditions given above and added to the system the Alpha and Gamma components. In this simulation, these last two bodies are treated as mass points. For comparison purposes, Fig. \ref{fig:beta anel} also illustrates the initial position of the particles that survived until the end of this simulation (red dots). Due to the gravitational influence of Alpha and Gamma, the stable region around the point $E_4$ decreased (angular amplitude $\sim$ 10.8$^{\circ}$), while the stable region around $E_2$ disappeared. We might also associate such reduction with the fact that points $E_2$ and $E_4$ have different topological structures, as we identified in the linear analysis of stability.

Finally, we opted not to perform simulations with a flat cloud of particles around Gamma, although it has equilibrium points away from its surface. Unlike the Beta body, which is further away from Alpha, Gamma is too close and due to its small mass has a weak gravitational field. This can be easily seen by comparing the geopotential intensity across the surface of Alpha and Gamma, as shown in Fig. \ref{fig:geo}. This energy computed for the largest component is at least 100 times stronger than that of the smallest component. Consequently, stable regions, if any, are strongly perturbed by Alpha, causing instability in the regions near the points. This contributes to a higher occurrence of particle collision and ejection during the numerical simulation.

\subsection{Guaranteed return speed}
\label{return speed}

As discussed earlier, zero-velocity curves delimit the location of equilibrium points in the gravitational field around the body. These contour lines produce energy values that limit the physical movement of a particle near the object, to a given Jacobi constant value. Thus, considering the body-fixed reference frame, the smallest value of the Jacobi constant ($J^*$), among those computed at the equilibrium point locations, taking into account the geopotential, produces zero-velocity curves that involve the entire asteroid in three-dimensional space, restricting the space that contains the body and space that does not contain the body. This surface, which satisfies $V(\pmb r) = J^*$, defines the rotational Roche Lobe on the body \citep{Scheeres2012}. Through this minimum energy, it is possible to obtain a lower limit of the guaranteed return speed, a velocity measured across the body surface that allows evaluating whether the movement of a particle is conditioned to the vicinity of the asteroid or not \citep{Scheeres2012}. The expression that computes this lower limit is given by
\begin{equation}
\label{eq: return speed}
v_{\rm ret} = \sqrt{-2(V(\pmb r)-J^*)},
\end{equation}
such this value will be set to zero if the evaluated geopotential on the body surface exceeds $J^*$. 

We are interested in mapping the behavior of guaranteed return speed across the surface of each component of the system 2001 SN$_{263}$. Therefore, let us assume the motion of a particle across the surface of the body under the energy $V(\pmb r)$. The particle can escape from the surface of the body if this energy is greater than $J^*$. However, if this energy is less than $J^*$ and the particle is moving inside the zero-velocity curve that surrounds the body, it will not be able to escape, and its movement will be restricted to the vicinity of the asteroid. Thus, under gravitational dynamics, a particle with a velocity lower than $v_{\rm ret}$ will be trapped inside the zero-velocity curve whose energy is $J^*$, and can only move across the surface of the body or in the environment very close to it. The $J^*$ values for the Alpha, Beta and Gamma components are $-0.70537$ m$^2$ s$^{-2}$, $-2.4999\times 10^{-2}$ m$^2$ s$^{-2}$ and $-1.1833\times 10^{-2}$ m$^2$ s$^{-2}$, respectively.

\begin{figure}
\begin{center}
\subfloat{\includegraphics*[trim = 0mm 7cm 0mm 0mm, width=\columnwidth, frame]{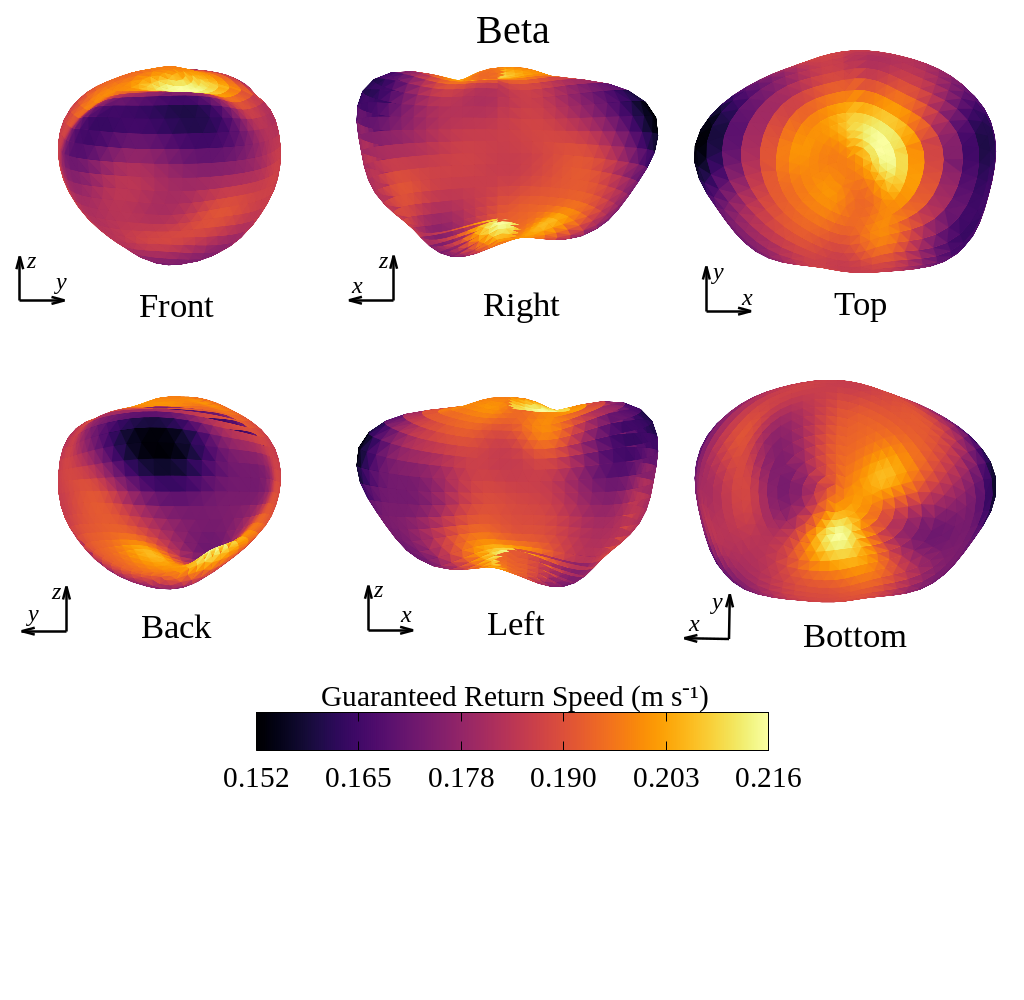}}\\
\subfloat{\includegraphics*[trim = 0mm 5.3cm 0mm 0mm, width=\columnwidth, frame]{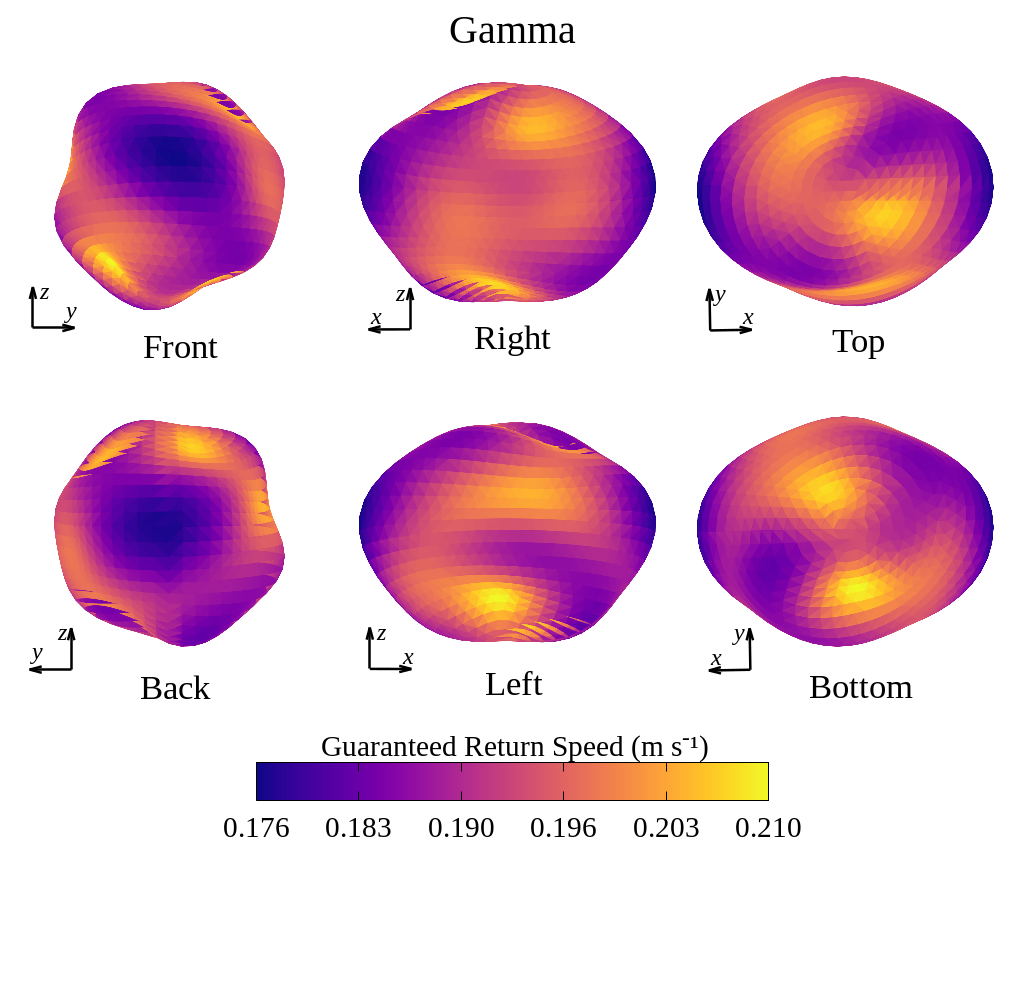}}
\end{center}
\caption{Guaranteed return speed $v_{\rm ret}$ computed across the surface of Beta and Gamma, respectively. Note that the colour scales are different for each body.}
\label{fig: return speed}
\end{figure}

In Fig. \ref{fig: pontos alpha} we present the rotational Roche Lobe about the Alpha body.
Note that the zero-velocity curve that encompasses the Alpha equilibrium points ends up crossing the body surface. This is a characteristic observed in spheroidal bodies with high rotational speed \citep{Scheeres2015, Scheeres2016}. Thus, it is always possible for a particle to escape from the Alpha surface, especially in regions at mid to upper latitudes whose energy intensity is greater than $J^*$, as shown in Fig. \ref{fig:geo} - Alpha. Consequently, we have $v_{\rm ret}=0$ throughout the body surface. So, Fig. \ref{fig: return speed} illustrates the return speed mapping across the Beta, and Gamma surfaces only.

Fig. \ref{fig: Beta and Gamma curves}, although showing a variation in the Jacobi constant, also records the rotational Roche Lobe of both Beta and Gamma. This refers to the curve whose $J^*$ is associated with the equilibrium point $E_3$ and is enveloping the entire body. And unlike for Alpa, this curve does not intersect the surface of the body.
The Beta component has the maximum values of $v_{\rm ret}$ in regions close to the poles (Top and Bottom views of Fig. \ref{fig: return speed} - Beta). While the minimum values are concentrated at the ends of the equatorial region, according to Front and Back views of Fig. \ref{fig: return speed} - Beta. However, the analysis of Fig. \ref{fig:speed} - Beta shows that a particle requires a speed of up to $0.153$ m s$^{-1}$ to boost its displacement from a higher point in the geopotential toward the lowest point. Note that this value is smaller than the lower limit of $v_{\rm ret}$ mapped in the polar regions (Top and Bottom views of Fig. \ref{fig: return speed} - Beta). At these locations a particle to escape from the Beta surface should have at least a speed of $0.216$ m s$^{-1}$. Therefore, a particle located within the velocity curve that encompasses Beta will be confined close to the body, as it does not have enough energy to escape.

Similar behavior occurs when analyzing Fig. \ref{fig: return speed} for the Gamma component. The mapped speeds across its surface, Fig. \ref{fig:speed} - Gamma, are at least about 35$\%$ smaller than the lower limits of $v_{\rm ret}$, Fig. \ref{fig: return speed} - Gamma.

\section{Mapping of impacts across the surface}
\label{mapa}

\begin{figure*}
\includegraphics[trim = -0.4cm -0.3cm 0cm 0cm,
width=13cm, frame]{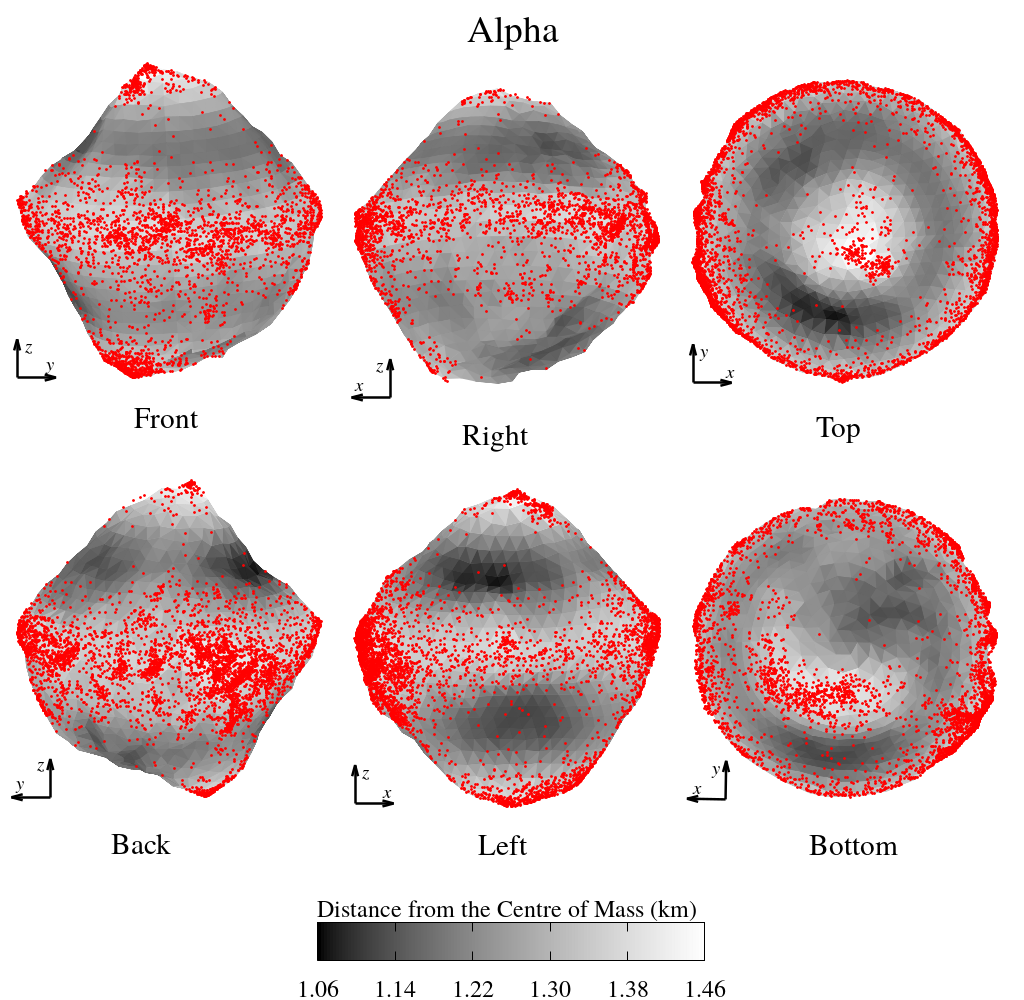}
    \caption{Mapping of the impact of particles across the surface of Alpha, from a cloud of particles initially orbiting this body.}
    \label{fig:mapa}
\end{figure*}

The works of \citet{kruger2003}, \citet{krivov2003}, and \citet{winter2018} discuss the collisions between planetary satellites and interplanetary dust particles that generate small materials, denominated ejecta, that initially tend to orbit the satellite. This same behavior is expected for smaller bodies, as in the case of asteroids.

\citet{Moura2019} considered a cloud of ejecta particles around the asteroid (16) Psyche and mapped the fall of these particles on the surface of the body. Then, they showed that certain regions tend to suffer more falls, a fact that is directly connected to the gravitational potential of an irregular object. Thus, we also investigated the distribution of the fall the ejecta particles on the surface of the system 2001 SN$_{263}$ bodies, with the interest of delimiting which regions suffer more impacts of particles and, possibly, to previse regions with an accumulation of materials.

We performed numerical simulations using the N-BoM integrator package again. The numerical integration of the equations of motion involves a system composed of Alpha, Beta, Gamma, and 10 thousand test particles. All the particles started with circular orbits and inclination from 0$^{\circ}$ to 90$^{\circ}$, distributed randomly. Values from 0$^{\circ}$ to 360$^{\circ}$ for the longitude of the ascending node and mean anomaly were randomly distributed among the particles, in order to create a cloud. The entire system was integrated for a timespan of 1 year, corresponding to $\sim58$ orbital periods of Beta and $\sim532$ orbital periods of Gamma, sufficient time for the proposed objective and validation of the results. In order to statistically evaluate the results of the simulations, we monitored throughout the integration period the particles that remained in orbit, those that were ejected, and those that collided with any of the system 2001 SN$_{263}$ bodies. In the event of a collision with any of the bodies, the impact position is recorded, and the particle is immediately removed from the simulation. As the goal is to map impacts across the surface of Alpha, Beta and Gamma, we plan individual simulations for each object: a cloud of particles is distributed in the vicinity each of the bodies, whose mass distribution is through mascons, whereas the gravitational influence of the other two bodies is accounted for by treating them as mass points.

In the simulations where the volume of Alpha is given by 62864 mascons, the particles were randomly distributed a few meters from the location of the equilibrium points, in a range from 1.5 km to 3.0 km. Statistically, we verified that about 68\% of the particles collided with the surface of Alpha. However, most of these impacts occurred in a timespan of less than 2.4 months, equivalent to $\sim511$ spin periods of Alpha. Fig. \ref{fig:mapa} shows the collisions mapped across the surface of Alpha, under different views. Considering that the particles were initially distributed in a cloud around the body, not all the regions on the surface were impacted. Fig. \ref{fig:mapa} identifies that the collisions occurred preferentially in the equatorial region of Alpha, with some impacts on the poles, according to Top and Bottom views. Since the color code gives the distance from the centre of mass of the body, we can observe that it is precisely in the regions with a higher altitude, those of lighter color, that the concentration of the collisions occurs although it is the region of the poles that exerts the greatest gravitational attraction, according to the Top and Bottom views of Fig. \ref{fig:acel} for Alpha, it is a fairly wide equatorial band that presents a larger number of impacts, exactly where the attraction is smaller. This occurrence may be related to the fact that the particles are also receiving the influence of the inner and outer bodies, which possibly contributes to the flow of collisions migrate to the equatorial region. Finally, we can conciliate the mapping of these falls with the slope angle mapping, which indicates the tendency of the movement of free particles on the surface of the object, as we saw in Section \ref{slope}. Thus, although most falls occur in the equatorial region, the variation of the slope angle shows that loose material on the surface of the body is prone to have its movement directed towards regions with low slope, that is, the poles (Top and Bottom views of Fig. \ref{fig:slope} for Alpha).

Investigating the simulations where the cloud of particles was distributed around Beta, whose gravitational attraction is given by 64259 mascons, we verified that a little more than 65\% of the particles fell on the surface of the body. While simulations with the cloud of particles around Gamma, modeled with 60342 mascons, resulted in a smaller number of impacts, below 45\%. In both cases, there was a homogeneous distribution of the impacts across the surface of the objects, excluding the existence of preferential regions.  Consequently, these impacts do not depend on locations with higher or lower altitude or even gravitational attraction. Thus, we will not present the graphs of these simulations, because they do not add any discrepant results.

\section{Final Comments}
\label{final}

In this study, we briefly discuss the shape of the triple system 2001 SN$_{263}$ components that were fitted to polyhedra with faces and vertices. Then, we investigate the gravitational potential of each object using the polyhedra method, proposed by \citet{wernerscheeres1996}. We define the geopotential \citep{Scheeres2016} in order to investigate the dynamic environment around and on the surface of the bodies.

We present the topographic characteristics in relation to the geometry and geopotential of each component. From the perspective of geometric topography, we discuss the results generated by the geometric altitude and tilt quantities, highlighting regions with high altitudes and the irregular relief of the surfaces of Alpha, Beta, and Gamma. The geopotential topography analysis included the mapping of the geopotential, slope, surface accelerations, and potential speed across the surfaces of the system 2001 SN$_{263}$. Such quantities allowed us to locate the highest and lowest points in the geopotential of the bodies. In addition, the intensity of this energy is at least 100 times greater on the surface of the largest component than in the other two components. The investigation of the slopes across the surface of Alpha suggests an accumulation of loose material around its equatorial region, enabling the collection by a space mission.

Next, the equilibrium points in the gravitational field of each object were located, and we also present the zero-velocity curves that illustrate the regions where these points can be found. We emphasize that Alpha has a relatively large number of equilibrium points, 12, which are located very close to its surface. We identified the topological structure of the equilibrium points of the system 2001 SN$_{263}$ through linear analysis of stability. In addition, we investigated the presence of stable regions around these equilibrium points. We distribute massless particles around each body and perform numerical simulations to determine the size of these regions. The results showed that only Beta has small stable regions around points $E_2$ and $E_4$. And that under the gravitational influence of the other two components the stable region around point $E_4$ decreased angularly, while the region around $E_2$ disappears.

Finally, we integrate a cloud of massless particles orbiting the vicinity of each component of the triple system to map the impacts across Alpha, Beta, and Gamma surfaces. We present a statistical analysis regarding the distribution of these impacts across the body surface, showing possible preferential regions. We emphasize that for Alpha the fall of the particles on its surface occurs mainly in the equatorial region. 

All the discussions presented in this study, whose objective was to analyze the surface characteristics and dynamical environment of the triple system 2001 SN$_{263}$, might help in the planning of the first Brazilian space mission. Moreover, it may contribute to the investigation of other multiple asteroid systems, and perhaps to motivate future space missions in order to understand these unique systems a little more.


\section*{Acknowledgements}

This study was financed in part by the Coordenação de Aperfeiçoamento de Pessoal de Nível Superior - Brasil (CAPES) - Finance Code 001, Fundação de Amparo à Pesquisa do Estado de São Paulo (FAPESP) - Proc. 2016/24561-0 and Proc. 2017/26855-3, Conselho Nacional de Desenvolvimento Científico e Tecnológico (CNPq) - Proc. 305210/2018-1. 

We also would like to thank Dr. T. M. Becker for all the help with the data.

\section*{ORCID iDs}
O. C. Winter \orcidicon{0000-0002-4901-3289} \href{https://orcid.org/0000-0002-4901-3289}{https://orcid.org/0000-0002-4901-3289}\\
G. Valvano \orcidicon{0000-0002-7905-1788} \href{https://orcid.org/0000-0002-7905-1788}{https://orcid.org/0000-0002-7905-1788}\\
T. S. Moura \orcidicon{0000-0002-3991-8738} \href{https://orcid.org/0000-0002-3991-8738}{https://orcid.org/0000-0002-3991-8738}\\
G. Borderes-Motta \orcidicon{0000-0002-4680-8414} \href{https://orcid.org/0000-0002-4680-8414}{https://orcid.org/0000-0002-4680-8414}\\
A. Amarante \orcidicon{0000-0002-9448-141X} \href{https://orcid.org/0000-0002-9448-141X}{https://orcid.org/0000-0002-9448-141X}\\
R. Sfair \orcidicon{0000-0002-4939-013X} \href{https://orcid.org/0000-0002-4939-013X}{https://orcid.org/0000-0002-4939-013X}\\



\bibliographystyle{mnras}
\bibliography{bilbli} 

\appendix
\section{Eigenvalues}
\label{eigenvalues}
Eigenvalues of the equilibrium points presented in Table \ref{table: loc}, referring to the triple system 2001 SN$_{263}$ components.
\begin{table*}
\centering
\caption{Eigenvalues ($\lambda_n \times10^{-4}$) of equilibrium points in the gravitational field of the triple system 2001 SN$_{263}$ components and their topological structures.}
\label{table:autovalores} 
\begin{tabular}{c|r|r|r|l}
\hline\hline
\multicolumn{5}{|c|}{Alpha}\\
\hline\hline
\multicolumn{1}{c}{{Point}} & \multicolumn{1}{c}{{$\lambda_{1,2}$}} & \multicolumn{1}{c}{{$\lambda_{3,4}$}} & \multicolumn{1}{c}{{$\lambda_{5,6}$}} & \multicolumn{1}{c}{{Topological structure}} \\
\hline
    $E_1$ & $\pm$ $4.552$ & $\pm$ $5.344i$ & $\pm$ $6.641i$ & Saddle-Centre-Centre\\
     \hline
    $E_2$ & $-2.255$ $\pm$ $3.534i$ & $2.255$ $\pm$ $3.534i$ & $\pm$ $6.093i$ & Sink-Source-Centre\\
     \hline
    $E_3$ & $\pm$ $2.653$ & $\pm$ $4.580i$ & $\pm$ $6.164i$ & Saddle-Centre-Centre\\  
     \hline
    $E_4$ & $-2.675$ $\pm$ $3.369i$ & $2.675$ $\pm$ $3.369i$ & $\pm$ $6.599i$ & Sink-Source-Centre\\ 
     \hline
    $E_5$ & $\pm$ $1.078$ & $\pm$ $10.20i$ & $\pm$ $6.421i$ & Saddle-Centre-Centre\\  
     \hline
    $E_6$ & $-3.108$ $\pm$ $4.222i$ & $3.108$ $\pm$ $4.222i$ & $\pm$ $5.966i$ & Sink-Source-Centre \\
     \hline
    $E_7$ & $\pm$ $7.103$ & $\pm$ $8.406i$ & $\pm$ $5.632i$ & Saddle-Centre-Centre\\    
     \hline
    $E_8$ & $-2.955$ $\pm$ $3.889i$ & $2.955$ $\pm$ $3.889i$ & $\pm$ $6.257i$ & Sink-Source-Centre\\ 
     \hline
    $E_9$ & $\pm$ $5.432$ & $\pm$ $6.026i$ & $\pm$ $6.718i$ & Saddle-Centre-Centre\\  
     \hline
    $E_{10}$ & $-3.076$ $\pm$ $4.082i$ & $3.076$ $\pm$ $4.082i$ & $\pm$ $6.126i$ & Sink-Source-Centre \\ 
     \hline
    $E_{11}$ & $\pm$ $4.612$ & $\pm$ $3.540i$ & $\pm$ $7.789i$ & Saddle-Centre-Centre\\ 
     \hline
    $E_{12}$ & $-2.599$ $\pm$ $4.037i$ & $2.599$ $\pm$ $4.037i$ & $\pm$ $5.731i$ & Sink-Source-Centre\\
\hline \hline
\multicolumn{5}{c}{Beta}\\
\hline \hline
\multicolumn{1}{c}{{Point}} & \multicolumn{1}{c}{{$\lambda_{1,2}$}} & \multicolumn{1}{c}{{$\lambda_{3,4}$}} & \multicolumn{1}{c}{{$\lambda_{5,6}$}} & \multicolumn{1}{c}{{Topological structure}} \\
\hline
    $E_1$ & $\pm$ $0.552$ & $\pm$ $1.367i$ & $\pm$ $1.347i$ & Saddle-Centre-Centre\\
     \hline
    $E_2$ & $-0.027$ $\pm$ $0.904i$ & $0.027$ $\pm$ $0.904i$ & $\pm$ $1.321i$ & Sink-Source-Centre\\
    \hline
    $E_3$ & $\pm$ $0.677$ & $\pm$ $1.398i$ & $\pm$ $1.372i$ & Saddle-Centre-Centre\\
     \hline
    $E_4$ & $\pm$ $1.064i$ & $\pm$ $0.710i$ & $\pm$ $1.319i$ & Centre-Centre-Centre\\
\hline \hline
\multicolumn{5}{c}{Gamma}\\
\hline\hline
\multicolumn{1}{c}{{Point}} & \multicolumn{1}{c}{{$\lambda_{1,2}$}} & \multicolumn{1}{c}{{$\lambda_{3,4}$}} & \multicolumn{1}{c}{{$\lambda_{5,6}$}} & \multicolumn{1}{c}{{Topological structure}} \\
\hline
    $E_1$ & $\pm$ $0.287$ & $\pm$ $1.090i$ & $\pm$ $1.077i$ & Saddle-Centre-Centre\\
     \hline
    $E_2$ & $\pm$ $1.022i$ & $\pm$ $0.291i$ & $\pm$ $1.065i$ & Centre-Centre-Centre\\
     \hline
    $E_3$ & $\pm$ $0.290$ & $\pm$ $1.090i$ & $\pm$ $1.077i$ & Saddle-Centre-Centre\\
     \hline
    $E_4$ & $\pm$ $1.022i$ & $\pm$ $0.294i$ & $\pm$ $1.065i$ & Centre-Centre-Centre\\
\hline \hline
\end{tabular}
\end{table*}

\newpage
\bsp	
\label{lastpage}
\end{document}